\documentstyle[11pt,epsfig]{article}
\newcommand{\resection}[1]{\setcounter{equation}{0}\section{#1}}

\newcommand{\scs}{\scriptsize}

\def\Im{{\rm Im\,}}

\def\Sq{S_{q^{-3}}}
\def\St{S_{\tilde{q}^{-3}}}
\def\qt{\tilde{q}}

\newcommand{\Bbb}{\cal}
\newcommand{\beq}{\begin{equation}}
\newcommand{\eeq}{\end{equation}}
\newcommand{\bdm}{\begin{displaymath}}
\newcommand{\edm}{\end{displaymath}}
\newcommand{\bea}{\begin{eqnarray}}
\newcommand{\eea}{\end{eqnarray}}

\newcommand{\tensor}{\otimes}

\newcommand{\ib}{\bar{\imath}}
\newcommand{\jb}{\bar{\jmath}}

\begin{document}
\setcounter{page}{0}
\topmargin 0pt
\newpage
\setcounter{page}{0}
\begin{titlepage}
\vspace{0.5cm}
\begin{flushright}{\bf SWAT 95-96/121}\\
{\bf PRELIMINARY}
\end{flushright}
\begin{center}
{\Large 
{Exact quantum S-matrices for solitons in simply-laced affine Toda
field theories}} \\
\vspace{1cm}
{\large Peter R. Johnson}\footnote{current address: CEA-Saclay, 
Service de Physique Th\'eorique, 91191 Gif-sur-Yvette Cedex, France} \\
\vspace{1.0cm}
{\em Department of Physics,\\
University of Wales at Swansea,\\
Singleton Park,\\
Swansea,\\
SA2 8PP, Wales, UK.} \\ \end{center}
\vspace{1cm}
%{\it Address for correspondence}: Dr. P.R. Johnson, CEA-Saclay
%adress given by footnote. E-mail: Johnson@spht.saclay.cea.fr,
%Telephone: (33) 1 69 08 70 06
%\vspace{0.5cm}
\setcounter{footnote}{0}\baselineskip=14pt 
\begin{center}{\bf Abstract} \\ \end{center}
We obtain exact solutions to the quantum S-matrices for solitons in
simply-laced affine Toda field theories, except for certain factors
of simple type which remain undetermined in some cases. 
These are found by postulating solutions which are consistent with the 
semi-classical limit, $\hbar\rightarrow 0$, and the known time delays for a 
classical two soliton interaction. This is done by a `$q$-deformation' 
procedure, to move from the classical time delay to the exact S-matrix, by
inserting a special function called the `regularised' quantum
dilogarithm, which only holds when $|q|=1$.
It is then checked that the solutions satisfy the
crossing, unitarity and bootstrap constraints of S-matrix
theory. These properties essentially follow from analogous properties
satisfied by the classical time delay. Furthermore, the lowest mass
breather S-matrices are computed by the bootstrap, and it is shown
that these agree with the particle S-matrices known already in the
affine Toda field theories, in all simply-laced cases. 

\vspace{0.5cm}
\noindent{\it MSC Classification codes}: 81Q20, 81R50, 81U20, 58F07

\vspace{0.5cm}
\end{titlepage}\baselineskip=14pt 

\resection{Introduction}
The affine Toda field theories are integrable systems, depending on a
Lie algebra $g$, which admit soliton solutions interpolating the
degenerate vacua in the potential, provided the coupling constant 
$\beta$ in the model is chosen to be purely imaginary. 
In the simply-laced cases, the
classical single soliton solutions can be grouped together into
species or equivalently associated with a node of the Dynkin diagram
of $g$. All the solitons of a particular species have the same mass,
but different topological charges. However there can be species with
the same mass.  The topological charges of these single solitons of a
particular species are contained in the weights of the fundamental
representation associated with that node. Furthermore, the fusing rule
$i+j\rightarrow k$ 
(Dorey's fusing rule \cite{Dorey1}), when two solitons of species $i$ and $j$ 
 fuse into a third single
soliton of species $k$, 
is always contained within the Clebsch-Gordon decomposition of
the tensor product of the two 
fundamental representations $V^i$ and $V^j$. These facts suggest that
the S-matrix for the collision of two solitons of species $i$ and $j$
in the quantum theory is given by the R-matrix of some affine quantum
group (up to a multiplicative scalar factor) intertwining the tensor
product of the two fundamental representations $V^i$ and $V^j$. This
means that the S-matrix must satisfy the quantum Yang-Baxter equation
with spectral parameter -- a necessary property of an S-matrix in an
integrable theory, since the infinite number of conservation laws
suggest that an $n$-particle S-matrix will factorise into 
products of the two-particle result. This can be done in two different
ways for the 3-soliton case, implying the quantum Yang-Baxter
equation.

This R-matrix approach to finding the S-matrix was first discussed in
\cite{H,H2} for the $A_n$ case, and there is also a discussion in
\cite{D}. 
The result for the scalar factor in the sine-Gordon case ($g=su(2)$)
 is the well
known Zamolodchikov-Zamolodchikov solution \cite{ZZ}, which is given
as an infinite product of gamma functions, or a double infinite
product (\ref{eq: full}). We can easily pass between these two
descriptions by using the Weierstrass product formula for the gamma
function.  There is the result, also
for sine-Gordon, 
due to Karowski, Thun, Truong and Weisz \cite{Karowski}, given as a neater
integral formula (\ref{eq: Karowski}), which, on expansion, 
can be shown to be the same as
the formula given by the Zamolodchikov's. 

The purpose of this paper is to announce solutions to the
scalar factors which are built out of a
special function $S_{q^{-h}}(w)$ called the `regularised' quantum
dilogarithm (\ref{eq: rqd}). This special function 
has recently been introduced by Faddeev \cite{Fad}. 
For sine-Gordon, we can easily reproduce the Karowski formula
\cite{Karowski}. These solutions have only so far 
appeared in \cite{thesis},
which has been circulated privately. It is not our intention to study
the quantum groups and R-matrices, or their relation with quantum
integrable systems, in any great detail. We shall take the R-matrices
that we shall need from the literature.

There are two advantages in writing the scalar factors in this
way. The first is that the semi-classical limit $\hbar\rightarrow 0$,
can be taken very efficiently. The overall effect of a classical two
soliton interaction is a time delay or phase shift. In \cite{FJKO}, it
is shown that the time delays for simply-laced affine Toda field
theories have a vertex operator origin, and are given by
$(2h/|\beta|^2)\log X^{jk}(\theta)$, where $X^{jk}(\theta)$
arises from the normal ordering of two vertex operators. It is well
known \cite{JW}, that the S-matrix $S^{jk}(\theta)$, must have the 
semi-classical limit when
$\hbar\rightarrow 0$, for ${\rm Re\ }(\theta)>0$,
\beq S^{jk}(\theta)\Bigl|_{\hbar\rightarrow 0}\sim
\exp\Bigl({2hi\over\hbar|\beta|^2}\int_0^\theta\,d\theta'\log
X^{jk}(\theta')\Bigr).\label{eq: intro}\eeq
Considerable manipulations are required to check this 
in the formalism involving infinite products of gamma functions
\cite{ZZ,H}, and also for the integral formula due to Karowski {\it et al} 
\cite{Karowski}, even for sine-Gordon.

This way of writing the solution means that new solutions can be
obtained by extrapolating back from the classical $X^{ij}(\theta)$, in
cases where solutions are not yet known (the $D$ and $E$ series of
algebras). However, we do require input from the R-matrices, which are
known for a few of the fundamental representations, in these cases, as
discussed in section 11.
The method in this paper also provides an alternative description of the $A_n$
theories \cite{H}, 
where the crossing, unitarity, and bootstrap
constraints are clearly 
demonstrated. The extrapolation can be thought of as 
`$q$-deforming' $X^{ij}(\theta)$ to some $X_q^{ij}(\theta)$, 
by the insertion of appropriate quantum
dilogarithms, and then the S-matrix is given by
$$S^{ij}(\theta)={X_q^{ij}(\theta)\over X_q^{ij}(-\theta)},$$
modulo the simple factors needed for the different topological
charges, which are provided by the R-matrix.

The second advantage is that the crossing, unitarity and bootstrap
properties of the S-matrix, which have to be checked, essentially
follow from the analogous properties satisfied by the classical
$X^{ij}(\theta)$. Again considerable manipulations with the infinite
products of gamma functions, which are avoided in the formalism
presented in this paper, have to be made in order to check
these properties, in the other formalism\cite{ZZ,H}. 
Because of these two simplifications it is possible
to essentially write down the general solution for an arbitrary
simply-laced algebra, modulo some special factors, and to check the
semi-classical limit, and the crossing, unitarity, and bootstrap
conditions in a general uniform manner.

However it should be noted that the quantum dilogarithm itself,
$S_{q^{-h}}(w)$, can be written as an infinite product of gamma
functions (\ref{eq: gammas}), or alternatively as a double infinite
product (\ref{eq: double2}), and this is how the pole structure of
the proposed solutions is studied. If we wanted to, 
this would allow us to make contact
with the previous solutions known in the $A_n$ case.

We shall also show in section 14, that the lowest mass breather
S-matrices are the same as the Toda particle
S-matrices\cite{BCDS,Dorey2}, for the real coupling Toda theories,
after the analytic continuation in the coupling $\beta\rightarrow
i\beta$. This is to be expected if the solutions found are indeed
correct, and if we interpret the lowest mass breather $b_i$, made up of a
bound state of a soliton with species $i$ and the anti-soliton
$\ib$, with the particle $i$
of the quantum field theory (which is still present in the imaginary
coupling theories). This is an important independent check on the
solutions. However it is remarkable to the author that the result
survives the analytic continuation $\beta\rightarrow i\beta$.
This result was known for the $su(2)$ case in \cite{Fad_Kor}, and
Gandenberger studied the $su(3)$ case \cite{Gandenberger}, using
Hollowood's formula \cite{H}. Here we will employ the general formula
for the particle S-matrix (\ref{eq: Doreys_formula})  due to Dorey 
\cite{Dorey2,Fring}, which holds for an arbitrary simply-laced Lie
algebra $g$.
 
\vskip 0.2in

\noindent {\it Contents}:\\ \\ 
1. Introduction \\
2. The affine Toda field theories and their classical soliton solutions\\
3. Time delays in affine Toda field theories and their S-matrix like
properties\\
4. Axiomatics of S-matrix theory: crossing, unitarity and the bootstrap\\
5. The quantum Yang-Baxter equation\\
6. The `regularised' quantum dilogarithm\\
7. The sine-Gordon solution (Zamolodchikov-Zamolodchikov)\\
8. The semi-classical limit for sine-Gordon\\
9. The $su(3)$ solution\\
10. The general solution\\
11. Comparison with the known R-matrices\\
12. The poles of the general solution\\
13. The general soliton bootstrap\\
14. The lowest mass breather bootstrap, and comparison with the Toda
particle S-matrices\\
15. Discussion and conclusions\\
\resection{The affine Toda field theories and their classical soliton
solutions}

The affine Toda field theories are based on a Lie algebra $g$, and
have a non-linear interaction term built from the root system of
$g$. Let $\alpha_i:\ i=1,\ldots,r$, where $r$ is the rank of $g$, be
the simple roots, and let $\alpha_0=-\psi$, where $\psi$ is the highest
root. Let $h$ be the Coxeter number of $g$. Expand $\psi$ in terms of
the simple roots
$${\psi\over\psi^2}=\sum_{i=1}^r m_i{\alpha_i\over\alpha_i^2},$$
defining $m_i$, $i=1,\ldots r$, which are positive integers. 
Define $m_0=1$, and then $\sum_{i=0}^r
m_i{\alpha_i\over\alpha_i^2}=0$. Now the affine Toda field theories,
for an $r$-component scalar field $u$, have equations of motion
\beq
{\partial^2 u\over\partial t^2}-{\partial^2 u\over\partial x^2}
+{4\mu^2\over\beta}\sum_{i=0}^r
m_i{\alpha_i\over\alpha_i^2}e^{\beta\alpha_i.u} =0.\label{eq: Toda}\eeq
Observe that any constant $u$ such that $e^{\beta u.\alpha_i}=1$ or
alternatively $u\in {2\pi\over |\beta|}\Lambda_W(g^\vee)$, the weight
lattice of the co-root algebra, is a solution, provided $\beta$ is
chosen to be purely imaginary. Thus we expect soliton solutions to
exist which smoothly interpolate these vacua at
$x\rightarrow\pm\infty$. Hollowood found some exact soliton solutions in
the $A_{n-1}$ case \cite{H3}, using the Hirota method. There is also a
discussion of these solutions, and excited modes around them,
discovered using the inverse scattering method in \cite{BJ}.

For $\lambda_i$ a fundamental weight, defined by ${2\lambda_i
\cdot\alpha_j\over\alpha_j^2}=\delta_{ij}$, Hollowood found in
\cite{H3}, for a constant $Q\in{\Bbb C}\,$,
\beq e^{-\beta\lambda_i\cdot u}={1+Q\omega^{ij}W\over
1+QW},\quad\omega=e^{2\pi i\over n},\label{eq: jumps}\eeq
where $W=e^{2\mu\sin({\pi j\over n})(e^{-\theta}x_+-e^\theta x_-)}$,
with $x_\pm=t\pm x$.

The integer $j$ denotes the {\it species} of soliton, which can be
associated with the fundamental representation $V^j$ of $g$, which is
defined to have highest weight $\lambda_j$. Indeed,
it has been checked explicitly in the $A_{n-1}$ case \cite{McGhee}
that the topological charges of a
soliton of species $j$ are weights of $V^j$. The topological charge is
defined to be the difference between $u$ as $x\rightarrow\infty$ and
$x\rightarrow-\infty$, and the charge of (\ref{eq: jumps}) jumps to
different values as we vary the phase of $Q$.
However there still remains the problem that, for $n\ge 4$,
not all the weights are filled by classical single soliton solutions,
and this begs the question whether there are some missing classical
solitons. In what follows we must assume that these missing classical 
solutions actually exist, or that states are created in the quantum theory, so
that a multiplet of solitons of species $j$ transform in the
fundamental representation $V^j$. Otherwise the representations would
have to be restricted to physical spaces,
 to fit in with what is known classically, and somehow
incorporated into the residues at the poles in the S-matrix, and the
unitarity sum, $S^{ij}(\theta)S^{ji}(-\theta)$, of the S-matrix.
These spaces would no longer be representations.

Soliton solutions for the other simply-laced cases were found by
Olive, Turok and Underwood \cite{OTUa,OTUb}. Again there is the
problem that there are many missing charges, but the charges of a
given species are still contained in the appropriate fundamental
representation, as in the $A_n$ case. Within their abstract formalism,
the true origin of the different sorts of species of soliton, each
species being associated with a node of the Dynkin diagram of $g$, was
discovered. It was shown that the solitons of a given species  have
the same mass, and crucially the solitons of species $i$ have the same
mass as the particle associated with the $i^{\rm th}$ node of the
Dynkin diagram in the quantum field theory, up to some global
renormalisation of the particle masses. Furthermore, a certain
coefficient, $X^{ij}(\theta)$,
which we shall call the interaction function, and which plays an important
r\^ole in the multi-soliton solutions, arose from the normal ordering 
of two vertex operators $F^i(z)$ and $F^j(w)$. Here $\theta$ is the
relative rapidity of the two solitons of species $i$ and $j$.
$$F^i(z)F^j(w)=X^{ij}(\theta):F^i(z)F^j(w):.$$
 Each vertex operator
$F^i(z)$ `creates' a soliton in the formalism. An explicit formula
is given for $X^{jk}(\theta)$:
\beq X^{jk}(\theta)=\prod_{p=1}^h\Bigl(1-e^\theta e^{{\pi i\over
h}(2p+{c(j)-c(k)\over 2})}\Bigr)^{\gamma_j\cdot\sigma^p\gamma_k}.\label{eq:
X} \eeq
Here $c(i)=\pm 1$ is the `colour' of the node $i$ of the Dynkin
diagram of $g$, which can be bi-coloured in a certain way. $\sigma$ is
a special element of the Weyl group known as the Coxeter element, 
$\sigma^h=1$, and
$\gamma_i=c(i)\alpha_i$. The Coxeter element $\sigma$ partitions the
root system into $r$ orbits of $h$ elements, with $\gamma_i$
a representative of each orbit, for $i=1,\ldots, r$.
Also $\gamma_j\cdot\sigma^p\gamma_k=\pm 2,\pm
1, 0$. These integers are known explicitly 
for each algebra $g$, and are
summarised in tables later on in this paper.

As discussed in \cite{OTUb}, 
$X^{jk}(\theta)$ has poles at some purely imaginary values of
$\theta$. These occur if $\gamma_j\cdot\sigma^p\gamma_k=-2$, or if 
$\gamma_j\cdot\sigma^p\gamma_k=-1$. The latter case is equivalent to 
$\gamma_j+\sigma^p\gamma_k=\sigma^q\gamma_r$, for some integers $q$
and $r$. This is Dorey's fusing rule for the fusing of particles in
the real coupling affine Toda field theories\cite{Dorey1,Fring},
$j+k\rightarrow r$. It is
no surprise that we get a classical fusing of solitons in this case,
in the sense that if we analytically continue the relative rapidity
$\theta$ of the two soliton solution, made up of species $j$ and $k$
solitons,  to this pole in $X^{jk}(\theta)$, and subject to
renormalising the constants $Q$ which occur in the solution, we get a
single soliton solution of species $k$. We expect that in the quantum
theory the solitons will couple according to this Dorey fusing rule.

The case $\gamma_j\cdot\sigma^p\gamma_k=-2$, implies that $\gamma_j+ 
\sigma^p\gamma_k=0$, or equivalently that the two solitons are
anti-solitons of each other. This pole in $X^{jk}(\theta)$ 
corresponds to the breather, and
actually lies at $\theta=i\pi$.

It should be noted that the zeroes of the interaction function,
$X^{jk}(\theta)$, are also important and lead to additional excited
modes of oscillation  around the standard single solitons\cite{BJ}.

\resection{Time delays in affine Toda field theories and their S-matrix like
properties}
The two soliton solution in the $A_{n-1}$ theories, of species $j$ and
$k$, is \cite{OTUb}
\beq
e^{-\beta\lambda_i\cdot u}={1+Q_j\omega^{ij}W_j+Q_k\omega^{ik}W_k+
X^{jk}(\theta)Q_jQ_k\omega^{i(j+k)}W_jW_k\over
1+Q_jW_j+Q_kW_k+
X^{jk}(\theta)Q_jQ_kW_jW_k}\label{eq: 2sol},\eeq
with $W_r=e^{2\mu\sin({\pi r\over n})(e^{-\theta_r}x_+-e^{\theta_r} x_-)}$,
and $X^{jk}(\theta)$ is defined by equation (\ref{eq: X}), with
$\theta=\theta_j-\theta_k$. Following the argument in \cite{FJKO},
suppose that the velocity of the $r^{\rm th}$ soliton is $v_r$, and
that $v_j>v_k$. We track the $j^{\rm th}$ soliton by looking at a
neighbourhood in space-time of $x\sim v_j t$. In this neighbourhood
$W_j$ is of order one, and for $\kappa>0$, 
\bea W_k\sim e^{\kappa(v_j-v_k)t}&\rightarrow &\infty,{\rm\ \ as\ }t\rightarrow \infty \cr\cr
&\rightarrow &0,{\rm\ \ as\ }t\rightarrow -\infty, \eea
so in the limit $t\rightarrow\infty$, the two soliton solution becomes
\bea e^{-\beta\lambda_i.u}&=&{Q_k\omega^{ik}W_k+X^{jk}(\theta)Q_j Q_k 
\omega^{i(j+k)}W_jW_k\over Q_k W_k+X^{jk}(\theta)Q_jQ_kW_jW_k}\cr\cr
&=&\omega^{ik}\Biggl({1+Q_jX^{jk}(\theta)\omega^{ij}W_j\over
1+Q_jX^{jk}(\theta)W_j}\Biggr),
\eea
and in the limit $t\rightarrow -\infty$, 
$$e^{-\beta\lambda_i.u}={1+Q_j\omega^{ij}W_j\over
1+Q_jW_j}.$$
We
see that the overall affect of soliton $k$ on the progress of soliton
$j$ through the interaction is $Q_j\rightarrow Q_jX^{jk}(\theta)$.
Since the position of the soliton is proportional to $\log|Q_j|$, the
time delay or phase shift is proportional to $\log X^{jk}(\theta)$.

This result is particularly  simple for the $A_{n-1}$ theories, where
the single solitons have only one power of $W$ in the numerator and
denominator of $e^{-\beta\lambda_i.u}$. However in the remaining
simply-laced theories there are higher powers of $W$, and the
coefficients of the intermediate powers are not determined explicitly
in the abstract formalism of Olive, Turok, and Underwood \cite{OTUa,
OTUb}. It is therefore fortunate that these intermediate powers do not
affect the asymptotic behaviour in time of the solutions (where we
have seen that $W\rightarrow 0$, or $W\rightarrow\infty$), and that
the time delay is determined completely be the ordering of the vertex
operators, giving rise to $X^{jk}(\theta)$. In \cite{FJKO}, it is
demonstrated that the time delay $\Delta(\theta)$, for all
simply-laced theories, is still (for ${\rm Re\ }(\theta)>0$)
$$\Delta(\theta)=-{2h\over |\beta|^2}\log X^{jk}(\theta).$$
In \cite{FJKO}, it is also shown that $X^{jk}(\theta)$ is real for
$\theta$ real, and $0<X^{jk}(\theta)<1$, so that the force between
distinguishable solitons must be attractive. 
In the following, we shall need the classical `unitarity' condition \cite{FJKO}
\beq X^{jk}(\theta)=X^{jk}(-\theta).\label{eq: class_unitarity}\eeq
Also in \cite{FJKO}, a
crossing property of $X^{jk}(\theta)$ is discussed (crossing has no
meaning classically) in anticipation of
its relevance to the crossing property of the S-matrix,
$S^{jk}(i\pi-\theta)=S^{\jb k}(\theta)$. Note that a smooth  analytic
continuation $\theta\rightarrow i\pi-\theta$ is implicit in this crossing.
We reproduce the argument here, since a modified form of the argument
will be needed for the crossing of the exact S-matrix, and it will be
useful to refer back.

Recall that the anti-soliton species $\jb$ to the species $j$ is
defined by $\gamma_j+\sigma^p\gamma_{\jb}=0$, in fact the integer $p$ which
achieves this is known to be $p=-\frac h2-\frac{c(j)-c(\jb)}4$, see 
\cite{Fring}. Then $X^{jk}(\theta)$ satisfies the following crossing
property:
\beq X^{jk}(\theta+ i\pi)=X^{\jb k}(\theta)^{-1} \label{eq: class_cross}.\eeq
The proof is:
\begin{eqnarray*}
X^{jk}(\theta+i\pi)&=&\prod_{p=1}^h\biggl(e^{-\theta}-e^{{i\pi\over
h}(2p+h+ {c(j)-c(k)\over
2})}\biggr)^{-\gamma_{\jb}\cdot\sigma^{p+\frac h2+{c(j)-c(\jb)\over
4}}\gamma_k} \cr\cr
&=&\prod_{p'=1}^h\biggl(e^{-\theta}-e^{{i\pi\over
h}(2p'+ {(c(\jb)-c(k))\over
2})}\biggr)^{-\gamma_{\jb}\cdot\sigma^{p'}\gamma_k} \cr\cr
&=&X^{\jb k}(\theta)^{-1}. \qquad\end{eqnarray*}
Note that we do not have the expected $X^{jk}(i\pi-\theta)=X^{\jb
k}(\theta)$. This is because, in checking the crossing property of the
semi-classical limit (\ref{eq: intro}), we must take the crossing
condition $\theta\rightarrow  i\pi +\theta$, rather than 
$\theta\rightarrow i\pi-\theta$, so that we do not analytically
continue through the imaginary $\theta$ axis \cite{Coleman}. During
the limit, poles accumulate on the imaginary axis, which becomes a
natural boundary when the limit is taken. There are two different
expressions for the semi-classical limit (\ref{eq: intro}), for 
${\rm Re\ }(\theta)>0$ and ${\rm Re\ }(\theta)<0$,
which are not analytic continuations of each other\footnote{I would
like to thank David Olive for discussions on this point.}. The best
that we can do is to check the crossing property by staying within  
${\rm Re\ }(\theta)>0$, say, by the analytic continuation 
$\theta\rightarrow i\pi+\theta$, and then relate the point $i\pi
+\theta$ to $i\pi-\theta$ by the Hermitian analyticity property 
$S(i\pi +\theta)^\dagger = S(i\pi - \theta)$, for $\theta$ real
\cite{Olive}. The
complex conjugation reverses the sign of $\log X^{\jb k}(\theta)$.

Now $X^{jk}(\theta)$ has other properties which are remeniscent of the
S-matrix. The poles of $X^{jk}(\theta)$ have already been discussed in
the context of the classical fusing of solitons, and
breathers. However it is also clear that the simple poles (due to
fusing) are in precisely the same positions on the physical strip as
the pole due to the fusing of particles in the exact particle Toda
S-matrix $S^{jk}(\theta)$ \cite{BCDS,Dorey2}. It should be noted that
the fusing angles $U^r_{jk},U^j_{kr}$, and $U^k_{rj}$, for the fusing
$j+k\rightarrow \bar{r}$, are the same for particles in the particle
S-matrices, and for ground state solitons in the ground state soliton
S-matrices. This is so, provided the classical soliton masses all
receive the same quantum corrections in the quantum theory, that is,
they are all rescaled by the same constant. For the simply-laced
theories, this is believed to be true, see \cite{H_correction} for the
$A_n$ case, and the same methods as in \cite{H_correction} are used for the
other simply-laced cases in \cite{MW}. However, for the
non-simply-laced cases, which are not discussed in this paper, it is
believed that the masses do not all receive the same correction \cite{MW}.

If the pole in
$X^{jk}(\theta)$ is simple and due to the fusing $j+k\rightarrow \bar{r}$, 
using the notation of section 4 for the fusing angles,
and referring forward to that section,
the Toda particle masses
$M_j$ obey the equation, from \cite{Fring},
$$q(1)\cdot\sigma^{p}(\gamma_j)=iM_j e^{-{i\pi\over
h}(2p+{(1-c(j))\over 2})},$$
here  $\sigma q(1)=\omega q(1)$.
Hence from the fusing rule
$\gamma_j+\sigma^p\gamma_k=\sigma^q\gamma_r$,
we have
\beq M_j+M_ke^{iU^r_{jk}}=M_{\bar{r}}e^{i\bar{U}^k_{jr}}, 
\label{eq: Masses} \eeq
where
$$U^r_{jk}=-\frac\pi h\biggl(2p+{c(j)-c(k)\over 2}\biggr),$$
and
$$\bar{U}^k_{jr}=-\frac\pi h\biggl(2q+{c(j)-c(r)\over 2}\biggr).$$
There are two poles in the variable $e^\theta$ in $X^{ij}(\theta)$ due
to the fusing $j+k\rightarrow r$, since there are precisely two
inequivalent values of $p$ and $q$, $p'$ and $q'$, such that
$$\gamma_j+\sigma^p\gamma_k=\sigma^q\gamma_r,\quad 1\leq p\leq h,$$
$$\gamma_j+\sigma^{p'}\gamma_k=\sigma^{q'}\gamma_r,\quad 1\leq p'\leq h,$$
with $p,p'$ and $q,q'$ related by \cite{Fring}
$p'=h-p+{c(k)-c(j)\over 2}$, and 
$q'=h-q+{c(k)-c(j)\over 2}$.
Now the poles in $X^{jk}(\theta)$ lie at 
$e^{-\theta}=e^{{\pi i\over h}(2p+{c(j)-c(k)\over 2})}$,
or \beq\theta=-{i\pi\over h}(2p+{c(j)-c(k)\over 2})+2\pi n i,
\label{eq: theta3} \eeq
where $n$ is an integer, and also at the position obtained by
replacing $p$ with $p'$. From equation (\ref{eq: Masses}), the pole on
the physical strip of the S-matrix lies at 
$\theta=iU^r_{jk}+2\pi m i$,
where $m$ is the integer which allows us to take
$0< {\rm Im}(\theta)\leq\pi$. Since $1\leq p\leq h$, we must have
$m=1$, and hence the pole lies at
\beq\theta={i\pi\over h}(2(h-p)+{c(k)-c(j)\over 2}).\label{eq: class_pole} \eeq
In order to have ${\rm Im}(\theta)\leq\pi$, we must also have 
$p\geq\frac h2+ {c(k)-c(j)\over 4}$,
and this is only possible for precisely one of the two allowed values of $p$
in the fusing rule, since suppose that
$p<\frac h2+ {c(k)-c(j)\over 4}$,
then for the alternative value $p'$, where 
$p'=h-p+{c(k)-c(j)\over 2}$, and
$p'>\frac h2+ {c(k)-c(j)\over 4}$,
we can choose this possibility for $p$, and the pole in the
S-matrix on the physical strip lies at
\beq\theta=iU^r_{jk}+2\pi i
={i\pi\over h}\biggl(2(h-p)+{c(k)-c(j)\over 2}\biggr). \eeq
From (\ref{eq: theta3}), we see that $X^{jk}(\theta)$ also has a
pole in this position, and we have established the result. 

$X^{jk}(\theta)$ also satisfies a classical  bootstrap equation. Define the
fusing angles $\bar{U}_{ab}^c$, as discussed in section 4, and suppose
that the fusing $j+r\rightarrow k$ is allowed, then we have the
classical bootstrap equation
\beq X^{ik}(\theta)=X^{ij}(\theta-i\bar{U}^r_{j\bar{k}})X^{ir}(\theta+
i\bar{U}^j_{r\bar{k}}).\label{eq: class_boot}\eeq
This is proved as follows:

\noindent The fusing angles 
$\bar{U}^r_{j\bar{k}}$ and $\bar{U}^j_{r\bar{k}}$ can be defined
through the equation
$$M_k=M_je^{-i\bar{U}^r_{j\bar{k}}}+M_re^{i\bar{U}^j_{r\bar{k}}}.$$
Again from \cite{Fring}, we have from
$\sigma^q\gamma_k=\gamma_j+\sigma^s\gamma_r$,
where $s$ is the choice out of the two inequivalent possibilities that
places the pole due to the fusing at $\theta=i{U}^{\bar{k}}_{jr}$
on the physical strip, that
\beq\bar{U}^r_{j\bar{k}}=\frac\pi h\biggl(-2q+{c(k)-c(j)\over
2}\biggr) \label{eq: bootfuse1}\eeq
and
\beq\bar{U}^j_{r\bar{k}}=-\frac\pi h\biggl(2(s-q)+{c(k)-c(r)\over
2}\biggr).\label{eq: bootfuse2}\eeq

We compute 
$$X^{ij}(\theta-i\bar{U}^r_{j\bar{k}})X^{ir}(\theta+i\bar{U}^j
_{r\bar{k}})$$
$$=\prod_{p,p'=1}^h\biggl(e^{-\theta}-e^{{\pi i\over
h}(2p+{c(i)-c(j)\over 2})-{\pi i\over h}(-2q+{c(k)-c(j)\over
2})}\biggl)^{\gamma_i\cdot\sigma^p\gamma_j}\biggl(e^{-\theta}-e^{{\pi
i\over h}(2p'+{c(i)-c(r)\over 2})-{\pi i\over h}(2(s-q)+{c(k)-c(r)\over 2})}\biggl)^{\gamma_i\cdot\sigma^{p'}\gamma_r}$$
if $u=p+q=p'-s+q$, then
$$=\prod_{p=1}^h\biggl(e^{-\theta}-e^{{\pi i\over
h}(2u+{c(i)-c(k)\over
2})}\biggl)^{\gamma_i\cdot(\sigma^p\gamma_j+\sigma^{p'}\gamma_r)},$$
but $\sigma^p\gamma_j+\sigma^{p'}\gamma_r=\sigma^{u-q}(\gamma_j+
\sigma^s\gamma_r)=\sigma^u\gamma_k$,
so this expression equals
$$\prod_{u=1}^h\biggl(e^{-\theta}-e^{{\pi i\over
h}(2u+{c(i)-c(k)\over
2})}\biggl)^{\gamma_i\cdot\sigma^u\gamma_k} = X^{ik}(\theta),$$
as required.

\resection{Axiomatics of S-matrix theory: crossing, unitarity and the
bootstrap}

Consider an S-matrix $S^{ab}(\theta): V_a\tensor V_b\rightarrow
V_b\tensor V_a$, where $\theta$ is the relative rapidity of the two
incoming solitons, and $V_a$ is a vector spaces of charges, associated
with a soliton of species $a$. Typically in what follows, $V_a$ will be
the $a$'th fundamental representation of a Lie algebra.

It must satisfy the following properties:

\noindent 1. Crossing\\

\beq S^{\bar{a}b}(\theta)=(1\tensor C_a).[\sigma\cdot
S^{ba}(i\pi-\theta)]^{t_2}\cdot\sigma\cdot(C_{\bar{a}}\tensor
1),\label{eq: crossing2}\eeq
where $\sigma$ is the twist map $\sigma(x\tensor y)=(y\tensor x)$,
$t_2$ means transpose in the second space. $C_a$ is the charge
conjugation matrix from $V_a$ to $V_{\bar{a}}$.

\noindent 2. Unitarity

\beq
S^{ab}(\theta)S^{ba}(-\theta)=1. \eeq

\noindent 3. Bootstrap

Suppose that there is a pole at $\theta=iU^c_{ab}$ due to the fusing 
$a+b\rightarrow\bar{c}$ in $S^{ab}(\theta)$, 
in the
direct channel, then we must have
\beq m_{\bar{c}}^2=m_c^2=m_a^2+m_b^2+2m_a m_b\cos(U^{c}_{a b})
\label{eq: masses1}. \eeq
It also follows, by crossing, that we are allowed the fusings 
$b c\rightarrow \bar{a}$, and $c a\rightarrow \bar{b}$, and hence
$$U^{c}_{ab}+U^{a}_{bc}+U^{b}_{ca}=2\pi.$$
It is also helpful if rewrite the mass equation (\ref{eq: masses1}) as
\beq
m_c=m_ae^{-i\bar{U}^{b}_{ac}}+m_be^{i\bar{U}^{a}_{bc}}\label{eq: masses2},\eeq
where
$\bar{U}^{b}_{ac}=\pi-U^{b}_{ac}$, etc. 
Note that
$U^{c}_{ab}=\bar{U}^b_{ac}+\bar{U}^a_{bc}$.
This implies that we can rewrite (\ref{eq: masses2}) as 
\beq m_ce^{i\bar{U}^b_{ac}}=m_a+m_be^{i{U}^c_{ab}} \label{eq: Xijpole}\eeq

Now, associated with a fusing $a b\rightarrow\bar{c}$, there is a
`fusing' of the spaces $$V_a\tensor V_b=V_{\bar{c}}\oplus\cdots$$
Let $P_{\bar{c}}$ denote the projection from $V_a\tensor V_b$ to
$V_{\bar{c}}$. The residue of $S^{ab}(\theta)$ must be
proportional to $P_{\bar{c}}$, and for any third soliton $d$, we must have
the bootstrap equation:
\beq S^{d\bar c}(\theta)=(P_{\bar{c}}\tensor 1)
(1\tensor S^{da}(\theta+i\bar{U}^b_{ac}))
(S^{db}(\theta-i\bar{U}^a_{bc})\tensor 1)(1\tensor
P_{\bar{c}}).\label{eq: bootstrap}\eeq

\resection{The quantum Yang-Baxter equation}
The quantum Yang-Baxter equation with spectral parameter is the
following equation for the R-matrix 
$$R^{ab}(x): V_a\tensor V_b\rightarrow V_b\tensor V_a:$$
\beq
(R^{bc}(\frac{x_b}{x_c})\tensor 1)(1\tensor R^{ac}(\frac{x_a}{x_c}))
(R^{ab}(\frac{x_a}{x_b})\tensor 1)=
(1\tensor R^{ab}(\frac{x_a}{x_b}))(R^{ac}(\frac{x_a}{x_c})\tensor 1)(1\tensor
R^{bc}(\frac{x_b}{x_c})),
\label{eq: QYBES}
\eeq
both sides being a linear map  $$V_a\tensor V_b\tensor V_c
\rightarrow V_c\tensor V_b\tensor V_a.$$
This is represented diagrammatically as
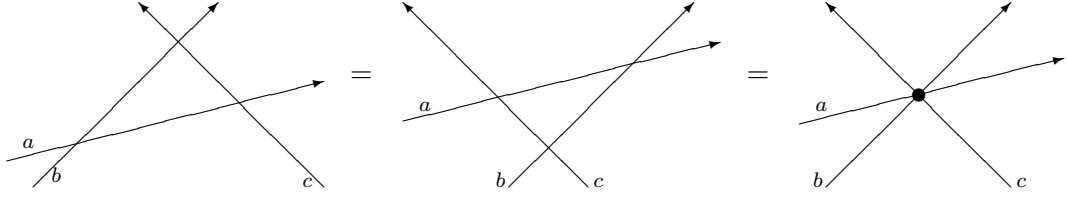
\begin{center}
\begin{picture}(270,120)(70,-20)
\put(10,10){\vector(1,1){70}}
\put(120,10){\vector(-1,1){70}}
\put(0,20){\vector(4,1){120}}
\put(130,50){$=$}
\put(190,10){\vector(1,1){70}}
\put(220,10){\vector(-1,1){70}}
\put(150,35){\vector(4,1){120}}
\put(17,12){$\scriptstyle b$}
\put(112,10){$\scriptstyle c$}
\put(6,25){$\scriptstyle a$}
\put(185,10){$\scriptstyle b$}
\put(222,10){$\scriptstyle c$}
\put(156,39){$\scriptstyle a$}
\put(280,50){$=$}
\put(310,10){\vector(1,1){70}}
\put(380,10){\vector(-1,1){70}}
\put(300,34){\vector(4,1){100}}
\put(345,45){\circle*{5}}
\put(305,10){$\scriptstyle b$}
\put(382,10){$\scriptstyle c$}
\put(306,39){$\scriptstyle a$}
\end{picture}
\end{center}
The affine quantum groups can provide some solutions for
$R^{ab}:V_a\tensor V_b\rightarrow V_b\tensor V_a$, for certain
specific representations $V_a$ and $V_b$ of the quantum group.

Suppose that $V_a\tensor V_b=\oplus\sum_{i=1}^n V'_{i}$, with each 
$V'_{i}$ appearing only once. The co-product of the affine quantum
group defines this decomposition, and tells us how to construct the
$q$-dependent projection matrices $P_i$, from $V_a\tensor V_b$ onto
$V'_i$. Also $\sum_{i=1}^n P_i=1$, and $P_n P_m=\delta_{nm}P_n$. The
R-matrix can be written
$$R^{ab}(x)=\sum_{i=1}^n\rho_i(x)P_i,$$
and the scalar functions $\rho_i(x)$ actually solved, since an
equivalent definition of the R-matrix is that the co-product commutes with
$R^{ab}(x)$. This is the celebrated work of Jimbo \cite{Jimbo}.

For $g=su(2)$, or sine-Gordon, it is possible to show, for example in
the spin half representation $V_1$, and for the so-called homogeneous
gradation, where the spectral parameter is associated only with the
zero'th root of the affine algebra, that
\beq R^{11}(x)=\bigr((x^{1/2}q^{1}-x^{-1/2}q^{-1})P_3+
(x^{-1/2}q^{1}-x^{1/2}q^{-1})P_0\bigl).\label{eq: Rprojs} \eeq
Here, $P_3$ projects onto the spin $\frac32$ representation, and $P_0$
the trivial representation.
We adopt the notation for writing $V\tensor W$ in
vector form, as, for $v_i$ the $i^{\rm th}$ entry in the vector in
$V$, and similarly for $w_i$, we write $v_i\tensor w_j\in V\tensor W$
as the $(2*(i-1)+j)^{\rm th}$ entry in the vector for $V\tensor W$.

For $q$ the standard deformation parameter in the quantum group 
$U_q(su(2))$, it is easy to show from the co-product that 
$$ P_0={1\over q+q^{-1}}\pmatrix{
0 &0 &0 &0 \cr 
0 &q^{-1}&-1 &0 \cr 
0 &-1&q &0 \cr
0 &0 &0 &0},$$
$$P_3=1-P_0={1\over q+q^{-1}}\pmatrix{
q+q^{-1} &0&0 &0 \cr
0 &q&1 &0\cr
0 &1&q^{-1}& 0 \cr
0 &0&0 &q+q^{-1} }.$$
Hence combining $P_0$ and $P_3$ into (\ref{eq: Rprojs}), we  have in
the homogeneous gradation
\beq R(x)=\pmatrix{x^{1/2}q-x^{-1/2}q^{-1}&0& 0&0 \cr
0&x^{1/2}(q-q^{-1})&x^{1/2}-x^{-1/2}&0 \cr
0&x^{1/2}-x^{-1/2}&x^{-1/2}(q-q^{-1})&0 \cr
0&0&0&x^{1/2}q-x^{-1/2}q^{-1} }.\label{eq: homog}\eeq

For the soliton S-matrices, we prefer to use the principal
gradation. For any simply-laced algebra $g$, we can move between the
homogeneous and principal gradations quite easily, by setting
$$\sigma_{12}=x^{T_3/h}\tensor 1,\qquad \sigma_{21}=1\tensor
x^{T_3/h},$$
where $T_3$ is the Cartan-subalgebra element of the maximal embedding
of an $su(2)$ subalgebra in $g$. Then from \cite{OT1,OT2},
$$R_P(x)=\sigma_{21}R_H(x^h)\sigma_{12}^{-1}.$$

Doing this for sine-Gordon gives
$$ R_{P}^{11}(x)=
                            \pmatrix{xq-x^{-1}q^{-1}&0&0&0 \cr
                                     0&q-q^{-1}&x-x^{-1}&0 \cr
                                     0&x-x^{-1}&q-q^{-1}&0 \cr
                                     0&0&0&xq-x^{-1}q^{-1}}.$$

Now it is possible to show, for example there is a proof in
\cite{D}, that $R^{ab}(x)$ satisfies the crossing condition
(\ref{eq: crossing2}), for $V_a$ and $V_b$ the fundamental
representations, in either the principal or homogeneous gradations.
This can be checked explicitly for sine-Gordon by the reader.

It is also possible to show, for $A_n$, that $R^{ab}(x)$ satisfies the
bootstrap equation (\ref{eq: bootstrap}), for $V_a$ and $V_b$ again
the fundamental representations. This is usually called `fusion' in
the literature. However for the $D$ and $E$ series of algebras, it is
only possible to show this for $R$ intertwining between tensor
products of larger spaces $W_a\supset V_a$, but with $W_1=V_1$ \cite{OW}. This
is an argument for saying that $R$-matrices do not exist for $R$
intertwining between the smaller spaces $V_a\tensor V_b$ (if either
$a$ or $b$ is different from $1$), which is vindicated by the method
for calculating $R$-matrices using projections, see section 11.

We also note that for $R^{11}(x)$ defined by (\ref{eq: Rprojs}) in
sine-Gordon, for the homogeneous gradation
\begin{eqnarray*}
R_H^{11}(x)R_H^{11}(x^{-1})&=&
(x^{1/2}q-x^{-1/2}q^{-1})(x^{-1/2}q-x^{1/2}q^{-1})(P_3+P_0)\cr\cr
&=&\frac{(1-xq^2)(1-x^{-1}q^2)}{q^2}.1. \end{eqnarray*}
The only change for the principal gradation is $x\rightarrow x^2$, and
 \beq
R_P^{11}(x)R_P^{11}(x^{-1})=
\frac{(1-x^2q^2)(1-x^{-2}q^2)}{q^2}.1.\label{eq: unitarity_P}\eeq

\resection{The `regularised' quantum dilogarithm}
The quantum dilogarithm \cite{Fad_Kas,Fad} is defined as
\beq
S_{q}(w)=\prod_{k=0}^\infty(1+q^{2n+1}w)=\exp\biggl({\sum_{k=1}^\infty{(-w)^k
\over k(q^k-q^{-k})}}\biggr)\label{eq: qd}. \eeq
It satisfies the key defining property
\beq
{S_q(qw)\over S_{q}(q^{-1}w)}={1\over 1+w} \label{eq: dilog_prop}. \eeq
The first expression converges for $|q|<1$, and the second for $|q|\neq 1$, and
$|w|<1$.
The classical dilogarithm, for $|w|<1$, is defined as 
\beq L_2(w)=-\int_{0}^w{dz\over z}\log(1-z)=\sum_{k=1}^\infty 
{w^{k}\over k^2}, \label{eq: cd} \eeq
so the second equation of (\ref{eq: qd}) is a sort of $q$-deformed exponential 
of the classical dilogarithm (\ref{eq: cd}). Furthermore in the limit $\epsilon
\rightarrow 0$, $\epsilon<0$ and $q=e^\epsilon$, it is easy to see that, to 
leading order
$$S_{q}(w)\sim e^{{1\over 2\epsilon}L_2(-w)},$$
and this further justifies the term `quantum dilogarithm' for (\ref{eq: qd}).

Unfortunately (\ref{eq: qd}) suffers from a serious defect which means that 
as it stands it 
cannot play a r\^{o}le in the integrable models discussed in this paper, 
simply because we expect that $|q|=1$. The
second expression in (\ref{eq: qd}) is seen to diverge, since 
if $q$ is a root of unity then a term in the series is infinite, otherwise
$q^{2k}$ becomes arbitrarily close to $1$ an infinite number of times, and so 
the series must be greater than any given bound. Nevertheless, the situation
can be repaired because the function, introduced in \cite{Fad},
\beq
\hat{S}_{q}(w)=\exp\biggl({{1\over 4}\int_{-\infty}^{\infty}{dx\over x}{(w)^{-ix}\over
\sinh(\pi x)\sinh(\mu x)}} \biggr),
\label{eq: rqd}
\eeq
where $q=e^{i\mu}$, and the contour goes above the pole at the origin,
satisfies 
\beq {\hat{S}_{q}(qw)\over\hat{S}_{q}(q^{-1}w)}=
{1\over 1+w}.\label{eq: qdp} \eeq
The same property as (\ref{eq: dilog_prop}).
The integral (\ref{eq: rqd}) converges if ${\rm Im}(\log(w))<\pi+\mu $, we
then use the above functional equation to define it for all $\log(w)$.
To show this property (\ref{eq: qdp}), we compute 
$${\hat{S}_{q}(qw)\over\hat{S}_{q}(q^{-1}w)}=
\exp\biggl({{1\over 2}\int_{-\infty}^{\infty}{dx\over x}{(w)^{-ix}\over
\sinh(\pi x)}} \biggr), $$
closing the contour in the upper-half plane if $|w|<1$, the 
lower-half plane if $|w|>1$, and
summing up the residue contributions from simple poles on the imaginary axis,
taking care of a residue contribution from the origin when closing the 
contour in the lower-half plane, 
as shown in \cite{Fad}.

We take (\ref{eq: rqd}) as a 
suitable definition of a `regularised' quantum dilogarithm which should
replace (\ref{eq: qd}), when  $|q|=1$. 
To further justify the term `dilogarithm', we
also show that the classical dilogarithm is obtained in the limit
$\mu\rightarrow 0$, this will also be important for the calculations
to follow.

The leading order behaviour in this limit is given by (for $|z|<1$),
closing the contour in the upper half plane,
\bea\log{\hat{S}_q(z)}&=&{{1\over 4\mu}\int_{-\infty}^\infty{dx\over x^2}
{z^{-ix}\over\sinh\pi x}} \cr \cr
&=&{{2\pi i\over 4\mu}\sum_{n=1}^\infty{z^n\over -n^2\pi(-1)^n}} \cr \cr
&=&{{-i\over 2\mu} L_2(-z)} \cr \cr
&=&{{i\over 2\mu}\int_0^{z}{dw\over w}\log(1+w)}. \label{eq: dilog_lim1} \eea
For $|z|>1$, closing the contour in the lower half plane,
\bea\log(\hat{S}_q(z))&=&{{1\over 4\mu}\int_{-\infty}^\infty{dx\over x^2}
{z^{-ix}\over\sinh\pi x}} \cr \cr
&=&{-{2\pi i\over 4\mu}\sum_{n=-1}^{-\infty}{z^n\over -n^2\pi(-1)^n}-2\pi i {\rm res}(0)} \cr \cr
&=&{i\over 2\mu}\sum_{n=1}^\infty{(-z)^{-n}\over n^2}+{i\over 4\mu}(\log z)^2 
\cr \cr
&=&{{i\over 2\mu} L_2(-z^{-1})}+{i\over 4\mu}(\log z)^2 \cr \cr
&=&{-{i\over 2\mu}\int_0^{z^{-1}}{dw\over w}\log(1+w)+{i\over 4\mu}
(\log z)^2}. \label{eq: dilog_lim2} \eea
Now observe that if we change the variable of integration of (\ref{eq: rqd})
by $x\rightarrow (\pi/\mu) x$, then
\beq \hat{S}_{q}(w)=\exp\biggl({{1\over 4}\int_{-\infty}^{\infty}
{dx\over x}{(w)^{-i{\pi\over\mu}x}\over
\sinh({\pi^2\over\mu} x)\sinh(\pi x)}} \biggr)  
= \hat{S}_{\tilde{q}}(w^{\pi/\mu}), \label{eq: duality} \eeq
where $\tilde{q}=e^{{\pi^2\over\mu}i}$.

We can also expand the function, in order to study its poles and
zeroes. Setting $e^{ip}=w$, we have
\beq
\hat{S}_{\mu}(w)=e^{-{ip^2\over 8\mu}-{i\over 24}(\mu+{\pi^2\over\mu})}\prod_{k,l=0}^{\infty}\biggl({p+\pi(2k+1)+\mu(2l+1)\over -p+\pi(2k+1)+\mu(2l+1)}\biggr)
\label{eq: double2}.
\eeq 
This can be written as an infinite product of gamma functions, using
the Weierstrass product formula for the gamma function. We can do this
in two different ways, by taking the product over $k$ or $l$. Thus
\bea
\hat{S}_{\mu}(w)&=&e^{-{ip^2\over 8\mu}-{i\over
24}(\mu+{\pi^2\over\mu})}\prod_{l=0}^{\infty}\biggl({\Gamma(-\frac
p{2\pi}-\frac1{2\pi} + \frac\mu{2\pi}(2l+1))\over 
\Gamma(\frac p{2\pi}-\frac1{2\pi} + \frac\mu{2\pi}(2l+1))}\biggr),\cr\cr
\hat{S}_{\mu}(w)&=&e^{-{ip^2\over 8\mu}-{i\over
24}(\mu+{\pi^2\over\mu})}\prod_{k=0}^{\infty}\biggl({\Gamma(-\frac
p{2\mu}-\frac1{2\mu} + \frac\pi{2\mu}(2k+1))\over 
\Gamma(\frac p{2\mu}-\frac1{2\mu} +
\frac\pi{2\mu}(2k+1))}\biggr).\label{eq: gammas} \eea
The duality property (\ref{eq: duality}) under
$\mu\rightarrow\pi^2/\mu$ can be rederived from (\ref{eq: double2}) 
by swapping $k$ and $l$ in
the product and multiplying the numerator and denominator by
$\pi/\mu$. 

We also note the property, 
\beq
\hat{S}_{\mu}(w)\hat{S}_{\mu}(w^{-1})=e^{{i(\log(w))^2\over 4\mu}-{i\over
12}(\mu+{\pi^2\over\mu})}, \label{eq: Sinv} \eeq
derived from computing the integral
$\hat{S}_{\mu}(w)\hat{S}_{\mu}(w^{-1})$, the contour becoming a small circle
around the orgin, so only the residue at the origin contributes.

In what follows we replace the notation $\hat{S}_q(z)$ with 
$S_q(z)$.

\resection{The sine-Gordon solution (Zamolodchikov-Zamolodchikov)}

For $g=su(2)$, we recall the equation for $R^{11}(\theta)$, as given 
in section 5,
$$S^{11}(\theta)=v(x)\pmatrix{xq-x^{-1}q^{-1}&0&0&0 \cr
                                         0&q-q^{-1}&x-x^{-1}&0 \cr
                                         0&x-x^{-1}&q-q^{-1}&0 \cr
                                         0&0&0&xq-x^{-1}q^{-1} }$$
$$=\pmatrix{S(\theta)&0&0&0 \cr
                         0&S_{R}(\theta)&S_{T}(\theta)&0 \cr
                         0&S_{T}(\theta)&S_{R}(\theta)&0 \cr
                         0&0&0&S(\theta) \cr}. $$
Here, following from the standard notation in \cite{ZZ}, $S(\theta)$
is the identical soliton-soliton process (or
anti-soliton--anti-soliton),
$S_R(\theta)$ the soliton--anti-soliton reflection process, 
and $S_T(\theta)$ the soliton--anti-soliton transmission process.
From the crossing property of $R^{11}(x)$ and (\ref{eq: unitarity_P}) in
section 5, the crossing and unitarity equations for $v(x)$ are
respectively
\beq
v(x)=v(x^{-1}q^{-1}), \label{eq: crossingv} \eeq
\beq v(x^{-1})v(x)={q^2\over(1-x^2q^2)(1-x^{-2}q^2)}.
\label{eq: unitarityv} \eeq
Here,
$$ x=e^{8\pi\theta\over\gamma},\quad q=e^{-{8\pi^2i\over\gamma}}=
-e^{-{8\pi^2i\over|\beta|^2}},\quad\gamma={|\beta|^2\over
1-{|\beta|^2\over 8\pi}},$$
and we see that the crossing condition $\theta\rightarrow i\pi-\theta$
implies $x\rightarrow x^{-1}q^{-1}$.

\noindent Now an infinite family of solutions to these equations, 
labelled by the integer $n$, is given by:
\def\Sq{S_{q^{-2}}}
\beq
v(x)={q\over(1-x^2q^2)}{X_q(x)\over X_q(x^{-1})}\label{eq: big_one} \eeq
where\beq X_q(x)={\Sq(e^{i\pi+2\pi ni}x^2q^2)\over\Sq(e^{i\pi+2\pi ni}x^2) }
\label{eq: sine-Gordon-2} \eeq
\par\noindent{\bf Proof}\par\noindent
We check crossing (\ref{eq: crossingv})
$$X_q(x^{-1}q^{-1})={\Sq(e^{i\pi+2\pi ni}x^{-2})\over\Sq(e^{i\pi+2\pi ni}x^{-2}q^{-2})}=X_q(x^{-1})^{-1}(1-x^{-2}), $$
in the last step we have used the property (\ref{eq: qdp}).
Similarly
$$X_q(xq)={\Sq(e^{i\pi+2\pi ni}x^2q^4)\over\Sq(e^{i\pi+2\pi ni}x^{2}q^{2})}
=X_q(x)^{-1}(1-x^{2}q^{2}). $$
Then
$$v(x^{-1}q^{-1})={q\over 1-x^{-2}}{X_q(x^{-1}q^{-1})\over X_q(xq)}=
{q\over 1-x^{2}q^2}{X_q(x)\over X_q(x^{-1})}=v(x),$$
as required.

Observe that the unitarity equation (\ref{eq: unitarityv}) follows 
automatically.

\vskip 0.5cm
\noindent{\bf Comparison with the Zamolodchikov-Zamolodchikov solution}

We will see that the above solution for $n=0$ is the same as the 
Zamolodchikov-Zamolodchikov solution \cite{ZZ}.
\par\noindent{\bf Proof}\par\noindent
We work with 
$$S_{T}(\theta)={q(x-x^{-1})\over 1-x^2q^2}v(x)=qx{\Sq(e^{i\pi}x^2q^2)
\Sq(e^{i\pi}x^{-2})\over \Sq(e^{i\pi}x^{-2}q^{-2})\Sq(e^{i\pi}x^2q^4)}.$$
Here the common intersection of ranges of $\theta$ for which each integral
in each of the four quantum dilogarithms separately converges is
$$\pi>\Im(\theta)>\pi-{\gamma\over 8}.$$

Then 
$$S_{T}(\theta)=qx\exp\biggl({1\over 4}\int_{-\infty}^{\infty}{dy\over y}
{(e^{i\pi}x^{2}q^{2})^{-iy}+ (e^{i\pi}x^{-2})^{-iy}-(e^{i\pi}x^{-2}q^{-2})^
{-iy} - (e^{i\pi}x^{2}q^{4})^{-iy}\over \sinh(\pi y)\sinh(\mu y)}\biggr)$$
$${\rm for}\quad q^{-2}=e^{i\mu},\qquad\mu={16\pi^2\over\gamma},$$
and
\bea
(e^{i\pi}x^{2}q^{2})^{-iy}+ &(e^{i\pi}x^{-2})^{-iy}&-(e^{i\pi}x^{-2}q^{-2})^
{-iy}- (e^{i\pi}x^{2}q^{4})^{-iy}\cr \cr 
&=&(e^{-16\pi i\theta y/\gamma-\mu y} - e^{16\pi i\theta y/\gamma +\mu y})
(1-e^{-\mu y})e^{\pi y} \cr \cr
&=&4\sinh(-16\pi i\theta y/\gamma-\mu y)\sinh({\mu y\over 2})e^{\pi y -{\mu 
y\over 2}}. \eea
Let
\bea
J=\log(S_{T}(\theta))-\log(qx)&=&{1\over 4}\int_{-\infty}^{\infty}{dy\over y}
{(e^{i\pi}x^{2}q^{2})^{-iy}+ (e^{i\pi}x^{-2})^{-iy}-(e^{i\pi}x^{-2}q^{-2})^
{-iy} - (e^{i\pi}x^{2}q^{4})^{-iy}\over \sinh(\pi y)\sinh(\mu y)} \cr \cr \cr
&=&-\int_{-\infty}^{\infty}{dy\over y}{\sinh(16\pi i\theta y/\gamma+\mu y)
e^{(\pi - {\mu\over 2})y}\over 2\cosh({\mu y \over 2})\sinh(\pi y)} 
\label{eq: 2.2}, \eea
and let $y\rightarrow -y$, then the contour passes below the pole at the 
origin. Moving the contour through this pole gives a positive residue 
contribution 
from the origin.
$$J=\int_{-\infty}^{\infty}{dy\over y}{\sinh(16\pi i\theta y/\gamma+\mu y)
e^{-(\pi - {\mu\over 2})y}\over 2\cosh({\mu y \over 2})\sinh(\pi y)} + 2\pi i.
{\rm 
res}(0),$$
where the contour here is the same as in (\ref{eq: 2.2}).

Now
$$2 \pi i.{\rm res}(0)=2\pi i {1\over 2\pi}({16\pi i\theta \over\gamma}+ 
{16\pi^{2} \over\gamma} ) ={-16\pi\theta \over\gamma}+{16\pi^{2}i\over\gamma},
$$
and
\bea
J&=&-\int_{-\infty}^{\infty}{dy\over y}{\sinh(16\pi i\theta y/\gamma+\mu y)
\over 2\cosh({\mu y \over 2})\sinh(\pi y)}{1\over 2}(e^{(\pi-\mu/2)y}-(e^{-(\pi-\mu/2)y}
+ 2\pi i.{\rm res}(0))) \cr \cr \cr
&=& {2\pi i.{\rm res}(0)\over 2} - \int_{-\infty}^{\infty}{dy\over y}{\sinh(16\pi i\theta y/\gamma+\mu y)
\sinh(\pi-\mu/2)y\over 2\cosh({\mu y \over 2})\sinh(\pi y)}. \eea
After the change of integration variable,  
$y\rightarrow {\gamma\over 16\pi^2}y$,
\beq
\log(S_{T}(\theta))-\log(qx) = {-8\pi\theta \over\gamma}+{8\pi^{2}i\over\gamma}+ {i\over 2}\int_{-\infty}^{\infty}{dy\over y}
{\sin((i-{\theta\over\pi})y)\sinh({\gamma\over 16\pi}-{1\over 2})y \over 
\cosh({y \over 2})\sinh({\gamma y\over 16\pi})}. \eeq
Since $\log(qx)={8\pi\theta \over\gamma}-{8\pi^{2}i\over\gamma}$,
we have the result
\beq S_{T}(\theta)=\exp\biggl({{i\over 2}\int_{-\infty}^{\infty}{dy\over y}
{\sin((i-{\theta\over\pi})y)\sinh({\gamma\over 16\pi}-{1\over 2})y \over 
\cosh({y \over 2})\sinh({\gamma y\over 16\pi})}}\biggr). 
\label{eq: Karowski}
\eeq
This is the solution due to Karowski {\it et al} \cite{Karowski},
and converges in
the stated range
$$\pi>\Im(\theta)>\pi-{\gamma\over 8}.$$
We also expand our result for $n=0$ in terms of an infinite double product,
using the expansion (\ref{eq: double2}) for the quantum dilogarithm
-- this is not strictly needed to establish the identity with the 
Zamolodchikov-Zamolodchikov solution \cite{ZZ}, 
since this is guaranteed through the Karowski integral formula above,
but the result, of course, agrees with \cite{ZZ}, after we move to the
description using gamma functions. 

\begin{eqnarray*}
S_{T}(\theta)&=&qx.e^{8\pi^2 i\over\gamma}e^{-8\pi\theta\over\gamma}.
\prod_{k,l=0}^{\infty}
\biggl({\pi(2k+2)+\mu(2l)-{16\pi\theta i\over\gamma} 
\over \pi(2k)+\mu(2l+2) +{16\pi\theta i\over\gamma}}\biggr)
\biggl({\pi(2k+2)+\mu(2l+1)+{16\pi\theta i\over\gamma} 
\over \pi(2k)+\mu(2l+1) -{16\pi\theta i\over\gamma}}\biggr) \cr \cr
&\times&\biggl({\pi(2k)+\mu(2l)-{16\pi\theta i\over\gamma} 
\over \pi(2k+2)+\mu(2l+2) +{16\pi\theta i\over\gamma}}\biggr)
\biggl({\pi(2k)+\mu(2l+3)+{16\pi\theta i\over\gamma} 
\over \pi(2k+2)+\mu(2l-1) -{16\pi\theta i\over\gamma}}\biggr),
\end{eqnarray*}
so
\bea
S_{T}(\theta)&=&
\prod_{k,l=0}^{\infty}
\biggl({k+1+{8\pi \over\gamma}2l-{8\theta i\over\gamma} 
\over k+{8\pi \over\gamma}(2l+2) +{8\theta i\over\gamma}}\biggr)
\biggl({k+1+{8\pi \over\gamma}(2l+1)+{8\theta i\over\gamma} 
\over k+{8\pi \over\gamma}(2l+1) -{8\theta i\over\gamma}}\biggr) \cr \cr
&\times&\biggl({k+{8\pi \over\gamma}(2l)-{8\theta i\over\gamma} 
\over k+1+{8\pi \over\gamma}(2l+2) +{8\theta i\over\gamma}}\biggr)
\biggl({k+{8\pi \over\gamma}(2l+3)+{8\theta i\over\gamma} 
\over k+1+{8\pi \over\gamma}(2l-1) -{8\theta i\over\gamma}}\biggr).
\label{eq: full}
\eea
\vskip 0.5cm

\noindent{\bf The reflectionless result}
\def\qt{\tilde{q}}

The reflection process, given by $S_{R}(\theta)$, vanishes at the values of the
coupling $q^2=1$, or ${8\pi\over\gamma}=N$, where $N$ is an integer.
We relate these reflectionless conditions on the coupling 
to $q$ a primitive even root 
of unity, in the hope that the expression (\ref{eq: big_one}) will simplify.
In order to do this we must pass to the dual description of
$S_{T}(\theta)$, using the duality property (\ref{eq: duality}) :
\beq 
S_{T}(\theta) = qx {S_{\qt^{-2}}(e^{-i\pi}\qt^{-2}e^\theta) 
S_{\qt^{-2}}(\qt^{-2}e^{-\theta})\over S_{\qt^{-2}}
(e^{-2i\pi}\qt^{-2}e^{\theta}) 
S_{\qt^{-2}} ({e^{i\pi}\qt^{-2}e^{-\theta}})}
\label{eq: dualdescription} \eeq
At these reflectionless values of the coupling 
$\qt^{-2}=e^{i\gamma\over 16}=e^{i\pi\over 2N}$.
Setting $2N=n$, and using the quantum dilogarithm property (\ref{eq: qdp}) for 
$S_{\qt^{-2}}(w)$:
$${S_{\qt^{-2}}(w)\over S_{\qt^{-2}}(we^{2\pi i\over n})}=
(1+we^{i\pi\over n}),$$
$${S_{\qt^{-2}}(we^{2\pi i\over n})\over S_{\qt^{-2}}(w
e^{4\pi i\over n})}=(1+we^{3\pi i\over n}),$$
$$\vdots$$
$${S_{\qt^{-2}}(we^{(n-2)\pi i\over n})\over S_{\qt^{-2}}(w
e^{i\pi})}=(1+we^{(n-1)\pi i\over n})$$
so
\beq
{S_{\qt^{-2}}(w)\over S_{\qt^{-2}}(we^{i\pi})}=
(1+we^{\pi i\over n})(1+we^{3\pi i\over n})\cdots(1+we^{(n-1)\pi i\over n}), 
\eeq
and 
\beq
{S_{\qt^{-2}}(we^{-i\pi})\over S_{\qt^{-2}}(w)}=
(1+we^{-\pi i\over n})(1+we^{-3\pi i\over n})
\cdots(1+we^{-(n-1)\pi i\over n}).\eeq 
Thus 
\bea S_{T}(\theta)&=&e^{iN\pi}e^{N\theta}{(1+e^{2\pi i\over n}e^{-\theta})
(1+e^{4\pi i\over n}e^{-\theta})\cdots(1+e^{n\pi i\over n}e^{-\theta})\over
(1+e^{2\pi i\over n}e^{\theta})(1+e^{4\pi i\over n}e^{\theta})\cdots
(1+e^{n\pi i\over n}e^{\theta})}  \cr \cr
&=&e^{iN\pi}{(1+e^{-i\pi\over N}e^{\theta})(1+e^{-2i\pi\over N}e^{\theta})
\cdots(1+e^{-Ni\pi\over N}e^{\theta})\over
(e^{-i\pi\over N}+e^{\theta})(e^{-2i\pi\over N}+e^{\theta})\cdots
(e^{-Ni\pi\over N}+e^{\theta})}. \eea
This expression was the first result for the sine-Gordon S-matrix  \cite{KKF},
originally obtained by extrapolating from the time delay,
$\Delta(\theta)=
\frac 8{|\beta|^2}\log\Bigl({1-e^\theta\over 1+e^\theta}\Bigr)^2$, by using
the formula
\beq i\int_{0}^{\theta}d\theta'\log\biggl({e^{\theta'}-1\over e^{\theta'}+1}
\biggr)^2=\int_{0}^{\pi}dx\log\biggl({e^{\theta}e^{-ix}+1\over e^{-ix}+
e^{\theta}}\biggr). \label{eq: extrap} \eeq
The full S-matrix (\ref{eq: full}), at arbitrary values of the
coupling, was later obtained by extrapolating further from this result in
\cite{ZZ}, and such an extrapolation was carried out explicitly in an
elegant way  in \cite{KT}.  However, we shall see that the method
involving quantum dilogarithms is a more efficient way of
extrapolating from the time delay to the full S-matrix, because it is
not clear, in general, how to find an analogue of (\ref{eq: extrap}) for
the other cases.

There are an infinite family of solutions  (\ref{eq: big_one}) labelled by the
integer $n$. These can all be written as the Zamolodchikov solution
(with $n=0$), multiplied by CDD factors, by repeated application of
(\ref{eq: qdp}), in the dual description.
\resection{The semi-classical limit for sine-Gordon}
We work with the soliton-soliton process
$$S(\theta)=-(xq-x^{-1}q^{-1})v(x)=x^{-1}{X_{q}(x)\over X_q(x^{-1})}.$$
Taking the limit as $\gamma\rightarrow 0$ of $\Sq(w)$ is problematical
since $q^{-2}=e^{16\pi^2i\over\gamma}$, and so we pass to the dual description
using property (\ref{eq: duality}) of the quantum dilogarithm. This limit
then corresponds to $\tilde{q}\rightarrow 1$, and it has already been shown
that we recover the classical dilogarithm in this limit.

\def\Sd{S_{\tilde{q}^{-2}}}
\def\qt{\tilde{q}}
With $\tilde{q}^{-2}=e^{i\gamma\over 16}$, we have
\bea X_{q}(x)&=&{\Sd((e^{i\pi+2\pi ni}x^2q^2)^{\gamma/16\pi})\over 
\Sd((e^{i\pi+2\pi ni}x^2)^{\gamma/16\pi})} \cr\cr\cr
&=&{\Sd(\qt^{-2-4n}e^{\theta}e^{-i\pi})\over 
\Sd(\qt^{-2-4n}e^{\theta})}. \eea
Taking ${\rm Re\ }(\theta)<0$,
the leading order behaviour of this limit is (from equation 
(\ref{eq: dilog_lim1}))
$$ \lim_{\gamma\rightarrow 0}\Sd(\qt^{-2-4n}e^{\theta})=\exp\biggl({{8i\over\gamma}
\int_0^{e^\theta}{dw\over w}\log(1+w)}\biggr), $$
and
$$ \lim_{\gamma\rightarrow 0}\Sd(\qt^{-2-4n}e^{\theta}e^{-i\pi})
=\exp\biggl({{8i\over\gamma}\int_0^{e^\theta}{dw\over w}\log(1-w)}\biggr), $$
and we see that
\beq
\lim_{\gamma\rightarrow 0}X_{q}(x)=\exp\biggl({{8i\over\gamma}\int_0^{e^\theta}{dw\over w}\log\biggl({1-w\over 1+w}\biggr)}\biggr). \label{eq: positive} \eeq
We see how the structure of the time delay for sine-Gordon, namely
$\Delta(\theta')={8\over\beta^2}\log\biggl({1-e^{\theta'}\over 
1+e^{\theta'}}\biggr)^2$,
enters $X_q(x)$, in fact $X_q(x)$ can be thought of as a sort of $q$-deformed 
$X^{ij}(\theta)$, the normal ordering coefficient of two classical 
vertex operators.

Now $$X_q(x^{-1})={\Sd(\qt^{-2-4n}e^{-\theta}e^{-i\pi})\over 
\Sd(\qt^{-2-4n}e^{-\theta})}, $$
and from equation (\ref{eq: dilog_lim2}),
$$\lim_{\gamma\rightarrow 0}\Sd(\qt^{-2-4n}e^{-\theta})
=\exp\biggl({-{8i\over\gamma}\int_0^{e^\theta}{dw\over w}\log(1+w)+{4i\over\gamma}\theta^2}\biggr),$$
$$\lim_{\gamma\rightarrow 0}\Sd(\qt^{-2-4n}e^{-\theta}e^{-i\pi})
=\exp\biggl({-{8i\over\gamma}\int_0^{e^\theta}{dw\over w}\log(1-w)+{4i\over\gamma}
(\theta+i\pi)^2}\biggr), $$
and
$$\lim_{\gamma\rightarrow 0}X_{q}(x^{-1})=\exp\biggl({-{8i\over\gamma}\int_0^{e^\theta}{dw\over w}\log\biggl({1-w\over 1+w}\biggr)}\biggr).e^{{4i\over\gamma}(-\pi^2+2\pi i\theta)}  \label{eq: positive2}. $$
Thus
\bea
\lim_{\gamma\rightarrow 0} 
x^{-1}{X_q(x)\over X_q(x^{-1})}&=&\exp\biggl({{16i\over\gamma}\int_0^{e^\theta}{dw\over w}\log\biggl({1-w\over 1+w}\biggr)}\biggr).x^{-1}e^{{8\pi\theta\over\gamma}+{4i\pi^2\over\gamma}} \cr \cr
&=&e^{4i\pi^2\over\gamma}\exp\biggl({{16i\over\gamma}\int_0^{e^\theta}{dw\over w}\log\biggl({1-w\over 1+w}\biggr)}\biggr) \cr \cr
&=&\exp\biggl({{16i\over\gamma}\int_1^{e^\theta}{dw\over w}\log\biggl({1-w\over 1+w}\biggr)}\biggr), \eea
since 
$$\int_0^1{dw\over w}\log\biggl({1-w\over 1+w}\biggr)=
\frac12\int_{-1}^1{dw\over w}\log\biggl({1-w\over
1+w}\biggr)=-{\pi^2\over 4}.$$
So, for ${\rm Re}(\theta)<0$,
\beq
\lim_{\gamma\rightarrow 0}S(\theta)=\exp\biggl({{8i\over\gamma}\int_0^{\theta}
d\theta'\log\biggl({1-e^{\theta'}\over 1+e^{\theta'}}\biggr)^2}\biggr)
\label{eq: lim-sine-Gordon1}. \eeq
This is the correct form of the semi-classical limit for $S(\theta)$.

\resection{The $su(3)$ solution}

We label the fundamental representations of $su(3)$ $\lambda_1$, by
$3$, and $\lambda_2$
by $\bar 3$, the trivial representation by $1$, the adjoint representation by
$8$.
From the decomposition $ 3\otimes 3=6\oplus{\bar{3}}$,
we have, in the homogeneous gradation
\beq R^{11}(x)=(x^{1/2}q-x^{-1/2}q^{-1})P^{33}_6 + (x^{-1/2}q-x^{1/2}q^{-1})
P^{33}_{{\bar{3}}} \label{eq: R11},\eeq 
where $P^{33}_6$ and $P^{33}_{\bar{3}}$ are the projections from 
$3\otimes 3\rightarrow 6$ and $3\otimes 3\rightarrow {\bar{3}}$
respectively. We also have
\beq P^{33}_6+P^{33}_{\bar{3}}=1, \label{eq: projs1} \eeq
and
\beq  (P^{33}_6)^2=P^{33}_6,\quad (P^{33}_{\bar{3}})^2=P^{33}_{\bar{3}},\quad
P^{33}_{\bar{3}}P^{33}_6=P^{33}_6 P^{33}_{\bar{3}}=0. \label{eq: projs2} \eeq
Similarly, from the decomposition $3\otimes {\bar{3}}=8\oplus 1$,
we have, in the homogeneous gradation,
\beq R^{12}(x_{12})=(x_{12}^{1/2}q^{3/2}-x_{12}^{-1/2}q^{-3/2})P^{3{\bar{3}}}_8 + (x_{12}^{-1/2}
q^{3/2}-x_{12}^{1/2}q^{-3/2})P^{3{\bar{3}}}_1 \label{eq: R12},\eeq
and
\beq R^{21}(x_{21})=(x_{21}^{1/2}q^{3/2}-x_{21}^{-1/2}q^{-3/2})P^{{\bar{3}}3}_8 + (x_{21}^{-1/2}
q^{3/2}-x_{21}^{1/2}q^{-3/2})P^{{\bar{3}}3}_1, \label{eq: R21} \eeq
where $x_{12}=x_{21}^{-1}$. Since $3\otimes {\bar{3}}\neq {\bar{3}}\otimes
3$ we must replace equations (\ref{eq: projs1}) and (\ref{eq: projs2})
by
$$ P^{{\bar{3}}3}_8 P^{3{\bar{3}}}_8+ P^{{\bar{3}}3}_1 P^{3{\bar{3}}}_1 = 1,$$
and
$$  P^{{\bar{3}}3}_1 P^{3{\bar{3}}}_8= P^{{\bar{3}}3}_8 P^{3{\bar{3}}}_1 =0.$$
Using the Ansatz
$S^{ij}(\theta)=v^{ij}(\theta)R^{ij}(x_{ij})$,
for the S-matrix, and $x=x_{12}=x_{11}=e^{4\pi\theta\over\gamma}$,
for the spectral parameter $x_{ij}$ in the principal gradation, and 
$q=e^{-{4\pi^2\over\gamma}i}$. $\gamma$ is defined as 
$$\gamma={|\beta|^2\over
1-{|\beta|^2\over 4\pi}}.$$

Now the unitarity condition is
$S^{ji}(-\theta)S^{ij}(\theta)=1$.
Using the Ansatz for the S-matrix, we then have
$$v^{ji}(-\theta)v^{ij}(\theta)R^{ji}(x_{ji})R^{ij}(x_{ij})=1.$$
We also impose $v^{ij}(\theta)=v^{ji}(\theta)$.
In this $su(3)$, case we compute 
\bea R^{11}(x^{-1})R^{11}(x)&=&{(1-xq^2)(1-x^{-1}q^2)\over q^2}, \cr\cr
R^{21}(x^{-1})R^{12}(x)&=&(x^{1/2}q^{3/2}-x^{-1/2}q^{-3/2})
(x^{-1/2}q^{3/2}-x^{1/2}q^{-3/2}). \label{eq: unitarityH} \eea
However the equations (\ref{eq: R11}),(\ref{eq: R12}) and 
(\ref{eq: R21}) were for the homogeneous gradation, to move to the
principal gradation we must conjugate by the matrix $\sigma_{12}$
$$R_P(x)=\sigma_{21}R_H(x^h)\sigma_{12}^{-1},$$
the only effect this will have on the unitarity equations
(\ref{eq: unitarityH}) is to replace $x$ by $x^3$, so that in the
principal gradation these equations read:
\bea R^{11}(x^{-1})R^{11}(x)&=&{(1-x^{3}q^2)(1-x^{-3}q^2)\over q^2} \cr\cr
R^{21}(x^{-1})R^{12}(x)&=&(x^{3/2}q^{3/2}-x^{-3/2}q^{-3/2})
(x^{-3/2}q^{3/2}-x^{3/2}q^{-3/2}). \label{eq: unitarityP} \eea
These equations in turn imply the following unitarity conditions for
the scalar factors $v^{11}(\theta)$ and  $v^{12}(\theta)$:
\bea v^{11}(-\theta)v^{11}(\theta)&=&{q^2\over
(1-x^{3}q^2)(1-x^{-3}q^2)}, \cr\cr
v^{12}(-\theta)v^{12}(\theta)&=&{1\over (x^{3/2}q^{3/2}-x^{-3/2}q^{-3/2})
(x^{-3/2}q^{3/2}-x^{3/2}q^{-3/2}) }. \label{eq: unitarityV} \eea
The following crossing condition must also be satisfied under the
crossing transformation $\theta\rightarrow i\pi-\theta$,
\beq
v^{11}(i\pi-\theta)=v^{21}(\theta)=v^{12}(\theta),
\label{eq: crossing2a}
\eeq
this is because with the normalisations chosen for $R^{ij}(x)$, fixed  by
equations (\ref{eq: R11}), (\ref{eq: R12}) and (\ref{eq: R21}), the
matrices  $R^{11}(x)$ and $R^{12}(x)$ cross
precisely into each other under the crossing transformation
$x\rightarrow x^{-1}q^{-1}$, and the
application  of the crossing
matrices,  with no extraneous $x$-dependent factors $(1-x^hq^p)$.
\def\Sq{S_{q^{-3}}}
\def\St{S_{\tilde{q}^{-3}}}
Bearing in mind that the classical time delays for $su(3)$ are given
by (\ref{eq: X}):
\bea
X^{11}(\theta)&=&{(1-e^\theta)^2\over(1-e^{2\pi i\over 3}e^\theta)
(1-e^{4\pi i\over 3}e^\theta) }, \cr \cr \cr
X^{12}(\theta)&=&{(1-e^{-{\pi i\over 3}}e^\theta)(1-e^{{\pi i\over 3}}e^\theta)
\over(1+e^\theta)^2}, \label{eq: Xs} \eea
we propose the following solution for $v^{11}(\theta)$,
\beq 
v^{11}(x)={-qe^{\theta/2}\over (1-x^3q^2)}{X_q^{11}(x)\over X_q^{11}(x^{-1})}, \label{eq: v11sol} \eeq
where
\beq
X_q^{11}(x)={\Sq(e^{i\pi}x^3q^3)\over\sqrt{\Sq(e^{i\pi}x^3q)
\Sq(e^{-i\pi}x^3q^{-1})} }.
\label{eq: Xq11} \eeq
By extending the semi-classical limit calculation as
$\gamma\rightarrow 0$ for sine-Gordon, section 8, we see that, for
${\rm Re}(\theta)< 0$, (also from (\ref{eq: dilog_lim1})), after
passing to the dual description using (\ref{eq: duality}), 
$$X_q^{11}(\theta)={\St(\qt^{-3}e^\theta e^{-i\pi})\over\sqrt{\St(\qt^{-3}e^\theta 
e^{-{i\pi\over 3}})
\St(\qt^{3}e^\theta e^{i\pi\over 3})}}, $$
in the limit $\gamma\rightarrow 0$,
$$\lim_{\gamma\rightarrow 0}X_{q}^{11}(\theta)\sim e^{{6i\over\gamma}
\int_{-\infty}^\theta d\theta'\log(X^{11}(\theta')^{1/2})},$$
and for ${\rm Re}(\theta)< 0$, from (\ref{eq: dilog_lim2})
$$\lim_{\gamma\rightarrow 0}X_{q}^{11}(-\theta)\sim e^{-{6i\over\gamma}
\int_{-\infty}^\theta d\theta'\log(X^{11}(-\theta')^{1/2})}.
e^{{6i\over 2\gamma}((-\theta-i\pi)^2-\frac12 (-\theta-\frac{i\pi}3)^2-\frac12 (-\theta+\frac{i\pi}3)^2)}.$$
We use the classical property $X^{ij}(\theta)=X^{ij}(-\theta)$, and
then
$$\lim_{\gamma\rightarrow 0}{X_q^{11}(\theta)\over
X_q^{11}(-\theta)}\sim e^{{6i\over\gamma}
\int_{-\infty}^\theta
d\theta'\log(X^{11}(\theta'))}.x^{3/2}e^{8\pi^2 i\over 3\gamma}, $$
the factor $x^{3/2}$ is cancelled in physical S-matrix elements. There
are the additional constant factors
$$e^{iA\over\gamma}=e^{{6i\over\gamma}\int_{-\infty}^0
d\theta'\log(X^{11}(\theta'))}e^{8\pi^2 i\over 3\gamma},$$
which have yet to be computed explicitly,
and are related to the number of bound
states in the channel under consideration. We have explicitly
$$\lim_{\gamma\rightarrow 0}S^{11}_{11}(\theta)=
\lim_{\gamma\rightarrow 0} e^{\theta/2}x^{-3/2}{X_q^{11}(\theta)\over X_q^{11}(-\theta)}
=e^{{iA\over\gamma}+{i6\over\gamma}
\int_{0}^\theta
d\theta'\log(X^{11}(\theta'))}.$$
Here, the lower indices denote certain topological charges,
and we recognize this as the correct semi-classical behaviour. 

Observe that in the solution for $v^{11}(\theta)$ (\ref{eq: v11sol}),
we trivially have the unitarity equation 
(\ref{eq: unitarityV}) for $v^{11}(\theta)$. There is
clearly a single-valuedness problem in the square-roots of the
quantum dilogarithms in the denominator of (\ref{eq: Xq11}).
This will turn out not to be a problem, the combination 
${X_q^{11}(\theta)/X_q^{11}(-\theta)}$,
which makes up
physical S-matrix elements is truly single-valued, as shall be
discussed below.

We calculate the solution $v^{12}(\theta)$ by crossing
$v^{11}(\theta)$, using equation (\ref{eq: crossing2a}):
\begin{eqnarray*}
X_q^{11}(x^{-1}q^{-1})&=&{\Sq(e^{i\pi}x^{-3})\over\sqrt{\Sq(e^{i\pi}x^{-3}q^{-2})
\Sq(e^{-i\pi}x^{-3}q^{-4})}} \cr \cr \cr
&=&\sqrt{(1+e^{-i\pi}x^{-3}q^{-1})}{\Sq(e^{i\pi}x^{-3})\over\sqrt{\Sq(e^{i\pi}x^{-3}q^{-2})\Sq(e^{-i\pi}x^{-3}q^{2})}},\end{eqnarray*}

\begin{eqnarray*}
X_q^{11}(xq)^{-1}&=&{\sqrt{\Sq(e^{i\pi}x^{3}q^{4})
\Sq(e^{-i\pi}x^{3}q^{2})}\over \Sq(e^{i\pi}x^{3}q^6)} \cr \cr \cr
&=&{\sqrt{(1+e^{i\pi}x^{3}q)}\over (1-x^3q^3)}
{\sqrt{\Sq(e^{i\pi}x^{3}q^{-2})
\Sq(e^{-i\pi}x^{3}q^{2})}\over \Sq(e^{i\pi}x^{3})}.\end{eqnarray*}
We define
\bea X_q^{12}(x)&=&{\sqrt{\Sq(e^{i\pi}x^3q^{-2})
\Sq(e^{-i\pi}x^3q^{2})}\over\Sq(e^{i\pi}x^3)} \cr \cr \cr
&=&{\sqrt{\St(\qt^{-3}e^\theta e^{{2i\pi\over 3}})
\St(\qt^{3}e^\theta e^{-{2i\pi\over 3}})}\over\St(\qt^{-3}e^\theta)},
\label{eq: Xq12} \eea
noting that in the same way as for the $X_q^{11}(\theta)$ case,
$X_q^{12}(\theta)$ reproduces the correct semi-classical limit based
on the time delay $X^{12}(\theta)$. We then have for $X_q^{11}(x)$
\begin{eqnarray*}
X_q^{11}(x^{-1}q^{-1})&=&\sqrt{(1+e^{-i\pi}x^{-3}q^{-1})}X_q^{12}(x^{-1})^{-1}
\cr \cr \cr
X_q^{11}(xq)^{-1}&=&{\sqrt{(1+e^{i\pi}x^{3}q)}\over (1-x^3q^3)}X_q^{12}(x).
\end{eqnarray*}
We note that the classical `crossing' equation for
$X^{ij}(\theta)$ which in this case reads \\
$X^{11}(i\pi-\theta)=X^{21}(-\theta)^{-1}$,
allows us to organise $X_q^{11}(i\pi-\theta)$ as
$X_q^{21}(-\theta)^{-1}$ times some extraneous $x$-dependent factors
obtained after repeatedly using the quantum dilogarithm property
(\ref{eq: qdp}).

In the general case this classical crossing property 
$X^{ij}(i\pi-\theta)=X^{\ib j}(-\theta)^{-1}$,
will always guarantee that $X_q^{ij}(i\pi-\theta)$ can be written as 
$X_q^{\ib j}(-\theta)^{-1}$ times some extraneous factors. 
The combination ${X_q^{ij}(\theta)/X_q^{ij}(-\theta)}$ will cross into 
${X_q^{\ib j}(\theta)/X_q^{\ib j}(-\theta)}$ times these
additional factors.
These extraneous
factors are important in the method.

Hence we have the solution
\bea
v^{11}(i\pi-\theta)=v^{21}(\theta)=v^{12}(\theta)&=&
-{qe^{-\theta/2}e^{i\pi/2}\over
(1-x^{-3}q^{-1})}{X_q^{11}(x^{-1}q^{-1})\over
X_q^{11}(xq)} \cr \cr \cr
&=&
-{qe^{-\theta/2}e^{i\pi/2}\over(1-x^{-3}q^{-1})}
{\sqrt{(1+e^{-i\pi}x^{-3}q^{-1})(1+e^{i\pi}x^3q)}\over
(1-x^3q^3)}{X_q^{12}(\theta)\over X_q^{12}(-\theta)} \cr \cr \cr
&=&{e^{-\theta/2}\over(x^{-3/2}q^{-3/2}-x^{3/2}q^{3/2})}{X_q^{12}(\theta)\over
X_q^{12}(-\theta)}. \eea
We see that the prefactor in this solution for $v^{12}(\theta)$, 
$(x^{3/2}q^{3/2}-x^{-3/2}q^{-3/2})^{-1}$, which we derived from crossing
$X_{q}^{11}(\theta)$ is precisely the prefactor which is needed to
ensure the unitarity condition (\ref{eq: unitarityV}) for
$v^{12}(\theta)$. We have also implicitly computed the prefactor 
$(1-x^3q^2)^{-1}$
in $v^{11}(\theta)$, (\ref{eq: v11sol}), since this factor when
crossed was cancelled by the factor 
$$\sqrt{(1+e^{-i\pi}x^{-3}q^{-1})(1+e^{i\pi}x^3q)},$$
which was generated by the crossing of
$X_q^{11}(\theta)/X_q^{11}(-\theta)$. Therefore all the prefactors in 
$v^{11}(\theta)$ and $v^{12}(\theta)$ which are required for unitarity
are generated by the crossing.

To summarise, the postulated solutions are the following:
\bea
v^{11}(\theta)&=&{-qe^{\theta/2}\over (1-x^3q^2)}{X_q^{11}(\theta)\over
X_q^{11}(-\theta)}, \cr \cr \cr
v^{12}(\theta)=v^{21}(\theta)&=&{e^{-\theta/2}\over(x^{-3/2}q^{-3/2}-x^{3/2}q^{3/2})}{X_q^{12}(\theta)\over
X_q^{12}(-\theta)} \label{eq: proposed1}. \eea
Where
\bea
X_q^{11}(\theta)&=&{\Sq(e^{i\pi}x^3q^3)\over\sqrt{\Sq(e^{i\pi}x^3q)
\Sq(e^{-i\pi}x^3q^{-1})}} \cr \cr \cr
X_q^{12}(\theta)&=&{\sqrt{\Sq(e^{i\pi}x^3q^{-2})
\Sq(e^{-i\pi}x^3q^{2})}\over\Sq(e^{i\pi}x^3)}. \label{eq: proposed2}
\eea
\vskip 10pt
Several remarks about these solutions are in order. We have shown that
they formally satisfy the equations (\ref{eq: unitarityV}) and
(\ref{eq: crossing2a}) and that they have the correct semi-classical
behaviour given by the time delays $X^{11}(\theta)$ and
$X^{12}(\theta)$ in the limit $\gamma\rightarrow 0$. But it is not yet clear that the solutions are
single-valued. An
obvious obstacle to single-valuedness is the presence of the
square-roots of the quantum dilogarithms in $X_q^{ij}(\theta)$, the
zeroes and poles of the square-rooted quantum dilogarithms would
apparently appear as branch points in the S-matrix, which are, of course,
unnacceptable. To overcome this problem, we use the fact that (\ref{eq: Sinv})
\beq S_\mu(w)S_\mu(w^{-1})=e^{i(\log w)^2\over 4\mu}.e^{-{i\over 12}
(\mu+{\pi^2\over\mu})} \label{eq: Sinv2}, \eeq
(the second constant factor is irrelevant),
and note that in the denominator of $X_q^{11}(\theta)$, the arguments
of the quantum dilogarithms are of the form $x^3w$ and $x^3w^{-1}$, for
$w=e^{i\pi}q$. Hence in the combination
${X_q^{11}(\theta)/X_q^{11}(-\theta)}$,
which is required for the S-matrix, the square-rooted factors combine
to give
\vskip 10pt
$${\sqrt{\Sq(e^{i\pi}x^{-3}q)\Sq(e^{-i\pi}x^{-3}q^{-1})\over
\Sq(e^{i\pi}x^{3}q)\Sq(e^{-i\pi}x^{3}q^{-1})}}={C(\theta)\over
\Sq(e^{i\pi}x^{3}q)\Sq(e^{-i\pi}x^{3}q^{-1})},$$
\vskip 10pt 
\noindent where $C(\theta)$ is a single-valued function, related to the
first factor in (\ref{eq: Sinv2}). We see that the obstacle to
single-valuedness has disappeared since there are no square-roots.

We can similarly repeat this discussion on single-valuedness for the
other S-matrix $S^{12}(\theta)$, $X_q^{12}(\theta)$ is given by
equation (\ref{eq: proposed2}), and it again satisfies the crucial 
property that the square-rooted quantum
dilogarithms in the numerator have arguments of the form $x^3w$ and
$x^3w^{-1}$, where $w=e^{i\pi}q^{-2}$.

Another remark that we wish to make about the solution concerns the
choice of the coupling constant independent phases $e^{\pm i\pi}$ in
the arguments of the quantum dilogarithms. These disappear from the
semi-classical limit when we use the duality transformation (\ref{eq:
duality}) to take
the limit $\gamma\rightarrow 0$. They obviously must be of the form 
$e^{i\pi+2\pi n i}$, where $n$ is an integer, since only then do we
reproduce the correct factors by crossing which are required for unitarity
(\ref{eq: unitarityV}), with a relative minus sign in each
factor. Suppose we choose the first phase in $X_q^{11}(\theta)$,
(\ref{eq: Xq11}), in the
denominator to be $e^{i\pi+2\pi n i}$, the second phase in the
denominator is then fixed to be $e^{-(i\pi+2\pi n i)}$, as required for
single-valuedness, and then the phase of the quantum dilogarithm in
the numerator is then fixed by requiring the soliton bootstrap to
work, see below. We then cross this solution over to
$X_q^{12}(\theta)$ and so all the coupling constant independent phases
in $X_q^{12}(\theta)$ are completely fixed. This shows that the solution is completely
determined up to the arbitary integer $n$. The choice in the integer
$n$ moves the positions of the simple poles, due to the fusing
solitons, around. This particular choice, $n=0$, is chosen so that a
pole lies precisely where we would expect it, at the same place as the
simple pole in the classical $X^{ij}(\theta)$, which leads to the
classical fusion of the two-soliton solution into a third
single-soliton solution, at $\theta=iU^{\bar{k}}_{ij}$, the fusing
angle corresponding to the process $i+j\rightarrow k$.

As in sine-Gordon, it is always possible to write the solution with 
$n\neq 0$, as the
solution with $n=0$, multiplied by the CDD factors, obtained by
repeated application of (\ref{eq: qdp}) in the dual description. 
\vskip 0.5cm

\noindent{\bf The poles} 

We label the weights of the fundamental representations $3$,$\bar{3}$
of $su(3)$ as follows:
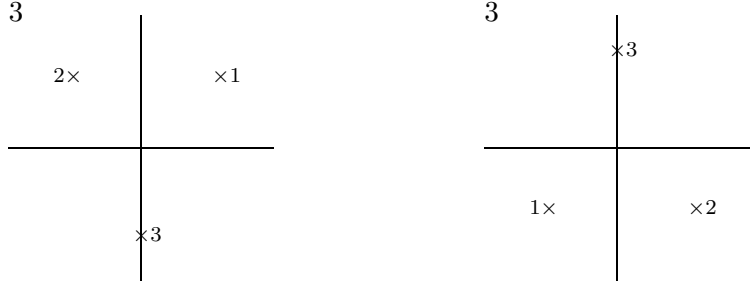
\begin{center}
\begin{picture}(270,120)(0,-20)
\put(0,50){\line(1,0){100}}
\put(50,0){\line(0,1){100}}
\put(47,15){\scs$\times$$3$}
\put(17,75){\scs$2$$\times$}
\put(77,75){\scs$\times$$1$}
\put(0,98){$3$}
\put(180,50){\line(1,0){100}}
\put(230,0){\line(0,1){100}}
\put(227,85){\scs$\times$$3$}
\put(180,98){$\bar{3}$}
\put(197,25){\scs$1$$\times$}
\put(257,25){\scs$\times$$2$}
\end{picture}
\end{center}
\noindent The poles of 
\beq {X_q^{11}(\theta)\over X_q^{11}(-\theta)}= 
C(\theta).{\Sq(e^{i\pi}x^3q^3)\Sq(e^{-i\pi} x^3q^{-3})\over
\Sq(e^{i\pi}x^3q)\Sq(e^{-i\pi}x^3q^{-1})}
\label{eq: polesof} \eeq \vskip 10pt
\noindent where $C(\theta)$ does not contain poles or zeroes, from the double
infinite product expansion of $\Sq(w)$ 
(\ref{eq: double2}), are at the following places:

The poles arising from the initially square-rooted factors (from
fusing solitons)
\bea
\theta &=&-i\biggl(\pi{(6l+2)\over 3}+{\gamma\over 6}(k+1)\biggr), \cr\cr
\theta &=&-i\biggl(\pi{(6l+4)\over 3}+{\gamma\over 6}(k)\biggr). \eea
Poles arising from the initially non-square-rooted factors
\bea \theta&=&i\biggl(2\pi l+{\gamma\over 6}(k+1)\biggr), \cr\cr
\theta &=&i\biggl(2\pi(l+1)+{\gamma\over 6}k\biggr), \eea
and we have zeroes at precisely minus these positions, where 
$k,l=0,1,2,\ldots$. On the physical strip these poles lie at precisely
$\theta=i{\gamma\over 6}(k+1)$.
There are no poles on the physical strip from the first set of
factors, and the zeroes lie at
$\theta=i\biggl({2\pi\over 3}+{\gamma\over 6}(k+1)\biggr)$.
There are no zeroes on the physical strip from the second set of factors.

Now consider actual S-matrix elements, rather than the solutions
$v^{11}(\theta)$ and $v^{12}(\theta)$. From explicit computations of
the matrices
$R^{11}(x)$ and $R^{12}(x)$, we have, (the lower indices denoting
topological charges, which are labelled by the diagrams above) 
\beq
S^{11}_{11}(\theta)=e^{\theta/2}x^{-3/2}
{X_q^{11}(\theta)\over X_q^{11}(-\theta)} \label{eq: S1111}, \eeq
and so the only poles on the physical strip are at
$\theta=i{{\gamma}(k+1)\over 6}$, $k=0,1,2,\ldots$. We interpret these as
crossed $S^{12}_{11}(\theta)$ breather poles, so that
$S^{12}_{11}(\theta)$ has direct channel breather poles at $\theta=
i\pi-i{{\gamma}(k+1)\over 6}$, $k=0,1,2,\ldots$. We do not expect to see
a pole due to the soliton fusion, $1+1\rightarrow 2$, since the
topological charges, or weights $1$ plus $1$ in the $3$ representation do not
give a weight in $\bar{3}$. We expect, of course, to see such a pole
in $S^{11}_{12\rightarrow 21}(\theta)$ (transmission),
since the weight $1$ in the representation
$3$ plus the weight $2$ in $3$ is equal to the weight $3$ in the
representation $\bar{3}$. Now
\beq S^{11}_{12\rightarrow 21}(\theta)
={\sinh({6\pi\theta\over\gamma})\over\sinh({6\pi\theta\over\gamma}-
{4\pi^2i\over\gamma})}x^{-3/2}e^{\theta/2}{X_q^{11}(\theta)\over
X_q^{11}(-\theta)}. \eeq
The prefactor contributes zeroes at 
$\theta={im\gamma\over 6},\  m\in {\Bbb Z}$, and poles at
 \beq \theta={2\pi i\over 3}+{i n\gamma\over 6},\quad
n\in {\Bbb Z}. \label{eq: 25} \eeq
Therefore the poles on the physical strip in $S^{11}_{12}(\theta)$ lie
at
$\theta={2\pi i\over 3}-{i n\gamma\over 6}$, $n=0,1,2,\ldots $.
The poles with $n>0$ in (\ref{eq: 25}) are cancelled by the
zeroes of $X_q^{11}(\theta)/X_q^{11}(-\theta)$, and the poles of 
$X_q^{11}(\theta)/X_q^{11}(-\theta)$ at $\theta=ik\gamma/6$,
$k=0,1,\ldots$, are cancelled by the zeroes of the prefactor at 
$\theta=i{m\gamma\over 6}$.\ 

The pole on the physical strip at $\theta={2\pi i\over 3}$
is due to the prefactor multiplying $\qquad$
$X_q^{11}(\theta)/X_q^{11}(-\theta)$, which is designed to ensure that
the unitarity condition is satisfied, and not
$X_q^{11}(\theta)/X_q^{11}(-\theta)$ itself. The factors due to the
fusing solitons (the ones that are square-rooted in
$X_q^{11}(\theta)$) do not apparently introduce poles onto the
physical strip. However the prefactors required for unitarity arise
from crossing $X_q^{11}(\theta)/X_q^{11}(-\theta)$ as explained
earlier. In fact for each $\sqrt{S(wx^3)S(w^{-1}x^3)}$ in the
denominator of $X_q^{ij}(\theta)$ we pick up a factor
$(1+wx^3)^{-1}$ for the prefactor in $v^{ij}(\theta)$, and for the
factor $S(wx^3)$ in the denominator of $X_q^{ij}(\theta)$ we also pick up a
factor $(1+wx^3)^{-1}$. This means that the simple poles in
$X^{ij}(\theta)$, which were previously explained in terms of the
explicit classical fusing of solitons, are also at the same place in
the S-matrix, provided they are not cancelled by zeroes in 
${X_q^{ij}(\theta)/X_q^{ij}(-\theta)}$,
as discussed in section 3.
\vskip 0.5cm

\noindent{\bf The soliton bootstrap}

We must check the internal consistency of the proposed result
(\ref{eq: proposed1}). We must check that the simple poles interpreted
as solitons do in fact generate S-matrix elements consistent with the
solutions so far derived. 
The method that is available for doing this is the so-called
bootstrap method, as discussed in section 4, which allows us to compute
new S-matrices from a given S-matrix. We can calculate the S-matrix
for scattering the state interpreted as the bound state in
the given S-matrix (specified by the pole) with one of the original states in
the  given S-matrix.
\vskip 0.5cm

\noindent{\bf The ground state soliton case}

We check this first for the ground state soliton, given by the pole 
at $\theta={2\pi i\over 3}$, in $S^{11}(\theta)$. We interpret it as pole due
to the soliton of species 2, corresponding to the fusing
$1+1\rightarrow 2$. 

We must check that the following matrix equation holds
\beq
S^{12}(\theta)=(I\otimes S^{11}(\theta+{i\pi\over
3}))(S^{11}(\theta-{i\pi\over 3})\otimes I)\Bigl |_{3\otimes\bar{3}},
\label{eq: 1sb} \eeq
where the notation means that we restrict the right-hand side which is
valued in the space $3\otimes 3\otimes 3$ to the
space $3\otimes\bar{3}$.
By using equation (\ref{eq: Sinv}), we write $v^{11}(\theta)$ and 
$v^{12}(\theta)$ in the form
\beq
v^{11}(\theta)={\St(\qt^{-3}e^\theta e^{-i\pi})\St(\qt^{3}e^\theta
e^{i\pi})\over\St(\qt^3 e^{-2\pi i+\theta}e^{i\pi\over
3})\St(\qt^{-3}e^\theta e^{-{i\pi\over 3}})},\eeq
and
\beq
v^{12}(\theta)={\St(\qt^3 e^{\theta}e^{-{2\pi i\over
3}})\St(\qt^{-3}e^\theta e^{{2\pi i\over 3}})\over\St(\qt^{-3}e^{-2\pi
i+\theta})\St(\qt^{3}e^\theta)},\eeq
and we strip off any possible constant factors, and factors of the
form $e^{\kappa\theta}$, in front of these expressions.

In order to check the bootstrap we compute
$$v^{11}(\theta+{i\pi\over 3})v^{11}(\theta-{i\pi\over 3}).$$ 
In order to be able to cancel the $\St(w)$ when we compute 
this, it is helpful if we
replace {\it all} $\St(\qt^{-3}w)$ with  $\St(\qt^{3}w)$, and in the
process pick up a factor of the form $(1+e^\theta e^{2\pi i p\over h})$,
each time we make the replacement. So we write
\bea
v^{11}(\theta)&=&{(1+e^\theta e^{-{i\pi\over 3}})\over 
(1+e^\theta e^{-i\pi})}.{\St(\qt^{3}e^\theta e^{-i\pi})\St(\qt^{3}e^\theta
e^{i\pi})\over\St(\qt^3 e^{-2\pi i+\theta}e^{i\pi\over
3})\St(\qt^{3}e^\theta e^{-{i\pi\over 3}})} \cr \cr \cr
&=&B^{11}(\theta)^{-1}\tilde{v}^{11}(\theta), \eea
where $$B^{11}(\theta)^{-1}={(1+e^\theta e^{-{i\pi\over 3}})\over 
(1+e^\theta e^{-i\pi})},$$
and similarly
\bea
v^{12}(\theta)&=&{(1+e^\theta)\over 
(1+e^\theta e^{{2\pi i\over 3}})}.{\St(\qt^3 e^{\theta}e^{-{2\pi i\over
3}})\St(\qt^{3}e^\theta e^{{2\pi i\over 3}})\over\St(\qt^{3}e^{-2\pi
i+\theta})\St(\qt^{3}e^\theta)}\cr\cr\cr
&=&B^{12}(\theta)^{-1}\tilde{v}^{12}(\theta), \eea
where $$B^{12}(\theta)^{-1}={(1+e^\theta)\over 
(1+e^\theta e^{{2\pi i\over 3}})}.$$
These equations serve to define $\tilde{v}^{ij}(\theta)$ and
$B^{ij}(\theta)$. $B^{ij}(\theta)$ can be considered as `half' of the classical
$X^{ij}(\theta)$, in the sense that 
\beq
X^{ij}(\theta)={\rm const}.B^{ij}(\theta).B^{ij}(-\theta). \eeq
For example, in $su(3)$,
$$ B^{11}(\theta) B^{11}(-\theta)=-e^{i\pi\over 3}{(1+e^\theta
e^{-i\pi})(1+e^\theta e^{i\pi})\over
(1+e^\theta e^{-{i\pi\over 3}}) (1+e^\theta e^{{i\pi\over 3}})}
=-e^{i\pi\over 3}X^{11}(\theta).$$
In the same way that the bootstrap equation is satisfied by the
classical $X^{ij}(\theta)$, equation (\ref{eq: class_boot}),
the bootstrap equation, in this example, is also satisfied
by only this `half' of $X^{ij}(\theta)$, namely $B^{ij}(\theta)$. 

For $su(3)$, we check the bootstrap equation for
$B^{ij}(\theta)$
$$B^{11}(\theta+\frac{i\pi} 3)B^{11}(\theta-\frac{i\pi} 3)=B^{12}(\theta),$$
$${(1+e^\theta e^{-{2\pi i\over 3}})\over(1+e^\theta)}
{(1+e^\theta e^{-{4\pi i\over 3}})\over(1+e^\theta e^{-{2\pi i\over
3}})}
={(1+e^\theta  e^{{2\pi i\over 3}})\over(1+e^\theta)}.$$
Now the bootstrap equation for the full classical $X^{ij}(\theta)$,
equation (\ref{eq: class_boot}), guarantees
that the bootstrap is satisfied by $\tilde{v}^{ij}(\theta)$ (although
we shall see this in detail below), except
for factors of the form $(1-x^hq^p)$, obtained by altering the
argument of $\St(w)$ by $e^{2\pi i}$. We have already dealt with the
factors of the form $(1+e^\theta e^{2\pi i p\over h})$, which are
required when we alter the argument by $\qt^{-6}$. Now
\bea
\tilde{v}^{11}(\theta+\frac{i\pi} 3)\tilde{v}^{11}(\theta-\frac{i\pi} 3)&=&
{\St(\qt^{3}e^\theta e^{-{2\pi i\over 3}})\St(\qt^{3}e^\theta
e^{4\pi i\over 3})\over\St(\qt^3 e^{-2\pi i+\theta}e^{2\pi i\over
3})\St(\qt^{3}e^\theta)}
{\St(\qt^{3}e^\theta e^{-{4\pi i\over 3}})\St(\qt^{3}e^\theta
e^{2\pi i\over 3})\over\St(\qt^3 e^{-2\pi
i+\theta})\St(\qt^{3}e^\theta e^{-{2\pi i\over
3}})}\cr\cr\cr
&=&{\St(\qt^{3}e^\theta e^{2\pi i\over 3})\St(\qt^{3}e^\theta
e^{4\pi i\over 3})\over\St(\qt^3 e^{-2\pi
i+\theta})\St(\qt^{3}e^\theta)}\cr\cr\cr
&=&{1\over 1-x^3q^{-1}}{\St(\qt^{3}e^\theta e^{2\pi i\over 3})\St(\qt^{3}e^\theta
e^{-{2\pi i\over 3}})\over\St(\qt^3 e^{-2\pi i+\theta})\St(\qt^{3}e^\theta)}\cr\cr\cr
&=&{1\over 1-x^3q^{-1}}\tilde{v}^{12}(\theta), \eea
and hence 
\beq{v}^{11}(\theta+\frac{i\pi} 3){v}^{11}(\theta-\frac{i\pi} 3)=
{1\over 1-x^3q^{-1}}{v}^{12}(\theta).\label{eq: andhence} \eeq
The factor that we have picked up here, $(1-x^3q^{-1})^{-1}$, is
accounted for by the discrepancy between the normalisation of 
$$(I\otimes
R^{11}(xq^{-1/3}))(R^{11}(xq^{1/3})\otimes I)\Bigl|_{3\otimes\bar{3}},$$
and $R^{12}(x)$ defined with no extraneous factors, that is with no
overall factors multiplying each element of the matrix. $R^{11}(x)$
must cross into $R^{12}(x)$ with no extraneous factors, since we have
imposed $v^{11}(i\pi - \theta)=v^{12}(\theta)$. By defining
$R^{12}(x)$ like this, to be in its lowest factorised form, we guarantee that
this crossing condition holds. It is possible to show
that 
\beq (x^{3/2}q^{-1/2}-x^{-3/2}q^{1/2})R^{12}(x)=(I\otimes
R^{11}(xq^{-1/3}))(R^{11}(xq^{1/3})\otimes I)\Bigl|_{3\otimes\bar{3}}
.\label{eq: star}\eeq
Hence the correct bootstrap equation for $S$, equation (\ref{eq: 1sb}),
follows.
\resection{The general solution}
\vskip 0.5cm

\noindent{\bf Properties and definition of $X_q^{ij}(\theta)$}

We shall discover that the solution to the soliton S-matrix is 
$$v^{ij}(\theta)=f(x).{X_q^{ij}(\theta)\over X_q^{ij}(-\theta)},$$
where we shall shortly define $X_q^{ij}(\theta)$, and
where $f(x)$ consists of products of the form $(1-x^hq^p)^{-1}$, and 
are related to the unitarity condition since
$v^{ij}(\theta)v^{ji}(-\theta)=f(x).f(x^{-1})$.

For $c(i)-c(j)\ge 0$, we define
\beq
X_q^{ij}(\theta)=\prod_{p=0}^{h-1}S_{q^{-h}}\biggl(e^{is_p\pi}x^h
q^{-{(c(i)-c(j))\over 2}}
q^{-2p+h}\biggr)^{\gamma_i\cdot\sigma^p\gamma_j\over 2}, 
\label{eq: Xqij1} \eeq
for $$x=e^{4\pi\theta\over\gamma}, \qquad
q=e^{-{4\pi^2i\over\gamma}}.$$
With $s_p=\pm 1$, $e^{is_p\pi}=e^{\pm i\pi}$ are certain phases to be
defined below. 

For $c(i)-c(j)=-2$, we define
\beq
X_q^{ij}(\theta)=\prod_{p=1}^{h}S_{q^{-h}}\biggl(e^{is_p\pi}x^h
q^{-{(c(i)-c(j))\over 2}}
q^{-2p+h}\biggr)^{\gamma_i\cdot\sigma^p\gamma_j\over 2}. 
 \label{eq: Xqij2}  \eeq
The differences in the ranges of $p$ in the two definitions are
designed to ensure that powers of $q$ in the arguments of $S_{q^{-h}}$
are between and including $h$ and $-h$. This in turn ensures that
poles due to fusing solitons and breathers are in the correct places.
The mod $2\pi i$ difference in the positions of the poles obtained by
replacing $p$ with $p+h$ in (\ref{eq: Xqij1}) and (\ref{eq: Xqij2}) is
important, since the S-matrices are not $2\pi i$ periodic. 

Observe that for $c(i)-c(j)=0$ and $i\neq j$, $\gamma_i.\gamma_j=0$,
so that the $p=0$ term of (\ref{eq: Xqij1}) and the $p=h$ term of 
(\ref{eq: Xqij2}) vanish. Hence (\ref{eq: Xqij1}) and 
(\ref{eq: Xqij2}) are  the same in this case. If $i=j$ however, there
is an important difference, and we must take (\ref{eq: Xqij1}).
\vskip 0.5cm

\noindent{\bf The semi-classical limit}

Reverting to the dual picture (\ref{eq: duality}), with $c(i)-c(j)\ge
0$, we have
\beq
X_q^{ij}(\theta)=\prod_{p=0}^{h-1}S_{\qt^{-h}}\biggl(\qt^{-s_p h}e^\theta
e^{{i\pi\over 2h}(c(i)-c(j))}e^{{2\pi i p\over h} -i\pi}\biggr)^
{\gamma_i\cdot\sigma^p\gamma_j\over 2} \eeq
where $$\qt^{-h}=e^{\gamma i\over 4h}.$$
It is easy to see that\footnote{Bearing in mind the sine-Gordon and
$su(3)$ cases.} 
$$\lim_{\gamma\rightarrow 0} X_q^{ij}(\theta)={\rm const}.e^{{2hi\over\gamma}\int_0^\theta
d\theta' \bigl(\log\bigl(X^{ij}(\theta')\bigr)^{1/2}\bigr)}.$$  
Hence 
\beq
\lim_{\gamma\rightarrow 0}{X_q^{ij}(\theta)\over
X_q^{ij}(-\theta)}={\rm const}.e^{{2hi\over\gamma}\int_0^\theta
d\theta' \log(X^{ij}(\theta'))}, \label{eq: semi-class} \eeq
where we have used the fact that $X^{ij}(\theta)=X^{ij}(-\theta)$. 
\vskip 0.5cm

\noindent{\bf Single-valuedness}

$${X_q^{ij}(\theta)\over
X_q^{ij}(-\theta)} {\rm\ \  is\ single\ valued. }
\qquad\qquad\qquad\qquad\qquad\qquad\qquad\qquad\qquad\qquad\qquad\qquad$$
{\bf Proof}

We only need to consider the factors with  $\gamma_i\cdot\sigma^p\gamma_j=-1$,
if $\gamma_i\cdot\sigma^p\gamma_j=1$, we cross the result so that these
factors are in the denominator. 

For $\gamma_i\cdot\sigma^p\gamma_j=-1$ and $$p'=h-p+{c(j)-c(i)\over
2},$$ then \cite{Fring},
$\gamma_i\cdot\sigma^{p'}\gamma_j=-1$. In fact all the values of $p$, such that
$\gamma_i\cdot\sigma^p\gamma_j=-1$, pair up in this way. If $p=p'$, then
$\gamma_i\cdot\sigma^p\gamma_j= -2$ (when $i=\jb$), or 
$\gamma_i\cdot\sigma^p\gamma_j=0$.

Hence for  $$T_p(\theta)=S_{\qt^{-h}}\biggl(\qt^{-s_p h}e^\theta
e^{{i\pi\over 2h}(c(i)-c(j))}e^{{2\pi i p\over h} -i\pi}\biggr)^{-1/2}
S_{\qt^{-h}}\biggl(\qt^{-s_p' h}e^\theta
e^{{i\pi\over 2h}(c(i)-c(j))}e^{{2\pi i p'\over h}
-i\pi}\biggr)^{-1/2},$$
the argument of the second quantum dilogarithm is 
$$\qt^{-s_{p'} h}e^\theta
e^{{i\pi\over 2h}(c(i)-c(j))}e^{{2\pi i p'\over h}
-i\pi}=\qquad\qquad\qquad\qquad$$
$$\qquad\qquad\qquad\qquad 
\qt^{-s_{p'}h}e^\theta e^{-{i\pi\over 2h}(c(i)-c(j))}e^{-{2\pi i p\over h}
+i\pi}.$$
We choose the phases, $s_p=\pm 1$, in such a way so that $s_p=-s_{p'}$.
Hence 
$$T_p(\theta)=S_{\qt^{-h}}\bigl(e^\theta w\bigr)^{-1/2}
S_{\qt^{-h}}\bigl(e^\theta w^{-1}\bigr)^{-1/2},$$
with $$w=\qt^{-s_p h}e^{{i\pi\over 2h}(c(i)-c(j))}e^{{2\pi ip\over
h}-i\pi}.$$
This completes the proof because (\ref{eq: Sinv})
$$S_{\qt^{-h}}\bigl(e^\theta w^{-1}\bigr)^{-1/2}=({\rm
single\ valued\ factor}).S_{\qt^{-h}}\bigl(e^{-\theta}
w\bigr)^{1/2}. $$
Then $${T_p(\theta)\over T_p(-\theta)}=({\rm single\ valued\ factor}).
{1\over S_{\qt^{-h}}\bigl(e^\theta w\bigr)S_{\qt^{-h}}\bigl(e^\theta
w^{-1}\bigr)},$$
which is single-valued.

${X_q^{ij}(\theta)/X_q^{ij}(-\theta)}$ is made up of products of
$T_p(\theta)/T_p(-\theta)$ with $p\ge \frac h2
+\frac{c(j)-c(i)}4$, say,
and the corresponding crossed factors, when
$\gamma_i\cdot\sigma^p\gamma_j=1$, and the already single-valued
factors when $\gamma_i\cdot\sigma^p\gamma_j=\pm 2$.  
\vskip 0.5cm

\noindent{\bf The phases $e^{is_p\pi}$} 

We define the set of integers $A^{ij}_{->}$ to be the set $p$ such
that $\gamma_i\cdot\sigma^p\gamma_j=-1$ and with 
$p\ge \frac h2 +\frac{c(j)-c(i)}4$.
If $c(i)-c(j)\ge 0$, we take $p\in\{0,\ldots,h-1\}$, otherwise 
$p\in\{1,\ldots,h\}$. 

Each $p$, such that
$\gamma_i\cdot\sigma^p\gamma_j=-1$,
pairs with another value of $p$, $p'$, where
$p'=h-p+{c(j)-c(i)\over 2}$,
and we have excluded the smaller of $\{p,p'\}$ from $A^{ij}_{->}$.
We similarly define the set of integers $A^{ij}_{-<}$ by excluding the
larger of $\{p,p'\}$.

By crossing we can also define the set $A_{+>}^{ij}$ in a similar way
but for those $p$ with $\gamma_i\cdot\sigma^p\gamma_j=1$ and the
larger of the two values $\{p,p'\}$ such that
$\gamma_i\cdot\sigma^p\gamma_j=\gamma_i\cdot\sigma^{p'}\gamma_j$.
And in the obvious way, we also define $A_{+<}^{ij}$. In fact 
$A^{ij}_{+<}$ and $A^{\ib j}_{->}$ are related by $p\in A^{ij}_{+<}$
if and only if $p+\frac h2 + \frac{c(i)-c(\ib)}4\in A^{\ib j}_{->}$,
where mod $h$ is implicitly not understood.
$A^{ij}_{+>}$ and $A^{\ib j}_{-<}$ are related by $p\in A^{ij}_{+>}$
if and only if $p-\frac h2 + \frac{c(i)-c(\ib)}4\in A^{\ib j}_{-<}$,
see equation (\ref{eq: below}).

We choose the phase $e^{is_p\pi}$ as follows:
$$s_p=-1,\qquad {\rm for\ }p\in A^{ij}_{->},$$
\beq s_p=+1,\qquad {\rm for\ }\gamma_i\cdot\sigma^p\gamma_j=-2.
\label{eq: rules} \eeq
For single-valuedness we must demand that 
$s_p=+1$, for $p\in A^{ij}_{-<}$.
The remaining values of $p$ give $\gamma_i\cdot\sigma^p\gamma_j=1,2$,
and the remaining $s_p$ are fixed by the crossing symmetry, since when
we cross an $X_q^{ij}(\theta)$, those factors with
$\gamma_i\cdot\sigma^p\gamma_j=1,2$ in $X_q^{ij}(\theta)$, are in the
denominator of $X_q^{\ib j}(\theta-i\pi)$, and are therefore treated
by the first set of rules (\ref{eq: rules}). They are fixed as
$$p'\in A^{\ib j}_{->},\qquad s_{p}=-1,$$
$$p'\in A^{\ib j}_{-<},\qquad s_{p}=+1,$$
$$\gamma_{\ib}\cdot\sigma^{p'}\gamma_j=-2,\qquad s_{p}=+1.$$
where $p'=p+\frac h2+{c(i)-c(\ib)\over 4}$ mod $h$.
This is because \cite{Fring},
$\gamma_i=-\sigma^{-\frac h2-{c(i)-c(\ib)\over 4}}\gamma_{\ib}$,
and therefore
\beq\gamma_i\cdot\sigma^p\gamma_j=-\gamma_{\ib}\cdot\sigma^{p+\frac h2+
{c(i)-c(\ib)\over 4}}\gamma_j.\label{eq: below} \eeq 
Note that these rules for $s_p$ mean that for $p$ in the lower half of
its range and $\gamma_i\cdot\sigma^p\gamma_j=\pm 1$, 
$s_p=-\gamma_i\cdot\sigma^p\gamma_j$.
In the upper half of the range, $s_p=\gamma_i\cdot\sigma^p\gamma_j$.
If $\gamma_i\cdot\sigma^p\gamma_j=\pm 2$, then $s_p=1$.
\vskip 0.5cm

\noindent{\bf Symmetry: $X_q^{ij}(\theta)=X_q^{ji}(\theta)$} \ 

We refer to the proof of the symmetry $X^{ij}(\theta)=X^{ji}(\theta)$ 
in the classical case, see \cite{FJKO}.

\noindent We again use 
$\gamma_i\cdot\sigma^p\gamma_j=\gamma_j\cdot\sigma^{p'}\gamma_i$,
with
$$2p+{c(i)-c(j)\over 2}=2p'+{c(j)-c(i)\over 2}.$$
If $c(i)-c(j)=0$, $p=p'$ and the result is trivially true.

\noindent If $c(i)-c(j)=2$, $p+1=p'$, and
\bea X_q^{ij}(\theta)&=&\prod_{p=0}^{h-1}S_{q^{-h}}\biggl(e^{is_p\pi}x^h
q^{-{(c(i)-c(j))\over 2}}
q^{-2p+h}\biggr)^{\gamma_i\cdot\sigma^p\gamma_j\over 2} \cr \cr \cr
&=&\prod_{p'=1}^{h}S_{q^{-h}}\biggl(e^{is_{p'-1}\pi}x^h
q^{-{(c(j)-c(i))\over 2}}
q^{-2p'+h}\biggr)^{\gamma_j\cdot\sigma^{p'}\gamma_i\over 2} \cr \cr
&=&X_q^{ji}(\theta) .\eea
Note that the phases $e^{is_p\pi}$ come out correctly, since in this
case $A^{ij}_{\pm >}+1=A^{ji}_{\pm >}$ and $A^{ij}_{\pm
<}+1=A^{ji}_{\pm <}$, where there is no implicit understanding that we
take $p$ mod $h$. If
$c(i)-c(j)=-2$, we swap $i$ and $j$ in the proof above. 
\vskip 0.5cm

\noindent{\bf The crossing property}

For 
\beq v^{ij}(\theta)=A^{ij}e^{c^{ij}\theta}{1\over\prod_{p\in
A_{->}^{ij}}\biggl(1-x^hq^{-{c(i)-c(j)\over 2}}q^{2h-2p}\biggr)}
{X_q^{ij}(\theta)\over X_q^{ij}(-\theta)}, \label{eq: vij} \eeq
with the aditional factor $(1-x^hq^h)^{-1}$ on the right-hand side, if
$i=\jb$.
Then we have 
$v^{ji}(i\pi-\theta)=v^{\ib j}(\theta)$.
Here $A^{ij}$ and $c^{ij}$ are certain constants. 

We also trivially have
$v^{ij}(\theta)=v^{ji}(\theta)$.

\noindent {\bf Proof}
 
We refer to the proof of the classical crossing property, equation
(\ref{eq: class_cross}, in section 3, 
$X^{ij}(\theta+i\pi)=X^{\ib j}(\theta)^{-1}$.
For $c(i)-c(j)\ge 0$,
$$X_q^{ij}(\theta)=\prod_{p=0}^{h-1}S_{q^{-h}}\biggl(e^{is_p\pi}x^h
q^{-{(c(i)-c(j))\over 2}}
q^{-2p+h}\biggr)^{\gamma_i\cdot\sigma^p\gamma_j\over 2}. $$
From \cite{Fring}
$$\gamma_i=-\sigma^{-\frac h2-{c(i)-c(\ib)\over 4}}\gamma_{\ib},$$
and we compute $X_q^{ij}(\theta+i\pi)$,
$$X_q^{ij}(\theta+i\pi)=\prod_{p=0}^{h-1}S_{q^{-h}}\biggl(e^{is_p\pi}x^h
q^{-{(c(i)-c(j))\over 2}}
q^{-2p}\biggr)^{-{\gamma_{\ib}\cdot\sigma^{p+\frac
h2+{c(i)-c(\ib)\over 4}}\gamma_j\over 2}}.$$
Let $p'=p+\frac h2+{c(i)-c(\ib)\over 4}=p+r$,
where $r$ is an integer and is positive,
$$=\prod_{p'=r}^{h-1+r}S_{q^{-h}}\biggl(e^{is_p\pi}x^h
q^{-{(c(\ib)-c(j))\over 2}}
q^{-2p'+h}\biggr)^{-{\gamma_{\ib}\cdot\sigma^{p'}\gamma_j\over
2}}. $$
For $c(\ib)-c(j)\ge 0$,
\bea &=&\prod_{p'=0}^{h-1}S_{q^{-h}}\biggl(e^{is_p\pi}x^h
q^{-{(c(\ib)-c(j))\over 2}}
q^{-2p'+h}\biggr)^{-{\gamma_{\ib}\cdot\sigma^{p'}\gamma_j\over
2}}.\prod_{p=0}^{r-1}\biggl(1-x^h q^{-{c(\ib)-c(j)\over
2}}q^{-2p}\biggr)^{\gamma_{\ib}\cdot\sigma^{p}\gamma_j\over 2} \cr \cr \cr
&=&X_q^{\ib j}(\theta)^{-1}.
\prod_{p=0}^{r-1}\biggl(1-x^h q^{-{c(\ib)-c(j)\over
2}}q^{-2p}\biggr)^{\gamma_{\ib}\cdot\sigma^{p}\gamma_j\over 2}. \eea

\noindent Note that the $s_p$ have been defined so that $s_p^{ij}=s_{p'}^{\ib
j}$, and the limits
of the range of $p$ are appropriate for the value of $c(i)-c(j)$ (see
the definitions of $A^{ij}_{\pm >}$ and  $A^{ij}_{\pm <}$, the
integers in these sets have been defined to lie in the appropriate
ranges) and the range of $p'$ appropriate for the value of $c(\ib)-c(j)$.
We also compute $X_q^{ij}(\theta-i\pi)$
\bea X_q^{ij}(\theta-i\pi)&=&\prod_{p=0}^{h-1}S_{q^{-h}}\biggl(e^{is_p\pi}x^h
q^{-{(c(i)-c(j))\over 2}}
q^{-2p+2h}\biggr)^{-{\gamma_{\ib}\cdot\sigma^{p+\frac
h2+{c(i)-c(\ib)\over 4}}\gamma_j\over 2}} \cr \cr \cr
&=&\prod_{p'=r}^{h-1+r}S_{q^{-h}}\biggl(e^{is_p\pi}x^h
q^{-{(c(\ib)-c(j))\over 2}}
q^{-2p'+3h}\biggr)^{-{\gamma_{\ib}\cdot\sigma^{p'}\gamma_j\over
2}}. \eea
For $c(\ib)-c(j)\ge 0$,
\bea &=&\prod_{p'=0}^{h-1}S_{q^{-h}}\biggl(e^{is_p\pi}x^h
q^{-{(c(\ib)-c(j))\over 2}}
q^{-2p'+h}\biggr)^{-{\gamma_{\ib}\cdot\sigma^{p'}\gamma_j\over
2}}.\prod_{p=r}^{h-1}\biggl(1-x^h q^{-{c(\ib)-c(j)\over
2}}q^{2h-2p}\biggr)^{-{\gamma_{\ib}\cdot\sigma^{p}\gamma_j}\over 2} \cr \cr \cr
&=&X_q^{\ib j}(\theta)^{-1}
\prod_{p=r}^{h-1}\biggl(1-x^h q^{-{c(\ib)-c(j)\over
2}}q^{2h-2p}\biggr)^{-{\gamma_{\ib}\cdot\sigma^{p}\gamma_j}\over 2}.
 \eea

\noindent Combining these two together
$${X_q^{ij}(i\pi-\theta)\over X_q^{ij}(-(i\pi-\theta))}=
{X_q^{\ib j}(\theta)\over X_q^{\ib j}(-\theta)}
 \prod_{p=r}^{h-1}\biggl(1-x^h q^{-{c(\ib)-c(j)\over
2}}q^{2h-2p}\biggr)^{\gamma_{\ib}\cdot\sigma^{p}\gamma_j\over 2}
\prod_{p=0}^{r-1}\biggl(1-x^{-h} q^{-{c(\ib)-c(j)\over
2}}q^{-2p}\biggr)^{\gamma_{\ib}\cdot\sigma^{p}\gamma_j\over 2},$$
letting $p'=h-p+{c(j)-c(\ib)\over 2}$ in the second product,
\bea 
&=&{X_q^{\ib j}(\theta)\over X_q^{\ib j}(-\theta)}
 \prod_{p=r}^{h-1}\biggl(1-x^h q^{-{c(\ib)-c(j)\over
2}}q^{2h-2p}\biggr)^{\gamma_{\ib}\cdot\sigma^{p}\gamma_j\over 2}
\prod_{p'=h+{c(j)-c(\ib)\over 2}-(r-1)}^{h+{c(j)-c(\ib)\over 2}}\biggl(1-x^{-h} q^{{c(\ib)-c(j)\over
2}}q^{2p'-2h}\biggr)^{\gamma_{\ib}\cdot\sigma^{p'}\gamma_j\over 2} \cr \cr
\cr
&=& Ae^{c\theta}{X_q^{\ib j}(\theta)\over X_q^{\ib j}(-\theta)}
{\prod_{p\in A^{\ib j}_{+>}}\biggl(1-x^{-h}q^{c(\ib)-c(j)\over
2}q^{-(2h-2p)}\biggr)\over 
\prod_{p\in A^{\ib j}_{->}}\biggl(1-x^{h}q^{-{c(\ib)-c(j)\over
2}}q^{2h-2p}\biggr)}, \label{eq: sdf} \eea
if $j=\ib$, there is the additional factor $(1-x^{-h})$ on the
right-hand side of equation (\ref{eq: sdf}), if $j=i$ then there is the
additional factor $(1-x^hq^h)^{-1}$.

Putting $f(\theta)$ as the numerator of the products in (\ref{eq: sdf}),
$$f(\theta)=\prod_{p\in A^{\ib j}_{+>}}\biggl(1-x^{-h}q^{c(\ib)-c(j)\over
2}q^{-(2h-2p)}\biggr).$$
If $p\in A_{+>}^{\ib j}$, then 
$p-\frac h2-{c(i)-c(\ib)\over 4}\in A_{-<}^{ij}$,
and thus
$$p'=\frac{3h}2 -p+{c(i)-c(\ib)\over 4} + {c(j)-c(i)\over 2}\in
A^{ij}_{->}.$$
Then
\bea f(i\pi-\theta)&=&\prod_{p'\in A^{ij}_{->}}\biggl(1-x^h
q^{c(\ib)-c(j)\over 2}q^{ -h + 3h -2p'-({c(\ib)+c(i)\over
2})+c(j)}\biggr) \cr \cr \cr
&=&\prod_{p'\in A^{ij}_{->}}\biggl(1-x^{h}q^{-{c(i)-c(j)\over
2}}q^{2h-2p'}\biggr).
 \eea
If $j=\ib$, the additional factor, $(1-x^{-h})$, should be included in
$f(\theta)$, when $\theta\rightarrow i\pi-\theta$, this factor becomes
$(1-x^hq^h)$. We see that this is the same as the factor present in
the denominator of equation (\ref{eq: sdf}), if $i=j$.

Hence we have established that for
\beq
v^{ij}(\theta)=A^{ij}e^{c^{ij}\theta}{1\over\prod_{p\in
A_{->}^{ij}}\biggl(1-x^hq^{-{c(i)-c(j)\over 2}}q^{2h-2p}\biggr)}
{X_q^{ij}(\theta)\over X_q^{ij}(-\theta)}, \label{eq: vij2} \eeq
with the aditional factor, $(1-x^hq^h)^{-1}$, when $i=\jb$,
then we have the crossing condition $
v^{ij}(i\pi-\theta)=v^{\ib j}(\theta)$.

The proof, giving the same formula for $v^{ij}$ (\ref{eq:
vij2}), is modified slightly for the case $c(\ib)-c(j)<0$, and then 
subsequently for the case $c(i)-c(j)<0$, but remains largely
unchanged. These cases are left to the reader. 

We note that for $p\in A^{ij}_{->}$, 
$p'=p-({c(j)-c(i)\over 2})\in A^{ji}_{->}$,
where mod $h$ is not implicit in this last equation.
Therefore
$$\prod_{p\in
A_{->}^{ij}}\biggl(1-x^hq^{-{c(i)-c(j)\over 2}}q^{2h-2p}\biggr)=
\prod_{p'\in
A_{->}^{ji}}\biggl(1-x^hq^{-{c(j)-c(i)\over 2}}q^{2h-2p'}\biggr).$$
Since we have already shown that $X_q^{ij}(\theta)=X_q^{ji}(\theta)$,
we have trivially established the symmetry 
$v^{ij}(\theta)=v^{ji}(\theta)$.
The correct crossing condition then follows for $v^{ij}(\theta)$,
$v^{ji}(i\pi-\theta)=v^{\ib j}(\theta)$.

\resection{Comparison with the known R-matrices}

We compare the proposed solutions (\ref{eq: vij}) with known $R$-matrices,
which have been computed in \cite{Delius1}. As
outlined in section 5, it is possible to compute $R^{ij}(\theta)$ in
terms of the projection matrices that project onto the representations
in the decomposition of $V_i\tensor V_j$ into representations. It is
only possible to do this when the multiplicity is one for each of the
representations in the decomposition. This does not happen for all
$i$, and $j$, in general. However the method always works for the
fundamental elements $R^{11}(x)$ (with slight alterations for $D_n$,
so we must use $R^{nn}(x)$ if $n$ is odd, and $R^{nn}(x)$, $R^{n,
n-1}(x)$, $R^{n-1,n-1}(x)$, and $R^{11}(x)$, if $n$ is even), and we
shall compare our conjectures in these cases. Then by application of
the soliton bootstrap we should be able to recover all the elements 
$S^{ij}(\theta)$, not explicitly needing all the other matrices 
$R^{ij}(\theta)$. 

{\noindent\bf Comparison with the fundamental element
$R^{11}(\theta)$}

Here we shall find that the fundamental element $v^{11}(\theta)$
 always has the correct unitarity condition given directly by the
prefactors from (\ref{eq: vij}), recall that these were derived from
the crossing symmetry.

\noindent $A_n$:

From \cite{Delius1}, equation (4.17), with $$<n>={x-q^n\over 1-xq^n},$$
in the homogeneous gradation,
$$R^{11}(x)=P_0 + <2>P_1,$$
where $P_0+P_1=1$, and $P_1$ projects onto the fundamental
representation $V_2$. We move to the principal gradation by replacing
$x$ with $x^h$ and conjugating by the matrix $\sigma(x)$. We also
change the normalisation by multiplying by $(1-x^h q^2)$\footnote{It
is crucial to do this because only then will this matrix be able to
cross precisely into another  matrix with no extraneous factors,
which must be normalized in the same way,
so that all the denominators in $<n>\ldots <m>$ are cleared. By
inspection this is the only
way of getting the matrices to cross into each other. }.We must adopt 
this normalisation, by clearing all the denominators of $<n>\ldots
<m>$, in all the $R$-matrices which are to be checked. This is an
important point, since if we had adopted the normalisation initially
given, we would have had the unitarity equation
$R^{11}(x)R^{11}(x^{-1})=1$, we would then have had to remove the
prefactor in $v^{11}(\theta)$, given by equation (\ref{eq: vij}), and
hence we would have generated factors on crossing $v^{11}(\theta)$, these
compensate for factors needed when we cross the matrix
$R^{11}(x)$, depending on the normalisation chosen for the crossed matrix.
These different ways of dealing with the factors are equivalent,
since it is only a question of normalisation, but it seems clearer to
arrange the normalisation so the crossing condition on the $v$ is
always
$v^{ij}(i\pi-\theta)=v^{\ib j}(\theta)$. This is not the approach used
in \cite{H}.

So, in the principal gradation, we have
$$R^{11}_P(x)=\sigma P_0 \sigma^{-1}(1-x^h q^2) + \sigma P_1
\sigma^{-1}(x^h-q^2),$$
note that at the pole in $v^{11}(\theta)$, corresponding to the fusing
$1+1\rightarrow 2$, $\theta={2\pi i\over h}$, or $(1-x^hq^2)=0$ so
that $R_P^{11}(x)$ projects onto the the fundamental representation
$V_2$.

We then compute 
\bea
 R_P^{11}(x)R_P^{11}(x^{-1})&=&\sigma (P_0(1-x^h
q^2)(1-x^{-h}q^2) + P_1 (x^h-q^2)(x^{-h}-q^2))\sigma^{-1} \cr\cr
&=&(1-x^hq^2)(1-x^{-h}q^2)\sigma(P_0+P_1)\sigma^{-1} \cr\cr
&=&(1-x^hq^2)(1-x^{-h}q^2).1 \eea

Now the prefactor in the $A_n$ theories in $v^{11}(\theta)$, given by
equation (\ref{eq: vij}), is (from the fusing $1+1\rightarrow 2$)
\footnote{The most direct route to finding the prefactors is to use
the fusing angle for the appropriate fusing, the prefactor must place a
pole at this fusing angle. These angles are available from affine Toda
particle S-matrix data\cite{Dorey2,BCDS}. }
$${1\over 1-x^h q^2},$$
so that 
$$v^{11}(x)v^{11}(x^{-1})={1\over (1-x^h q^2)(1-x^{-h}q^2)},$$
and the above calculation for $R_P^{11}(x)R_P^{11}(x^{-1})$ shows that
$S^{11}(\theta)S^{11}(-\theta)=1$, as required.\ \ 

\noindent $D_n$, $h=2n-2$,

For $n$ even we need to check $R^{11}(x)$, and we may as well also
check $R^{11}(x)$ for $n$ odd. So for all $n$, from \cite{Delius1},
equation (4.49), (noting that there is a printing error in passing
from (4.48) to (4.49), making equation (4.49) slightly incorrect), we
have

\beq R^{11}(x)=P_0 + <2>P_1 + <2><h>P_2, \label{eq: Dn} \eeq
$P_1$ projects onto $V_2$ and $P_2$ onto the trivial representation
(for breathers). Hence, after adopting the special normalisation needed
for crossing, as has been discussed above,
$$R_P^{11}(x)=\sigma P_0\sigma^{-1}(1-x^hq^2)(1-x^hq^h)+ 
\sigma P_1\sigma^{-1}(x^h-q^2)(1-x^hq^h) + \sigma P_2 \sigma^{-1}(x^h-q^2)(x^h-q^h),$$
and
\bea
R^{11}_P(x)R^{11}_P(x^{-1})&=&(1-x^hq^2)(1-x^hq^h).(1-x^{-h}q^2)(1-x^{-h}q^h)
\sigma(P_0+P_1+P_2)\sigma^{-1}\cr\cr
&=&(1-x^hq^2)(1-x^hq^h).(1-x^{-h}q^2)(1-x^{-h}q^h).1 \eea

The fusings are $1+1\rightarrow 0, 2$, and the prefactor for
$v^{11}(\theta)$, from (\ref{eq: vij}), is
$${1\over(1-x^h q^2)(1-x^h q^h)},$$
and this agrees with $R^{11}_P(x)R^{11}_P(x^{-1})$.

For $n$ even we need to check $R^{n n}(x)$ and $R^{n-1 n}(x)$, from
(4.44) and (4.45) of \cite{Delius1},
$$R^{n n}(x)=P_0'+\sum_{a=1}^{\frac n2}\prod_{i=1}^a<4i-2>P_{n-2a}$$
where here $P_{n-2a}$ projects onto the space $V_{n-2a}$, and the
fusings are 
$n+n\rightarrow 0,2,4,\ldots, n-2$, and the prefactor to $v^{n n}(\theta)$
is, from equation (\ref{eq: vij}),
$${1\over (1-x^h q^2)(1-x^h q^6)\ldots (1-x^h q^h)},$$
this agrees exactly with the factors derived from 
$R_P^{n n}(x)R_P^{n n}(x^{-1})$, provided we normalise $R_P^{n n}(x)$
in the manner that has already been discussed.

For $n$ even, we also have 
$$R^{n,n-1}(x)=P_0'+\sum_{a=1}^{\frac n2-1}\prod_{i=1}^a<4i>P_{n-2a-1},$$
and again $P_{n-2a-1}$ projects onto $V_{n-2a-1}$, the fusings are
$n+(n-1) \rightarrow 1,3,5,\ldots n-3$, and hence the prefactor to 
$v^{n,n-1}(\theta)$, from equation (\ref{eq: vij}), is
$${1\over (1-x^h q^4)(1-x^h q^8)\ldots (1-x^h q^{h-2})},$$
this agrees with the $R^{n n-1}(x)$.

For $n$ odd, we must check $R^{n n}(x)$, this is the fundamental
element in that case.
$$R^{n n}(x)=P_0'+\sum_{a=1}^{\frac {n-1}2}\prod_{i=1}^a<4i-2>P_{n-2a}$$
where here $P_{n-2a}$ projects onto the space $V_{n-2a}$, the fusings
are $n+n\rightarrow 1,3,5,\ldots n-2$, and the prefactor to $v^{nn}(\theta)$  
is 
$${1\over (1-x^hq^2)(1-x^h q^6)(1-x^h q^{10})\ldots (1-x^h q^{h-2})},$$
this agrees with $R^{n n}(x)$. \ \ 

\noindent $E_6$:

Equation (4.55) from \cite{Delius1} is 
$$R^{11}(x)=P_0 + <2>P_1 + <2><8> P_2,$$
$P_1$ projects onto $V_3$, and $P_2$ projects onto $V_6$. The fusings
are
$1+1\rightarrow 2, 6$, implying the prefactor to $v^{11}(\theta)$
$${1\over(1-x^h q^2)(1-x^h q^8)},$$
this agrees with $R^{11}(x)$. \ \ 

\noindent $E_7$:
Equation (89) of \cite{Delius2}, gives
$$R^{11}(x)=P_0 + <2>P_1 +<2><10>P_2 + <2><10><18>P_3,$$
where $P_1$ projects onto $V_4$, $P_2$ onto $V_2$ and $P_3$ onto
$V_0$. The fusings are $1+1\rightarrow 0, 2, 4$, and these imply the
prefactor to $v^{11}(\theta)$
$${1\over (1-x^h q^2)(1-x^h q^{10})(1-x^h q^{18})}.\qquad$$
Unfortunately the case $E_8$ has not been worked out in
\cite{Delius1,Delius2}, and so we cannot check the fundamental element of
$E_8$. However in principle, given the work \cite{Delius1,Delius2},
it should not be too difficult to find 
$R^{11}(x)$, also in this case, and check to see if it agrees with
our $v^{11}(\theta)$.
\vskip 0.5cm

{\noindent\bf Comparison with other elements}

\noindent $A_n$:

All fundamental representations $V_i$ and $V_j$ are amenable to the
method and, for $N={\rm min}(j,n-i)$,
$$R^{ij}(x)=P_0 + <2+i-j>P_1 + <2+i-j><4+i-j>P_2 \qquad\qquad\qquad\qquad$$ 
$$ + 
<2+i-j><4+i-j><6+i-j>P_3 + \cdots $$ 
\beq \qquad\qquad\qquad\qquad\qquad
+<2+i-j><4+i-j>\ldots<2N+i-j>P_N, \label{eq: An}\eeq
where $P_N$ projects onto $V_{i+j}$.
We observe that 
$$R^{i1}(\theta)=P_0 + <1+i>P_1,$$
implying after the appropriate normalising that
$$R_P^{11}(x)R_P^{11}(x^{-1})=(1-x^h q^{1+i})(1-x^{-h}q^{1+i})1,$$ 
and this agrees with the prefactor to $v^{1i}(\theta)$, namely
$(1-x^h q^{1+i})^{-1}$.

However, if there are more than two representations in the
decomposition of $V_i\tensor V_j$, we do not get agreement with
the prefactor in $v^{ij}(\theta)$, and further corrections to
$v^{ij}(\theta)$ will be needed.
For example
$$R^{22}(x)=P_0 + <2>P_1 + <2><4>P_2,$$
from $R_P^{22}(x)R_P^{22}(x^{-1})$ 
this implies the prefactor to $v^{22}(\theta)$,
$((1-x^h q^2)(1-x^h q^4))^{-1}$,
but the prefactor is, from (\ref{eq: vij})
$(1-x^h q^4)^{-1}$,
and so we require a `correction' $(1-x^h q^2)^{-1}$ to
$v^{22}(\theta)$. We must then separately cross over this extra factor
when we compute $v^{2 2}(i\pi-\theta)=v^{\bar{2} 2}(\theta)=v^{n-1,2}(\theta)$.
We will now see whether an extra factor is required for $v^{n-1,
2}(\theta)$, and if it is consistent with the crossed factor
previously found for $v^{22}(\theta)$.
$$R^{n-1,2}(\theta)=P_0+<h-2>P_1 + <h-2><h>P_2,$$
the prefactor already present in $v^{n-1,2}(\theta)$ is 
$(1-x^h q^h)^{-1}$,
and so an extra factor is needed, and must be $(1-x^h q^{h-2})^{-1}$.
When crossed, $x\rightarrow x^{-1}q^{-1}$, this extra factor becomes
\beq {1\over 1-x^{-h}q^{-2}}={-x^h q^2\over (1-x^h q^2)}.
\label{eq: extrafac} \eeq
The extra  $-x^h q^2$ is not important, and this (\ref{eq: extrafac}) 
is essentially
the extra prefactor needed for $v^{22}(\theta)$. 

These corrections are consistent with
the bootstrap, but this is harder to see because of the differences in
the normalisation of $R^{ij}(x)$, defined in its lowest factorised
form, and $R^{ij}(x)$ defined by fusion, as discussed for $su(3)$.
However, in some cases, there is no difference in this normalisation,
for example in $A_6$,
$$(1\tensor R^{12}(\theta+{i\pi\over 7}))(R^{12}(\theta-\frac{i\pi}7)\tensor 1)\bigl |_{V_2\tensor V_2}= R^{22}(\theta),$$
where here, $R^{ij}(\theta)$ must have the special normalisation given by
clearing the denominators of $<n>..<m>$ in equation (\ref{eq: An}). By
counting powers of $(1-x^hq^p)$ on both sides of this equation, we see
that the normalisations agree. Both sides have a power two.

We check the bootstrap equation for $v^{ij}(\theta)$, in this specific
case:

From (\ref{eq: vij2p}),
\bea
\tilde{v}^{12}(\theta)&=&{S_{\qt^{-h}}(\qt^h e^\theta e^{6\pi i\over 7})
S_{\qt^{-h}}(\qt^h e^\theta e^{-{6\pi i\over 7}}) \over
S_{\qt^{-h}}(\qt^h e^{-2\pi i+\theta} e^{4\pi i\over 7})
S_{\qt^{-h}}(\qt^h e^{\theta} e^{-{4\pi i\over 7}})} \cr\cr\cr
\tilde{v}^{22}(\theta)&=&{S_{\qt^{-h}}(\qt^h e^\theta e^{-i\pi})
S_{\qt^{-h}}(\qt^h e^\theta e^{i\pi})\over
S_{\qt^{-h}}(\qt^h e^{-2\pi i+\theta} e^{3\pi i\over 7})
S_{\qt^{-h}}(\qt^h e^{\theta} e^{-{3\pi i\over 7}})}, \eea
and
$$\tilde{v}^{12}(\theta+\frac{i\pi}7)
\tilde{v}^{12}(\theta-\frac{i\pi}7)\qquad\qquad\qquad\qquad\qquad\qquad\qquad\qquad\qquad\qquad\qquad\qquad$$
\bea
&=&{S_{\qt^{-h}}(\qt^h e^\theta e^{i\pi})
S_{\qt^{-h}}(\qt^h e^\theta e^{-{5\pi i\over 7}}) \over
S_{\qt^{-h}}(\qt^h e^{-2\pi i+\theta}e^{5\pi i\over 7})
S_{\qt^{-h}}(\qt^h e^{\theta}e^{-{3\pi i\over 7}})}
{S_{\qt^{-h}}(\qt^h e^\theta e^{5\pi i\over 7})
S_{\qt^{-h}}(\qt^h e^\theta e^{-\pi i}) \over
S_{\qt^{-h}}(\qt^h e^{-2\pi i+\theta}e^{3\pi i\over 7})
S_{\qt^{-h}}(\qt^h e^{\theta}e^{-{5\pi i\over 7}})} \cr\cr\cr
&=&{S_{\qt^{-h}}(\qt^h e^\theta e^{\pi i})
S_{\qt^{-h}}(\qt^h e^\theta e^{-\pi i}) \over
S_{\qt^{-h}}(\qt^h e^{-2\pi i+\theta}e^{3\pi i\over 7})
S_{\qt^{-h}}(\qt^h e^{\theta}e^{-{3\pi i\over 7}})}.{1\over 1-x^h q^2}
\cr\cr\cr
&=&\tilde{v}^{22}(\theta){1\over 1-x^h q^2}.\eea
This indicates that we must correct
$v^{22}(\theta)$ by 
$(1-x^h q^2)^{-1}$,  
and this, of course, agrees with what we said before.  
We can read off these additional corrections required for
$v^{ij}(\theta)$, directly from (\ref{eq: An}). For simplicity we restrict
ourselves to $A_6$, although it is immediately obvious how to extend
this to $A_n$. We write $a$ for the factor $(1-x^h q^a)^{-1}$.
\begin{center}
\begin{tabular}{||l|c|c|c|c|c|c||} \hline\hline
$i\backslash j$ &  $1$ & $2$ & $3$ & $4$ & $5$ & $6$ \\ \hline
$1$           &$\scriptstyle\surd$     & $\scriptstyle\surd$    & $\scriptstyle\surd$    &  $\scriptstyle \surd$   &  $\scriptstyle \surd$   & $\scriptstyle \surd$    \\ \hline
$2$         & $\scriptstyle -$ & $\scriptstyle 2$ &
$\scriptstyle 3$ & $\scriptstyle 4 $& $\scriptstyle 5$ &  $\scriptstyle \surd$
\\ \hline
$3$         & $\scriptstyle -$ & $\scriptstyle -$& $\scriptstyle
2\ 4$ & $\scriptstyle 3\ 
5$ & $\scriptstyle 4$ & $\scriptstyle \surd$\\ \hline
$4$         & $\scriptstyle -$ & $\scriptstyle -$ &$\scriptstyle -$ &
$\scriptstyle 2\ 
4$ & $\scriptstyle 3$ & $\scriptstyle\surd$ \\ \hline 
$5$         & $\scriptstyle -$ & $\scriptstyle - $& $\scriptstyle -$&
$\scriptstyle -$  & $\scriptstyle 2$ & $\scriptstyle\surd$
 \\ \hline 
$6$           &  $\scriptstyle -$   &   $\scriptstyle -$   &  $\scriptstyle -$  &  $\scriptstyle -$    &  $\scriptstyle -$   &  $\scriptstyle\surd$   \\ \hline\hline
\end{tabular}
\end{center}
\begin{center}{\rm Table 1: Corrections to $v^{ij}(\theta)$ for $A_6$}
\end{center}
For the $A_n$ theories, the normalisation of the fused R-matrices can
be determined completely \cite{H2}, and we can show that this table of
corrections is consistent with the bootstrap on the $v^{ij}(\theta)$.
This will be sketched in section 13. For the $D$ and $E$ series, there
will be similar corrections, which have not yet been determined. The
reason for this is that, so far, we cannot determine the
normalisation of the fused R-matrices. The trick that was used for $A_n$,
cannot be repeated for the other algebras.
\resection{The poles of the general solution}

$v^{ij}(\theta)$, as defined by equation (\ref{eq: vij}), has simple poles on
the physical strip at the following places:

\noindent For $p\in A^{ij}_{->}$, $$\theta=i\biggl(2\pi-{2\pi p\over
h}-{\pi\over 2h}(c(i)-c(j))-{\gamma n\over 2h}\biggr),\qquad
n=0,1,2,\ldots$$
If $i=\jb$, we have breather poles at 
$$\theta=i\pi-i{\gamma n\over 2h},\qquad n=1,2,3,\ldots $$
If $i=j$, then we have crossed breather poles at
$$\theta=i{\gamma n\over 2h},\qquad n=1,2,3,\ldots $$

\noindent {\bf Proof}

\noindent With the specific choices of the phases $e^{is_p\pi}$, that have
already  been specified, we have
$$
{X_q^{ij}(\theta)\over X_q^{ij}(-\theta)}=Ae^{c\theta}\prod_{p\in A^{ij}_{->}}
S_{\qt^{-h}}\bigr(\qt^h e^\theta e^{{i\pi\over 2h}(c(i)-c(j))}e^{{2\pi i
p\over h}-i\pi}\bigl)^{-1}
S_{\qt^{-h}}\bigr(\qt^{-h} e^\theta e^{-{i\pi\over 2h}(c(i)-c(j))}e^{-{2\pi i
p\over h}+i\pi}\bigl)^{-1} \qquad\qquad\qquad\qquad $$
$$\qquad\qquad\qquad\qquad \times\prod_{p\in A^{\ib j}_{->}}
S_{\qt^{-h}}\bigr(\qt^h e^{-i\pi+\theta} e^{{i\pi\over 2h}(c(\ib)-c(j))}e^{{2\pi i
p\over h}-i\pi}\bigl)
S_{\qt^{-h}}\bigr(\qt^{-h} e^{i\pi+\theta} e^{-{i\pi\over 2h}(c(\ib)-c(j))}e^{-{2\pi i
p\over h}+i\pi}\bigl)$$ 
\beq\qquad\qquad\qquad\qquad\qquad\qquad\qquad\qquad\times{S_{\qt^{-h}}\bigl(\qt^{-h}e^\theta e^{-i\pi}\bigr)\over 
S_{\qt^{-h}}\bigl(\qt^{-h}e^{-\theta} e^{-i\pi}\bigr)}\biggl |_
{{\rm if\ }i=j} 
{S_{\qt^{-h}}\bigl(\qt^{-h}e^{-\theta}\bigr)\over 
S_{\qt^{-h}}\bigl(\qt^{-h}e^{\theta}\bigr)}\biggl |_
{{\rm if\ }i=\jb},  \label{eq: Xqij3} \eeq
where $A$ and $c$ are certain, irrelevant, constants.

This form (\ref{eq: Xqij3}) for 
${X_q^{ij}(\theta)/X_q^{ij}(-\theta)}$ is an alternative method of
seeing the crossing property (\ref{eq: sdf}). 

We can also write 
$$
v^{ij}(\theta)=A'e^{{c'}\theta}\prod_{p\in A^{ij}_{->}}
S_{\qt^{-h}}\bigr(\qt^h e^{-2\pi i + \theta} e^{{i\pi\over 2h}(c(i)-c(j))}e^{{2\pi i
p\over h}-i\pi}\bigl)^{-1}
S_{\qt^{-h}}\bigr(\qt^{-h} e^\theta e^{-{i\pi\over 2h}(c(i)-c(j))}e^{-{2\pi i
p\over h}+i\pi}\bigl)^{-1} \qquad$$
$$\qquad\times\prod_{p\in A^{\ib j}_{->}}
S_{\qt^{-h}}\bigr(\qt^h e^{-i\pi+\theta} e^{{i\pi\over 2h}(c(\ib)-c(j))}e^{{2\pi i
p\over h}-i\pi}\bigl)
S_{\qt^{-h}}\bigr(\qt^{-h} e^{i\pi+\theta} e^{-{i\pi\over 2h}(c(\ib)-c(j))}e^{-{2\pi i
p\over h}+i\pi}\bigl)$$ 
\beq\qquad\qquad\qquad\qquad\times{S_{\qt^{-h}}\bigl(\qt^{-h}e^\theta
e^{-i\pi}\bigr)S_{\qt^{-h}}\bigl(\qt^{h}e^{\theta}
e^{i\pi}\bigr)\over 1} \biggl |_ {{\rm if\ }i=j}
{1\over
S_{\qt^{-h}}\bigl(\qt^{-h}e^{-2\pi i + \theta}\bigr)
S_{\qt^{h}}\bigl(\qt^{h}e^{\theta}\bigr)
}\biggl |_ {{\rm
if\ }i=\jb}, \label{eq: vij2p} \eeq 

Consider the
contribution in (\ref{eq: Xqij3}) from $p\in A^{ij}_{->}$, from
(\ref{eq: double2}), we have simple poles at
$$\theta=-i\biggl(\pi(2k)+{2\pi p\over h}+{\pi\over 2h}(c(i)-c(j)) +
 {\gamma\over 4h}(2l)\biggr),$$
and at
$$\theta=-i\biggl(\pi(2k+2)-{2\pi p\over h}-{\pi\over 2h}(c(i)-c(j)) +
 {\gamma\over 4h}(2l+2)\biggr),$$
and zeroes at precisely minus these positions, where
$k,l=0,1,2,\ldots$ On the physical strip, there are no poles, and
zeroes at 
$$\theta=i\biggl({2\pi p\over h}+{\pi\over 2h}(c(i)-c(j)) +
 {\gamma\over 2h}l\biggr),$$
and 
$$\theta=i\biggl(2\pi-{2\pi p\over h}-{\pi\over 2h}(c(i)-c(j)) +
 {\gamma\over 2h}(l+1)\biggr).$$
In fact since $p\ge\frac h2 + \frac{c(j)-c(i)}4$, the zeroes from
the first set are actually not on the physical strip.

\noindent For 
$$v^{ij}(\theta)=A^{ij}e^{c^{ij}\theta}{1\over\prod_{p\in
A_{->}^{ij}}\biggl(1-x^hq^{-{c(i)-c(j)\over 2}}q^{2h-2p}\biggr)}
{X_q^{ij}(\theta)\over X_q^{ij}(-\theta)}.$$
The prefactor contributes simple poles at 
$$e^{4\pi\theta h\over\gamma}e^{{2\pi i n}}=e^{-{4\pi^2
i\over\gamma}{c(i)-c(j)\over 2}}e^{-{4\pi^2 i\over\gamma}(2p-2h)},$$
or  
$$\theta=i\biggl(2\pi-{2\pi p\over h}-{\pi\over 2h}(c(i)-c(j)) -
 {\gamma\over 2h}n\biggr),\qquad n\in {\Bbb Z}. $$
Therefore the poles with $n=-1,-2,\ldots$ are cancelled by the zeroes,
and we are left with poles at
$$\theta=i\biggl(2\pi-{2\pi p\over h}-{\pi\over 2h}(c(i)-c(j)) -
 {\gamma\over 2h}n\biggr),\qquad n=0,1,2,\ldots $$
in $v^{ij}(\theta)$.

In fact we have chosen the phases $e^{is_p\pi}$ in such a way so that
there is a pole at $n=0$ corresponding to the ground state fusing
soliton. This is at precisely the same position as the pole present in
the classical $X^{ij}(\theta)$, see section 3.

Now consider the contribution from $p\in A^{\ib j}_{->}$ in (\ref{eq:
Xqij3}), these poles are essentially in the crossed position of the poles
from $v^{\ib j}(\theta)$.

On the physical strip the poles are at 
\bea \theta&=&-i\pi+i\biggl({2\pi p\over h}+{\pi\over 2h}(c(\ib)-c(j)) +
 {\gamma\over 2h}l\biggr) \cr
&=&i\pi-i\biggl(2\pi-{2\pi p\over h}-{\pi\over 2h}(c(\ib)-c(j)) -
 {\gamma\over 2h}l\biggr),\qquad l=0,1,2,\ldots \eea

\noindent If $i=\jb$, there are no poles in (\ref{eq: Xqij3}) on the physical
strip, but there are zeroes at
$$\theta=i\biggl(\pi(2k+1)+{\gamma\over 2h}l\biggr),$$
for $k,l=0,1,2,\ldots$, so that there is a zero at
$\theta=i\pi$. From the prefactor $(1-x^hq^h)^{-1}$ in the equation 
(\ref{eq: vij}) for $v^{ij}(\theta)$, there are poles at 
$$\theta=i\pi-i{\gamma\over 2h}n,\qquad n\in {\Bbb Z}.$$
The zero at $\theta=i\pi$ cancels the pole at $\theta=i\pi$ and we therefore
have breather poles at 
$$\theta=i\pi-i{\gamma \over 2h}n,\qquad n=1,2,3,\ldots$$
If $i=j$, we pick up crossed breather poles in $v^{ij}(\theta)$ at
$$\theta=i{\gamma\over 2h}(l+1),\qquad l=0,1,2,\ldots $$
with no contribution from the prefactor.

\resection{The general soliton bootstrap}

Starting from $v^{11}(\theta)$, given by equation (\ref{eq: vij2p}),
we follow through the bootstrap, comparing the result obtained with
the $v^{ij}(\theta)$, defined by (\ref{eq: vij2p}), with the
additional corrections of the form $(1-x^hq^p)^{-1}$ discussed in the
$A_n$ case. Following the $su(3)$ case, discussed in section 9, the
coefficients $B^{ij}(\theta)$ are defined by turning all the 
$\tilde{q}^{-h}$'s in the arguments of the $S_{\tilde{q}^{-h}}(w)$'s
into $\tilde{q}^{h}$.

Define the block $$(x)_\theta={\sinh(\frac\theta 2 + \frac{i\pi
x}{2h})\over\sinh(\frac\theta 2 - \frac{i\pi x}{2h})}.$$
We start with $A_6$. 

$$A_6,\qquad\qquad c(1)=-1,\qquad $$
\begin{center}
\begin{tabular}{||l|r|r|r|r|r|r|r|r|l||} \hline\hline
$p,q$           & $0$ & $1$ & $2$ & $3$ & $4$ & $5$ & $6$ & $7$ & $h=7$ \\ \hline
$\gamma_1\cdot\sigma^p\gamma_1$ & $2$ & $-1$ & $0$ & $0$ &$0$ & $0$ & $-1$
&  & 
$p'=7-p$ \\ \hline
$\lambda_1\cdot\sigma^q\gamma_1$ & $-1$ & $0$& $0$  & $0$  & $0$ &
$0$ & $1$ &  &
$q'=6-q$  \\ \hline
$s_p$             & $1$ & $1$ & $0$  & $0$ & $0$ & $0$ & $-1$ &  &   \\ \hline\hline
\end{tabular}
\end{center}

We calculate $B^{ij}(\theta)$ by taking the values of $p$,  such that
$s_p=1$, from the table above. For each value of $p$, we include the factor
\beq (1+e^{{\pi i\over h}(2p+{c(i)-c(j)\over 2})-i\pi}e^\theta)^{{\rm sign}
(\gamma_i\cdot\sigma^p\gamma_j)1} \label{eq: strange} \eeq
in $B^{ij}(\theta)$.

So $$B^{11}(\theta)={(1+e^{i\pi}e^\theta)\over 
(1+e^{-{5\pi i\over 7}}e^\theta)}.$$
For $B^{12}(\theta)$, we have

$$A_6,\qquad\qquad c(1)=-1,\ c(2)=1, \qquad $$
\begin{center}
\begin{tabular}{||l|r|r|r|r|r|r|r|r|l||} \hline\hline
$p,q$           & $0$ & $1$ & $2$ & $3$ & $4$ & $5$ & $6$ & $7$ & $h=7$ \\ \hline
$\gamma_1\cdot\sigma^p\gamma_2$ & & $1$ & $-1$ & $0$ & $0$ &$0$ & $-1$ & $1$
&  
$p'=8-p$ \\ \hline
$\lambda_1\cdot\sigma^q\gamma_2$ & & $-1$ & $0$ & $0$  & $0$  & $0$ &
$1$ & $0$ &
$q'=7-q$  \\ \hline
$s_p$            & & $-1$ & $1$ &   &  &  & $-1$ & $1$ &    \\ \hline\hline
\end{tabular}
\end{center}
and therefore, from the values of $p$ such that $s_p=1$,
$$B^{12}(\theta)={(1+e^{6\pi i\over 7}e^\theta)\over 
(1+e^{-{4\pi i\over 7}}e^\theta)},$$
we check the bootstrap, from the fusing $1+1\rightarrow 2$, as follows,
\bea
B^{11}(\theta-\frac{i\pi}7)B^{11}(\theta+\frac{i\pi}7)&=&
{(1+e^{6\pi i\over 7}e^\theta)\over (1+e^{-{6\pi i\over 7}}e^\theta)}
{(1+e^{-{6\pi i\over 7}}e^\theta)\over (1+e^{-{4\pi i\over
7}}e^\theta)} \cr\cr\cr
&=&{(1+e^{{6\pi i\over 7}}e^\theta)\over (1+e^{-{4\pi i\over
7}}e^\theta)}=B^{12}(\theta). \eea
In this case the bootstrap checks out, and no corrections are needed.

Now, from the fusing $1+1\rightarrow 2$, we compute $B^{22}(\theta)$. 
\bea
B^{12}(\theta+\frac{i\pi}7)B^{12}(\theta-\frac{i\pi}7)&=&
{(1+e^{i\pi}e^\theta)\over (1+e^{-{3\pi i\over 7}}e^\theta)}
{(1+e^{{5\pi i\over 7}}e^\theta)\over (1+e^{-{5\pi i\over
7}}e^\theta)} \cr\cr\cr
&=&{(1+e^{i\pi}e^\theta)\over (1+e^{-{3\pi i\over 7}}e^\theta)}
{\sinh(\frac\theta 2- {2\pi i\over 14})\over 
\sinh(\frac\theta 2+ {2\pi i\over 14})} \cr\cr\cr
&=&B^{22}(\theta)(2)^{-1}_\theta, \eea
but
$$B^{22}(\theta)=
{(1+e^{i\pi}e^\theta)\over (1+e^{-{3\pi i\over 7}}e^\theta)}.$$
Hence we have to `correct' $v^{22}(\theta)$, as given in equation
(\ref{eq: vij2p}), by $(2)_\theta$, since
$v^{ij}(\theta)=B^{ij}(\theta)^{-1}\tilde{v}^{ij}(\theta)$. This
introduces another pole at $\theta={2\pi i\over 7}$ on the physical
strip. Observe that there is a double pole in this position in the
particle S-matrix $S^{22}(\theta)=\{1\}_\theta\{3\}_\theta$, which is
not explained in terms of fusings, but as a Landau
singularity \cite{BCDS}.

We continue with the bootstrap on $B^{ij}(\theta)$. For the fusing 
$1+2\rightarrow 3$,
$$B^{13}(\theta)={(1+e^{{5\pi i\over 7}}e^\theta)\over (1+e^{-{3\pi i\over
7}}e^\theta)},$$ 
\bea
B^{11}(\theta-\frac{2\pi i}7)B^{12}(\theta+\frac{i\pi}7)&=&
{(1+e^{5\pi i\over 7}e^\theta)\over (1+e^{-{i\pi}}e^\theta)}
{(1+e^{{i\pi}}e^\theta)\over (1+e^{-{3\pi i\over
7}}e^\theta)} \cr\cr\cr
&=&{(1+e^{5\pi i\over 7}e^\theta)\over (1+e^{-{3\pi i\over 7}}e^\theta)}
\cr\cr\cr
&=&B^{13}(\theta), \eea
so the bootstrap checks out, without introducing any extraneous
factors.

We compute $B^{32}(\theta)$ from the fusing $1+2\rightarrow 3$,
$$B^{32}(\theta)={(1+e^{6\pi i\over 7}e^\theta)\over (1+e^{-{2\pi i\over 7}}e^\theta)},$$
\bea
B^{12}(\theta+\frac{2\pi i}7)B^{22}(\theta-\frac{i\pi}7)&=&
{(1+e^{8\pi i\over 7}e^\theta)\over (1+e^{-{2\pi i\over 7}}e^\theta)}
{(1+e^{{6\pi i\over 7}}e^\theta)\over (1+e^{-{4\pi i\over
7}}e^\theta)} {(1+e^{{4\pi i\over 7}}e^\theta)\over (1+e^{-{6\pi i\over
7}}e^\theta)} 
\cr\cr\cr
&=&{(1+e^{6\pi i\over 7}e^\theta)\over (1+e^{-{2\pi i\over
7}}e^\theta)}
{(1+e^{4\pi i\over 7}e^\theta)\over (1+e^{-{4\pi i\over 7}}e^\theta)}
=B^{32}(\theta)(3)_\theta^{-1}, \eea
here we have already included the $(2)_\theta^{-1}$ in
$B^{22}(\theta)$.

Consider the fusing $1+3\rightarrow 4$, after including the correction
$(3)_\theta^{-1}$ in $B^{32}(\theta)$,
\bea
B^{32}(\theta+\frac{\pi i}7)B^{12}(\theta-\frac{3\pi i}7)&=&
{(1+e^{i\pi}e^\theta)\over (1+e^{-{\pi i\over 7}}e^\theta)}
{(1+e^{{5\pi i\over 7}}e^\theta)\over (1+e^{-{3\pi i\over
7}}e^\theta)} {(1+e^{{3\pi i\over 7}}e^\theta)\over (1+e^{-{i\pi}}e^\theta)} 
\cr\cr\cr
&=&{(1+e^{5\pi i\over 7}e^\theta)\over (1+e^{-{\pi i\over
7}}e^\theta)}
{(1+e^{3\pi i\over 7}e^\theta)\over (1+e^{-{3\pi i\over 7}}e^\theta)}
=B^{42}(\theta)(4)_\theta^{-1}, \eea
where
$$B^{42}(\theta)={(1+e^{5\pi i\over 7}e^\theta)\over (1+e^{-{\pi i\over
7}}e^\theta)}.$$
However, we must demand that (with the corrections)
$v^{32}(i\pi-\theta)=v^{\bar{3}2}(\theta)=v^{42}(\theta)$, and we must
check that the corrections cross over appropriately. It is easy to show
$(3)_{i\pi-\theta}=(4)_\theta$, as required.

Continuing with the bootstrap with the fusing $1+2\rightarrow 3$, we
compute $B^{33}(\theta)$,
$$B^{33}(\theta)={(1+e^{i\pi}e^\theta)\over (1+e^{-{\pi i\over
7}}e^\theta)}.$$
But 
\bea
B^{23}(\theta+\frac{\pi i}7)B^{13}(\theta-\frac{2\pi i}7)&=&
{(1+e^{i\pi}e^\theta)\over (1+e^{-{\pi i\over 7}}e^\theta)}
{(1+e^{{5\pi i\over 7}}e^\theta)\over (1+e^{-{3\pi i\over
7}}e^\theta)} {(1+e^{{3\pi i\over 7}}e^\theta)\over (1+e^{-{5\pi
i\over 7}}e^\theta)} \cr\cr\cr
&=&B^{33}(\theta)(2)_\theta^{-1}(4)_\theta^{-1}. \eea

For the fusing $3+1\rightarrow 4$, we compute $B^{34}(\theta)$,
$$B^{34}(\theta)={(1+e^{6\pi i\over 7}e^\theta)\over (1+e^\theta)}$$
and
\bea
B^{33}(\theta-\frac{\pi i}7)B^{31}(\theta+\frac{3\pi i}7)&=&
{(1+e^{6\pi i\over 7}e^\theta)\over (1+e^{-{2\pi i\over 7}}e^\theta)}
{(1+e^{{4\pi i\over 7}}e^\theta)\over (1+e^{-{4\pi i\over
7}}e^\theta)} {(1+e^{{2\pi i\over 7}}e^\theta)\over (1+e^{-{6\pi
i\over 7}}e^\theta)} {(1+e^{{8\pi i\over 7}}e^\theta)\over (1+e^\theta)}
\cr\cr\cr
&=&B^{34}(\theta)(3)_\theta^{-1}(5)_\theta^{-1}. \eea

We proceed in this manner and find all the corrections in Table 2b. We
summarise the $B^{ij}(\theta)$ for $A_6$ in Table 2a. The notation for
this is that we write 
$${ a\ b \ldots d \over e\ f \ldots g} $$ for
$${(1+e^{ia\over h}e^{\theta})(1+e^{ib\over h}e^{\theta})\ldots
(1+e^{id\over h}e^{\theta})\over
(1+e^{ie\over h}e^{\theta})(1+e^{if\over h}e^{\theta})\ldots
(1+e^{ig\over h}e^{\theta})}.$$

Note the remarkable fact that Table 2b is of the same form as Table 1,
so that the pole on the physical strip due to a $(p)_\theta$ from
Table 2b is doubled up by a pole from the factor $(1-x^hq^{p})^{-1}$,
from Table 1.

In tables 3a,b to 6a,b, we present the cases $D_5,E_6,E_7$ and
$E_8$ respectively, computed by following through the bootstrap. We
show both the $B^{ij}(\theta)$ and the corrections.
\vskip 0.5cm
\pagebreak
$$A_6, \qquad\qquad h=7,$$
\begin{center}
\begin{picture}(270,30)(0,0)
\put(0,20){\circle*{4}}
\put(2,20){\line(1,0){16}}
\put(20,20){\circle{4}}
\put(22,20){\line(1,0){16}}
\put(40,20){\circle*{4}}
\put(42,20){\line(1,0){16}}
\put(60,20){\circle{4}}
\put(62,20){\line(1,0){16}}
\put(80,20){\circle*{4}}
\put(82,20){\line(1,0){16}}
\put(100,20){\circle{4}}
\put(-2,10){$\scriptstyle 1$}
\put(18,10){$\scriptstyle 2$}
\put(38,10){$\scriptstyle 3$}
\put(58,10){$\scriptstyle 4$}
\put(78,10){$\scriptstyle 5$}
\put(98,10){$\scriptstyle 6$}
\end{picture}\end{center}
\vskip 0.5cm
\begin{center}
\begin{tabular}{||l|c|c|c|c|c|c||} \hline\hline
$B^{ij}(\theta)$: $i\backslash j$ &   $1$ & $ 2$ & $3$ & $4$ & $5$ & $6$ \\ \hline
$1$& $\scriptstyle {7 \over -5} $ & $\scriptstyle {6\over -4}$ &
$\scriptstyle {5\over -3}$ &$\scriptstyle {4\over -2}$ & $\scriptstyle {3\over -1}$ &$\scriptstyle {2\over 0}$
\\ \hline
$2$ & $\scriptstyle -$ & $\scriptstyle {7\over -3} $& $\scriptstyle
{6\over -2}$& $\scriptstyle {5\over -1}$ & $\scriptstyle {4\over 0}$ &$\scriptstyle {3\over -1}$\\ \hline
$3$ & $\scriptstyle -$ & $\scriptstyle -$ & $\scriptstyle  {7 \over
-1}$ & $\scriptstyle {6\over 0}$ & $\scriptstyle {5\over -1}$ &
$\scriptstyle {4\over -2}$ \\ \hline
$4$ &$\scriptstyle -$ &$\scriptstyle -$ &$\scriptstyle -$&
$\scriptstyle{7\over -1}$ & $\scriptstyle {6\over -2}$ &$\scriptstyle
{5\over -3}$ \\ \hline
$5$ &$\scriptstyle -$ &$\scriptstyle -$ &$\scriptstyle -$&
$\scriptstyle - $ & $\scriptstyle {7\over -3}$ &$\scriptstyle
{6\over -4}$ \\ \hline
$6$ &$\scriptstyle -$ &$\scriptstyle -$ &$\scriptstyle -$&
$\scriptstyle - $ & $\scriptstyle - $ &$\scriptstyle
{7\over -5}$ \\ \hline\hline
\end{tabular}
\end{center}
\begin{center} {Table 2a: $B^{ij}$ for $A_6$}
\end{center}
\vskip 0.5cm
\noindent (the $\surd$ denotes that the bootstrap is correct
and no corrections are needed)
\vskip 0.5cm
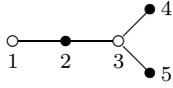
\begin{center}\begin{tabular}{||l|c|c|c|c|c|c||} \hline\hline
$i\backslash j$ &  $1$ & $2$ & $3$ & $4$ & $5$ & $6$ \\ \hline
$1$           &$\scriptstyle\surd$     & $\scriptstyle\surd$    & $\scriptstyle\surd$    &  $\scriptstyle \surd$   &  $\scriptstyle \surd$   & $\scriptstyle \surd$    \\ \hline
$2$         & $\scriptstyle -$ & $\scriptstyle (2)_\theta$ &
$\scriptstyle (3)_\theta$ & $\scriptstyle (4)_\theta $& $\scriptstyle
(5)_\theta$ &  $\scriptstyle \surd$
\\ \hline
$3$         & $\scriptstyle -$ & $\scriptstyle -$& $\scriptstyle (2)_\theta (4)_\theta$ & $\scriptstyle (3)_\theta
(5)_\theta$ & $\scriptstyle (4)_\theta$ & $\scriptstyle \surd$\\ \hline
$4$         & $\scriptstyle -$ & $\scriptstyle -$ &$\scriptstyle -$ & $\scriptstyle (2)_\theta
(4)_\theta$ & $\scriptstyle (3)_\theta$ & $\scriptstyle\surd$ \\ \hline 
$5$         & $\scriptstyle -$ & $\scriptstyle - $& $\scriptstyle -$&
$\scriptstyle -$  & $\scriptstyle (2)_\theta$ & $\scriptstyle\surd$
 \\ \hline 
$6$           &  $\scriptstyle -$   &   $\scriptstyle -$   &  $\scriptstyle -$  &  $\scriptstyle -$    &  $\scriptstyle -$   &  $\scriptstyle\surd$   \\ \hline\hline
\end{tabular}
\end{center}
\begin{center} {Table 2b: Corrections to $v^{ij}(\theta)$ for $A_6$}
\end{center}
\pagebreak

$$D_5,\qquad h=8$$
\begin{center}
\begin{picture}(270,30)(0,0)
\put(0,20){\circle{4}}
\put(2,20){\line(1,0){16}}
\put(20,20){\circle*{4}}
\put(22,20){\line(1,0){16}}
\put(40,20){\circle{4}}
\put(42,22){\line(1,1){10}}
\put(52,32){\circle*{4}}
\put(42,18){\line(1,-1){10}}
\put(52,8){\circle*{4}}
\put(-2,10){$\scriptstyle 1$}
\put(18,10){$\scriptstyle 2$}
\put(38,10){$\scriptstyle 3$}
\put(56,30){$\scriptstyle 4$}
\put(56,5){$\scriptstyle 5$}
\end{picture}
\end{center}

\begin{center}
\begin{tabular}{||l|c|c|c|c|c||} \hline\hline
$B^{ij}(\theta)$: $i\backslash j$ & $1$ & $2$ & $3$ & $4$ &$5$\\ \hline
$1$& $\scriptstyle {2\ 8\over 0\ -6} $ & $\scriptstyle {3\ 7\over -1\ -5}$ & $\scriptstyle {6\over -2}$ &$\scriptstyle {5\over -3}$
& $\scriptstyle {5\over -3}$\\ \hline
$2$ & $\scriptstyle -$ & $\scriptstyle {8\over 0} $& $\scriptstyle  {5\ 7\over -1\ -3}$& $\scriptstyle {6\over -2}$ & $\scriptstyle {6\over -2}$\\ \hline
$3$ & $\scriptstyle -$ & $\scriptstyle -$ & $\scriptstyle  {6\ 8\over 0\ -2}$ & $\scriptstyle {7\over -1}$ & $\scriptstyle {7\over -1}$ \\ \hline
$4$ &$\scriptstyle -$ &$\scriptstyle -$ &$\scriptstyle -$&  $\scriptstyle{4\ 8\over -2\ -6}$ &  $\scriptstyle{2\ 6\over 0\ -4}$\\ \hline
$5$ &$\scriptstyle -$&$\scriptstyle -$&$\scriptstyle -$&$\scriptstyle -$& $\scriptstyle {4\ 8\over -2\ -6}$ \\ \hline\hline 
\end{tabular}
\end{center}
\begin{center} {Table 3a} \end{center}
\vskip 0.5cm
\begin{center}
\begin{tabular}{||l|c|c|c|c|c||} \hline\hline
$i\backslash j$ &   $1$ & $2$ & $3$ & $4$ & $5$\\ \hline
$1$& $\scriptstyle \surd$ & $\scriptstyle \surd$ & $\scriptstyle (4)$ & $\scriptstyle \surd$ & $\scriptstyle \surd$ \\ \hline
$2$ & $\scriptstyle -$ & $\scriptstyle (2)(4)(6)$ & $\scriptstyle  (3)(5) $& $\scriptstyle (4)$ & $\scriptstyle (4)$ \\ \hline
$3$ & $\scriptstyle -$ & $\scriptstyle -$ & $\scriptstyle (2)(4)^2(6)$ & $\scriptstyle (3)(5)$ & $\scriptstyle (3)(5)$ \\ \hline
$4$ &$\scriptstyle -$ &$\scriptstyle -$ &$\scriptstyle -$ & $\scriptstyle \surd$ & $\scriptstyle \surd$ \\ \hline
$5$ & $\scriptstyle -$ &$\scriptstyle -$ &$\scriptstyle -$ &$\scriptstyle -$ & $\scriptstyle \surd$ \\ \hline\hline
\end{tabular}
\end{center}
\begin{center} Table 3b \end{center}
\pagebreak
$$E_6,\qquad h=12
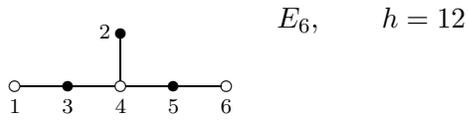
$$
\begin{center}
\begin{picture}(270,30)(0,0)
\put(0,20){\circle{4}}
\put(2,20){\line(1,0){16}}
\put(20,20){\circle*{4}}
\put(22,20){\line(1,0){16}}
\put(40,20){\circle{4}}
\put(42,20){\line(1,0){16}}
\put(60,20){\circle*{4}}
\put(62,20){\line(1,0){16}}
\put(80,20){\circle{4}}
\put(40,22){\line(0,1){16}}
\put(40,40){\circle*{4}}
\put(32,38){$\scriptstyle 2$}
\put(-2,10){$\scriptstyle 1$}
\put(18,10){$\scriptstyle 3$}
\put(38,10){$\scriptstyle 4$}
\put(58,10){$\scriptstyle 5$}
\put(78,10){$\scriptstyle 6$}
\end{picture}
\end{center}
\begin{center}
\begin{tabular}{||l|c|c|c|c|c|c||} \hline\hline
$B^{ij}(\theta)$: $i\backslash j$ &   $1$ & $2$ & $3$ & $4$ &$5$ &$6$\\ \hline
$1$& $\scriptstyle {6\ 12\over -4\ -10} $ & $\scriptstyle {5\ 9\over -3\ -7}$ & $\scriptstyle {7\ 11\over -3\
-9}$ &$\scriptstyle {10\over -2}$
& $\scriptstyle {3\ 9\over -1\ -5}$ & $\scriptstyle {2\ 8\over 0\ -6}$\\ \hline
$2$ & $\scriptstyle -$ & $\scriptstyle {2\ 8\ 12\over 0\ -4\ -10} $& $\scriptstyle  {10\over -2}$& 
$\scriptstyle{7\ 11\over -1\ -5}$ & $\scriptstyle {10\over -2}$ & $\scriptstyle {5\ 9\over -3\ -7}$ \\ \hline
$3$ & $\scriptstyle -$ & $\scriptstyle -$ & $\scriptstyle  {6\ 12\over -2\ -4}$ & $\scriptstyle {9\ 11\over -1\ -3}$ & 
$\scriptstyle{8\ 10\over 0\ -6}$ & $\scriptstyle {3\ 9\over -1\ -5}$ \\ \hline
$4$ &$\scriptstyle -$ &$\scriptstyle -$ &$\scriptstyle -$&  $\scriptstyle{8\ 10\ 12\over 0\ -2\ -4}$ &  $\scriptstyle{9\ 11\over -1\
-3}$ & $\scriptstyle {10\over -2}$ \\ \hline
$5$ &$\scriptstyle -$&$\scriptstyle -$&$\scriptstyle -$&$\scriptstyle -$& $\scriptstyle {6\ 12\over -2\ -4}$ & $\scriptstyle {7\ 11\over -3\ -9}$ \\ \hline 
$6$ &$\scriptstyle -$&$\scriptstyle -$&$\scriptstyle -$&$\scriptstyle -$& $\scriptstyle -$ & $\scriptstyle {6\ 12\over -4\ -10}$ \\ \hline\hline 
\end{tabular}
\end{center}
\begin{center} Table 4a
\end{center}
\vskip 0.5cm
\begin{center}\begin{tabular}{||l|c|c|c|c|c|c||} \hline\hline
$i\backslash j$ &   $1$ & $2$ & $3$ & $4$ & $5$ & $6$\\ \hline
$1$& $\scriptstyle \surd$ & $\scriptstyle \surd$ & $\scriptstyle (7)$ & $\scriptstyle (4)(6)(8)$ & $\scriptstyle (5)$ & $\scriptstyle \surd$ \\ \hline
$2$ & $\scriptstyle -$ & $\scriptstyle (6)$ & $\scriptstyle (4)(6)(8)$ & $\scriptstyle (3)(5)(7)(9)$ & $\scriptstyle (4)(6)(8)$ &
$\surd$ \\ \hline
$3$ & $\scriptstyle -$ & $\scriptstyle -$ & $\scriptstyle (2)(4)(6)(8)$ & $\scriptstyle (3)(5)^2(7)^2(9)$ &
$\scriptstyle(4)(6)(8)(10)$ & $\scriptstyle (5)$\\ \hline
$4$ &$\scriptstyle -$ &$\scriptstyle -$ &$\scriptstyle -$ & $\scriptstyle (2)(4)^2(6)^3(8)^2(10)$ & $\scriptstyle (3)(5)^2(7)^2(9)$ & 
$\scriptstyle(4)(6)(8)$ \\ \hline
$5$ & $\scriptstyle -$ &$\scriptstyle -$ &$\scriptstyle -$ &$\scriptstyle -$ & $\scriptstyle (2)(4)(6)(8)$ & $\scriptstyle (7)$\\ \hline
$6$ & $\scriptstyle -$ &$\scriptstyle -$ &$\scriptstyle -$ &$\scriptstyle -$ & $\scriptstyle -$ & $\scriptstyle \surd$\\ \hline
\end{tabular}
\end{center}
\begin{center} Table 4b
\end{center}
\pagebreak
$$E_7,\qquad h=18
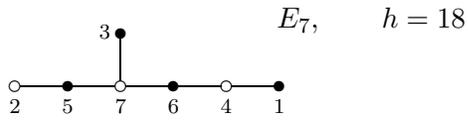
$$
\begin{center}
\begin{picture}(270,30)(0,0)
\put(0,20){\circle{4}}
\put(2,20){\line(1,0){16}}
\put(20,20){\circle*{4}}
\put(22,20){\line(1,0){16}}
\put(40,20){\circle{4}}
\put(42,20){\line(1,0){16}}
\put(60,20){\circle*{4}}
\put(62,20){\line(1,0){16}}
\put(80,20){\circle{4}}
\put(82,20){\line(1,0){16}}
\put(100,20){\circle*{4}}
\put(40,22){\line(0,1){16}}
\put(40,40){\circle*{4}}
\put(32,38){$\scriptstyle 3$}
\put(-2,10){$\scriptstyle 2$}
\put(18,10){$\scriptstyle 5$}
\put(38,10){$\scriptstyle 7$}
\put(58,10){$\scriptstyle 6$}
\put(78,10){$\scriptstyle 4$}
\put(98,10){$\scriptstyle 1$}
\end{picture}
\end{center}
\begin{center}
\begin{tabular}{||l|c|c|c|c|c|c|c||} \hline\hline
$B^{ij}(\theta)$: $i\backslash j$ &   $1$ & $2$ & $3$ & $4$ &$5$ &$6$
& $7$\\ \hline
$1$& $\scriptstyle \scriptstyle{2\ 10\ 18\over 0\ -8\ -16} $ & $\scriptstyle {7\ 13\over -5\ -11}$ & $\scriptstyle {6\
10\ 14\over -4\
-8\ -12}$ &$\scriptstyle {3\ 11\ 17\over -1\ -7\ -15}$
& $\scriptstyle {8\ 14\over -4\ -10}$ & $\scriptstyle {4\ 12\ 16\over -2\ -6\ -14}$ & $\scriptstyle {15\over
-3}$ \\ \hline
$2$ & $\scriptstyle -$ & $\scriptstyle {2\ 8\ 12\ 18\over 0\ -6\ -10\ -16} $& $\scriptstyle  {5\ 11\ 15\over
-3\ -7\ -13}$& 
$\scriptstyle{8\ 14\over -4\ -10}$ & $\scriptstyle {3\ 13\ 17\over -1\ -5\ -15}$ & $\scriptstyle {15\over
-3}$ & $\scriptstyle {10\ 16\over -2\ -8}$ \\ \hline
$3$ & $\scriptstyle -$ & $\scriptstyle -$ & $\scriptstyle  {2\ 14\ 18\over 0\ -4\ -16}$ & $\scriptstyle {15\over -3}$ & 
$\scriptstyle{10\ 16\over -2\ -8}$ & $\scriptstyle {8\ 12\ 16\over -2\ -6\ -10}$ & $\scriptstyle {13\
17\over -5\ -1}$  \\ \hline
$4$ &$\scriptstyle -$ &$\scriptstyle -$ &$\scriptstyle -$&  $\scriptstyle{4\ 10\ 12\ 18\over 0\ -6\ -8\ -14}$ &  $\scriptstyle{7\ 13\
15\over -3\ -5\ -11}$ & $\scriptstyle {11\ 17\over -1\ -7}$ & 
$\scriptstyle{14\ 16\over -2\ -4}$  \\ \hline
$5$ &$\scriptstyle -$&$\scriptstyle -$&$\scriptstyle -$&$\scriptstyle -$& $\scriptstyle {12\ 18\over 0\ -6}$ & $\scriptstyle {14\ 16\over -2\ -4}$ &
$\scriptstyle{11\ 15\ 17\over -1\ -3\ -7}$ \\ \hline 
$6$ &$\scriptstyle -$&$\scriptstyle -$&$\scriptstyle -$&$\scriptstyle -$& $\scriptstyle -$ & $\scriptstyle {10\ 14\ 18\over 0\ -4\ -8}$ & $\scriptstyle {13\ 15\
17\over -1\ -3\ -5}$ \\ \hline
$7$ &$\scriptstyle -$&$\scriptstyle -$&$\scriptstyle -$&$\scriptstyle -$& $\scriptstyle -$ & $\scriptstyle -$ & $\scriptstyle {12\ 14\ 16\ 18\over 0\ -2\ -4\ -6}$
\\ \hline\hline
\end{tabular}
\end{center}
\begin{center} Table 5a
\end{center}
\vskip 0.5cm
\begin{center}\begin{tabular}{||l|c|c|c|c|c|} \hline\hline
$i\backslash j$ &   $1$ & $2$ & $3$ & $4$ & $5$ \\ \hline
$1$& $\scriptstyle \surd$ & $\scriptstyle \surd$ & $\scriptstyle \surd$ & $\scriptstyle (9)$ & $\scriptstyle (6)(12)$ \\ \hline
$2$ & $\scriptstyle -$ & $\scriptstyle \surd$ & $\scriptstyle (9)$ & $\scriptstyle (6)(12)$ & $\scriptstyle (7)(9)(11)$\\ \hline
$3$ & $\scriptstyle -$ & $\scriptstyle -$ & $\scriptstyle (6)(8)(10)(12)$ & $\scriptstyle (5)(7)(9)(11)(13)$ &
$\scriptstyle(4)(6)(8)(10)(12)(14)$\\ \hline
$4$ &$\scriptstyle -$ &$\scriptstyle -$ &$\scriptstyle -$ & $\scriptstyle (2)(8)(10)(16)$ & $\scriptstyle (5)(7)(9)(11)(13)$\\ \hline
$5$ & $\scriptstyle -$ &$\scriptstyle -$ &$\scriptstyle -$ &$\scriptstyle -$ & $\scriptstyle (2)(4)(6)(8)^2(10)^2(12)(14)(16)$ 
\\ \hline
$6$ & $\scriptstyle -$ &$\scriptstyle -$ &$\scriptstyle -$ &$\scriptstyle -$ & $\scriptstyle -$   \\ \hline
$7$ & $\scriptstyle -$ &$\scriptstyle -$ &$\scriptstyle -$ &$\scriptstyle -$ & $\scriptstyle -$  \\ \hline\hline
\end{tabular}
\begin{tabular}{|l|c|c||} \hline\hline
$i\backslash j$ & $6$ & $7$ \\ \hline
$1$ & $\scriptstyle (8)(10)$ & $\scriptstyle (5)(7)(9)(11)$ \\ \hline
$2$ & $\scriptstyle (5)(7)(9)(11)(13)$ & $\scriptstyle (4)(6)(8)(10)(12)(14)$ \\ \hline
$3$ & $\scriptstyle (4)(6)(8)(10)(12)(14)$ & $\scriptstyle (3)(5)(7)^2(9)^2(11)^2(13)(15)$ \\
\hline
$4$ & $\scriptstyle (3)(5)(7)(9)^2(11)(13)(15)$ & $\scriptstyle (4)(6)^2(8)^2(10)^2(12)^2(14)$
\\ \hline 
$5$ & $\scriptstyle (4)(6)^2(8)^2(10)^2(12)^2(14)$ & $\scriptstyle (3)(5)^2(7)^2(9)^3(11)^2(13)^2(15)$
\\ \hline
$6$ & $\scriptstyle (2)(4)(6)^2(8)^2(10)^2(12)^2(14)(16)$ &
$\scriptstyle(3)(5)^2(7)^3(9)^3(11)^3(13)^2(15)$ \\ \hline
$7$ & $\scriptstyle -$ & $\scriptstyle \scriptstyle (2)(4)^2(6)^3(8)^4(10)^4(12)^3(14)^2(16)$ \\ \hline\hline
\end{tabular}
\end{center}
\begin{center} Table 5b
\end{center}
\pagebreak

$$E_8,\qquad h=30
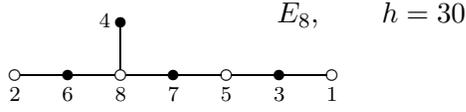
$$
\begin{center}
\begin{picture}(270,30)(0,0)
\put(0,20){\circle{4}}
\put(2,20){\line(1,0){16}}
\put(20,20){\circle*{4}}
\put(22,20){\line(1,0){16}}
\put(40,20){\circle{4}}
\put(42,20){\line(1,0){16}}
\put(60,20){\circle*{4}}
\put(62,20){\line(1,0){16}}
\put(80,20){\circle{4}}
\put(82,20){\line(1,0){16}}
\put(100,20){\circle*{4}}
\put(102,20){\line(1,0){16}}
\put(120,20){\circle{4}}
\put(40,22){\line(0,1){16}}
\put(40,40){\circle*{4}}
\put(32,38){$\scriptstyle 4$}
\put(-2,10){$\scriptstyle 2$}
\put(18,10){$\scriptstyle 6$}
\put(38,10){$\scriptstyle 8$}
\put(58,10){$\scriptstyle 7$}
\put(78,10){$\scriptstyle 5$}
\put(98,10){$\scriptstyle 3$}
\put(118,10){$\scriptstyle 1$}
\end{picture}
\end{center}

\noindent\begin{tabular}{||l|c|c|c|c|} \hline\hline
$B^{ij}(\theta)$: $i\backslash j$ &   $1$ & $2$ & $3$ & $4$\\ \hline
$1$& $\scriptstyle {30\ 20\ 12\ 2\over 0\ -10\ -18\ -28}$ & $\scriptstyle {24\ 18\ 14\ 8\over
-22\ -16\ -12\ -6}$ & $\scriptstyle {3\ 13\ 21\ 29\over -1\ -9\ -17\ -27}$ & 
$\scriptstyle {7\ 11\ 17\ 21\ 25\over -5\ -9\ -13\ -19\ -23}$ \\ \hline
$2$ & $\scriptstyle -$ & $\scriptstyle {2\ 8\ 14\ 20\ 24\ 30\over 0 -6\ -10\ -16\ -22\ -28}$ & 
$\scriptstyle {9\ 19\ 25\over -5\ -11\ -21}$ & $\scriptstyle {5\ 23\ 27\over -3\ -7\ -25}$ \\
\hline
$3$ & $\scriptstyle -$ & $\scriptstyle -$ & $\scriptstyle {4\ 12\ 14\
20\ 22\ 30\over 0\ -8\ -10\ -16\ -18\ -26}$ &
$\scriptstyle {16\ 26\over -4\ -14}$ \\ \hline
$4$ & $\scriptstyle -$ & $\scriptstyle -$ & $\scriptstyle -$ & $\scriptstyle {2\ 12\ 16\ 20\ 26\ 30\over 0\ -4\ -10\ -14\
-18\ -28}$ \\ \hline
$5$ & $\scriptstyle -$ & $\scriptstyle -$ & $\scriptstyle -$ & $\scriptstyle -$ \\ \hline
$6$ & $\scriptstyle -$ & $\scriptstyle -$ & $\scriptstyle -$ & $\scriptstyle -$ \\ \hline
$7$ & $\scriptstyle -$ & $\scriptstyle -$ & $-\scriptstyle $ & $\scriptstyle -$ \\ \hline
$8$ & $\scriptstyle -$ & $\scriptstyle -$ & $\scriptstyle -$ & $\scriptstyle -$ \\ \hline\hline
\end{tabular}

\noindent\begin{tabular}{|c|c|c|c||} \hline\hline
$5$ & $6$ & $7$ & $8$\\ \hline
$\scriptstyle {4\ 14\ 22\ 28\over -2\ -8\ -16\ -26}$ & $\scriptstyle {9\ 19\ 25\over -5\ -11\
-21}$ & $\scriptstyle {5\ 23\ 27\over -3\ -7\ -25}$ & $\scriptstyle {16\ 26\over -4\ -14}$ \\
\hline
$\scriptstyle {16\ 26\over -4\ -14}$ & $\scriptstyle {3\ 13\ 19\ 25\ 29\over -1\ -5\ -11\ -17\
-27}$ & $\scriptstyle {11\ 17\ 21\ 27\over -3\ -9\ -13\ -19}$ & $\scriptstyle {22\ 28\over -2\
-8}$\\ \hline
$\scriptstyle {5\ 13\ 21\ 23\ 29\over -1\ -7\ -9\ -17\ -25}$ & $\scriptstyle {8\ 18\ 24\ 26\over -4\
-6\ -12\ -22}$ & $\scriptstyle{22\ 28\over -2\ -8}$ & $\scriptstyle {17\ 25\ 27\over -3\ -5\ -13}$ \\
\hline
$\scriptstyle {9\ 19\ 23\ 27\over -3\ -7\ -11\ -21}$ & $\scriptstyle {22\ 28\over -2\ -8}$ & $\scriptstyle {14\ 18\ 24\
28\over -2\ -6\ -12\ -16}$ & $\scriptstyle {21\ 25\ 29\over -1\ -5\ -9}$ \\ \hline
$\scriptstyle {12\ 20\ 22\ 30\over 0\ -8\ -10\ -18}$ & $\scriptstyle {17\ 25\ 27\over -3\ -5\
-13}$ & $\scriptstyle {21\ 25\
29\over -1\ -5\ -9}$ & $\scriptstyle {18\ 24\ 26\ 28\over -2\ -4\ -6\ -12}$ \\ \hline
$\scriptstyle -$ & $\scriptstyle {14\ 20\ 24\ 30\over 0\ -6\ -10\ -16}$ & $\scriptstyle {16\ 22\ 26\ 28\over -2\ -4\ -8\ -14}$ 
& $\scriptstyle {21\ 23\ 27\ 29\over -1\ -3\ -7\ -9}$ \\ \hline
$\scriptstyle -$ & $\scriptstyle -$ & $\scriptstyle {20\ 24\ 26\ 30\over 0\ -4\ -6\ -10}$ & $\scriptstyle {19\ 23\ 25\ 27\ 29\over -1\
-3\ -5\ -7\ -11}$ \\ \hline
$\scriptstyle -$ & $\scriptstyle -$ & $\scriptstyle -$ & $\scriptstyle {20\ 22\ 24\ 26\ 28\ 30\over 0\ -2\ -4\ -6\ -8\
-10}$ \\ \hline\hline
\end{tabular}
\begin{center} Table 6a
\end{center}

\noindent\begin{tabular}{||l|c|c|c|c|c|c|} \hline\hline
$i\backslash j$ &   $1$ & $2$ & $3$ & $4$ & $5$ & $6$\\ \hline
$1$ & $\scriptstyle\surd$ & $\scriptstyle\surd$ & $\scriptstyle 11$ &
$\scriptstyle(15)$ & $\scriptstyle 10\ 12$ & $\scriptstyle 7\ 13\
(15)$ \\ \hline
$2$ & $\scriptstyle -$ & $\scriptstyle 12$ & $\scriptstyle 7\ 13\
(15)$ & $\scriptstyle 9\ 11\ 13\ (15)$ & $\scriptstyle 6\ 8\ 10\ 12\ 14$ & 
$\scriptstyle 7\ 9\ 11\ 13\ (15)$ \\ \hline
$3$ & $\scriptstyle -$ & $\scriptstyle -$ & $\scriptstyle 2\ 10\ 12$ &
$\scriptstyle 6\ 8\ 10\ 12\ 14$ & $\scriptstyle 3\ 9\ 11^2\ 13\ (15)$ & 
$\scriptstyle 6\ 8\ 10\ 12\ 14^2$ \\ \hline
$4$ & $\scriptstyle -$ & $\scriptstyle -$ & $\scriptstyle -$ & $\scriptstyle 6\ 8\ 10\ 12\ 14$ & $\scriptstyle 5\ 7\ 9\ 11\ 13^2\ (15)^2$ & 
$\scriptstyle 4\ 6\ 8\ 10^2\ 12^2\ 14^2$ \\ \hline
$5$ & $\scriptstyle -$ & $\scriptstyle -$ & $\scriptstyle -$ & $\scriptstyle -$ & $\scriptstyle 2\ 4\ 6\ 8\ 10^2\ 12^2\ 14^2$ &
$\scriptstyle 5\ 7^2\ 9^2\ 11^2\ 13^2\ (15)^3$ \\ \hline
$6$ & $\scriptstyle -$ & $\scriptstyle -$ & $\scriptstyle -$ & $\scriptstyle -$ & $\scriptstyle -$ & $\scriptstyle 2\ 4\ 6\ 8^2\ 10^2\ 12^3\ 14^2$ \\ \hline
$7$ & $\scriptstyle -$ & $\scriptstyle -$ & $\scriptstyle -$ & $\scriptstyle -$ & $\scriptstyle -$ & $\scriptstyle -$\\ \hline
$8$ & $\scriptstyle -$ & $\scriptstyle -$ & $\scriptstyle -$ & $\scriptstyle -$ & $\scriptstyle -$ & $\scriptstyle -$\\ \hline
\end{tabular}

\noindent \begin{tabular}{|c|c||} \hline\hline
$7$ & $8$\\ \hline
$\scriptstyle 9\ 11\ 13\ (15)$ & $\scriptstyle 6\ 8\ 10\ 12\ 14$ \\
\hline
$\scriptstyle 5\ 7\ 9\ 11\ 13\
(15)^2$ & $\scriptstyle 4\ 6\ 8\ 10^2\ 12^2\ 14^2$ \\ \hline
$\scriptstyle 4\ 6\ 8\ 10^2\ 12^2\ 14^2$ & $\scriptstyle  5\ 7^2\ 9^2\
11^2\  13^2\ (15)^3$ \\ \hline
$\scriptstyle 4\ 6\ 8^2\ 10^2\ 12^2\ 14^2$ & $\scriptstyle 3\ 5\ 7^2\
9^2\ 11^3\ 13^3\ (15)^3$ \\ \hline
$\scriptstyle 3\ 5\ 7^2\ 9^2\ 11^3\ 13^3\ (15)^3$ &
$\scriptstyle 4\ 6^2\ 8^3\ 10^3\ 12^3\ 14^4$ \\ \hline
$\scriptstyle 4\ 6^2\ 8^2\ 10^3\ 12^3\ 14^3$ & $\scriptstyle 3\ 5^2\
7^2\ 9^3\ 11^4\ 13^4\ (15)^4$ \\ \hline
$\scriptstyle 2\ 4\ 6^2\ 8^3\ 10^3\ 12^4\ 14^4$
& $\scriptstyle 3\ 5^2\ 7^3\ 9^4\ 11^4\ 13^5\ (15)^5$ \\ \hline
$\scriptstyle -$ & $\scriptstyle 2\ 4^2\ 6^3\ 8^4\ 10^5\
12^6\ 14^6$ \\ \hline
\end{tabular}
\begin{center} Table 6b: with $x=(x)(h-x)$ \end{center}
\pagebreak

{\noindent\bf The soliton bootstrap completed: the bootstrap equation
for $\tilde{v}^{ij}(\theta)$}

We remark that with $\tilde{v}^{ij}(\theta)$ defined through equation
(\ref{eq: vij2p}) and $v^{ij}(\theta)=B^{ij}(\theta)^{-1}
\tilde{v}^{ij}(\theta)$, the bootstrap equation for
$\tilde{v}^{ij}(\theta)$ follows from the classical bootstrap equation
(\ref{eq: class_boot}), except for possibly adjusting the arguments of
$S_{\tilde{q}^{-h}}(w)$ by $e^{2\pi i}$. This adjustment introduces
factors of the form $(1-x^hq^p)^{-1}$ into the bootstrap equation,
which have already been discussed for the $A_n$ case, in section
11. For these $A_n$ theories, the extra factors fit in precisely with
the factors which we can compute when we fuse the R-matrices. For
$A_{h-1}$, from (\ref{eq: An}), we have
\bea
R^{11}(x)&=&(1-x^hq^2)P_0+(x^h-q^2)P_2\cr\cr
R^{12}(x)&=&(1-x^hq^3)P_{0'}+(x^h-q^3)P_3,\eea
where $P_2$ and $P_3$ project onto the fundamental representations 
$V_2$ and $V_3$ respectively. Both $R^{11}$ and $R^{12}$ are of order
one. However, if we fuse $R^{11}(x)$, using the fusing $1+1\rightarrow
2$,
\beq 
g(x)R^{12}(x)=(P_2\tensor 1)(1\tensor R^{11}(xq^{-1/h}))
(R^{11}(xq^{1/h})\tensor 1)(1\tensor P_2),\label{eq: fuse1}\eeq
by comparing orders, we see that $g(x)$ must be a non-trivial factor
with order one. Now $R^{11}(1)=(1-q^2)1$, 
$R^{11}(q^{2/h})=(1-q^4)P_0$, and $P_2P_0=0$, so
that if we set $x=q^{1/h}$ in (\ref{eq: fuse1}), then the $(P_2\tensor 1)$
in the left most position acts on the $(R^{11}(q^{2/h})\tensor 1)$
giving zero, so that
$g(q^{1/h})=0$, and we infer that $g(x)=(1-x^hq^{-1})$. Note that
this agrees with (\ref{eq: star}), for $su(3)$ or $h=2$.

For $A_6$, we note that, from (\ref{eq: vij2p})
\bea
\tilde{v}^{11}(\theta)&=&{S_{\qt^{-h}}(\qt^h e^\theta e^{-i\pi})
S_{\qt^{-h}}(\qt^h e^\theta e^{i\pi})\over
S_{\qt^{-h}}(\qt^h e^{-2\pi i+\theta} e^{5\pi i\over 7})
S_{\qt^{-h}}(\qt^h e^{\theta} e^{-{5\pi i\over 7}})} \cr\cr\cr
\tilde{v}^{12}(\theta)&=&{S_{\qt^{-h}}(\qt^h e^\theta e^{6\pi i\over 7})
S_{\qt^{-h}}(\qt^h e^\theta e^{-{6\pi i\over 7}}) \over
S_{\qt^{-h}}(\qt^h e^{-2\pi i+\theta} e^{4\pi i\over 7})
S_{\qt^{-h}}(\qt^h e^{\theta} e^{-{4\pi i\over 7}})}, \eea
and we check $\tilde{v}^{11}(\theta+\frac{i\pi}7)
\tilde{v}^{11}(\theta-\frac{i\pi}7)=f(x)\tilde{v}^{12}(\theta)$

$$\tilde{v}^{11}(\theta+\frac{i\pi}7)
\tilde{v}^{11}(\theta-\frac{i\pi}7)\qquad\qquad\qquad\qquad\qquad\qquad\qquad\qquad\qquad\qquad\qquad\qquad$$
\bea
&=&
{S_{\qt^{-h}}(\qt^h e^\theta e^{8\pi i\over 7})
S_{\qt^{-h}}(\qt^h e^\theta e^{-{6\pi i\over 7}}) \over
S_{\qt^{-h}}(\qt^h e^{-2\pi i+\theta}e^{6\pi i\over 7})
S_{\qt^{-h}}(\qt^h e^{\theta}e^{-{4\pi i\over 7}})} 
{S_{\qt^{-h}}(\qt^h e^\theta e^{6\pi i\over 7})
S_{\qt^{-h}}(\qt^h e^\theta e^{-{8\pi i\over 7}}) \over
S_{\qt^{-h}}(\qt^h e^{-2\pi i+\theta}e^{4\pi i\over 7})
S_{\qt^{-h}}(\qt^h e^{\theta}e^{-{6\pi i\over 7}})}\cr\cr\cr
&=&{S_{\qt^{-h}}(\qt^h e^\theta e^{6\pi i\over 7})
S_{\qt^{-h}}(\qt^h e^\theta e^{{8\pi i\over 7}}) \over
S_{\qt^{-h}}(\qt^h e^{-2\pi i+\theta}e^{4\pi i\over 7})
S_{\qt^{-h}}(\qt^h e^{\theta}e^{-{4\pi i\over 7}})} \cr\cr\cr
&=&\tilde{v}^{12}(\theta){1\over 1-x^h q^{-1}}.\eea

Now consider a case where we have to use the corrections in Table
1. For example, 
\bea
\tilde{v}^{32}(\theta)&=&{S_{\qt^{-h}}(\qt^h e^\theta e^{6\pi i\over 7})
S_{\qt^{-h}}(\qt^h e^\theta e^{-{6\pi i\over 7}}) \over
S_{\qt^{-h}}(\qt^h e^{-2\pi i+\theta} e^{2\pi i\over 7})
S_{\qt^{-h}}(\qt^h e^{\theta} e^{-{2\pi i\over 7}})} \cr\cr\cr
\tilde{v}^{42}(\theta)&=&{S_{\qt^{-h}}(\qt^h e^\theta e^{5\pi i\over 7})
S_{\qt^{-h}}(\qt^h e^\theta e^{-{5\pi i\over 7}}) \over
S_{\qt^{-h}}(\qt^h e^{-2\pi i+\theta} e^{\pi i\over 7})
S_{\qt^{-h}}(\qt^h e^{\theta} e^{-{\pi i\over 7}})}, \eea
and for the fusing $1+3\rightarrow 4$, we check
$\tilde{v}^{32}(\theta+\frac{\pi i}7)
\tilde{v}^{12}(\theta-\frac{3\pi i}7)=f(x)\tilde{v}^{42}(\theta)$

$$\tilde{v}^{32}(\theta+\frac{\pi i}7)
\tilde{v}^{12}(\theta-\frac{3\pi i}7)\qquad\qquad\qquad\qquad\qquad\qquad\qquad\qquad\qquad\qquad\qquad\qquad$$
\bea
&=&{S_{\qt^{-h}}(\qt^h e^\theta e^{\pi i})
S_{\qt^{-h}}(\qt^h e^\theta e^{-{5\pi i\over 7}}) \over
S_{\qt^{-h}}(\qt^h e^{-2\pi i+\theta}e^{3\pi i\over 7})
S_{\qt^{-h}}(\qt^h e^{\theta}e^{-{\pi i\over 7}})}
{S_{\qt^{-h}}(\qt^h e^\theta e^{3\pi i\over 7})
S_{\qt^{-h}}(\qt^h e^\theta e^{-{9\pi i\over 7}}) \over
S_{\qt^{-h}}(\qt^h e^{-2\pi i+\theta}e^{\pi i\over 7})
S_{\qt^{-h}}(\qt^h e^{\theta}e^{-{\pi i}})} \cr\cr\cr
&=&{S_{\qt^{-h}}(\qt^h e^\theta e^{-{5\pi i\over 7}})
S_{\qt^{-h}}(\qt^h e^\theta e^{-{9\pi i\over 7}}) \over
S_{\qt^{-h}}(\qt^h e^{-2\pi i+\theta}e^{\pi i\over 7})
S_{\qt^{-h}}(\qt^h e^{\theta}e^{-{\pi i\over 7}})}{1\over
1-x^h}{1\over 1-x^hq^4}\cr\cr\cr
&=&\tilde{v}^{42}(\theta){(1-x^h q^{2})\over (1-x^h)(1-x^hq^4)}.
\label{eq: fusev} \eea

For the fusion
\beq
g(x)R^{42}(x)=(I\tensor R^{32}(xq^{-1/7}))(R^{12}(xq^{3/7})\tensor
1)\biggl|_{V_4\tensor V_2}\label{eq: fuse2}, \eeq
Hollowood gives a prescription \cite{H2} for the zeroes of $R^{42}(x)$, defined
by setting $g(x)=1$, throughout the fusion procedure. This is found by
repeatedly applying the argument (\ref{eq: fuse1}).

It is argued that the zeroes of $R^{ab}(x)$, for $b\ge a$, are at
$$x=q^{-\frac1h(a+b-2j-2k)},\qquad j=1,2,\ldots, a,\quad k=1,2,\ldots, b-1,$$
so the four zeroes of $R^{32}(x)$ are at 
$$x=q^{-\frac1h},\quad q^{\frac1h},\quad q^{\frac1h},\quad q^{\frac3h}.$$
The zero of $R^{12}(x)$ is at $x=q^{\frac1h}$, and the zeroes of
$R^{42}(x)$ are at
$$x=q^{-\frac2h},\quad 1,\quad 1,\quad q^{\frac2h},\quad
q^{\frac2h},\quad q^{\frac4h}.$$
In the fusion (\ref{eq: fuse2}), the zeroes of $R^{32}(x)$ and
$R^{12}(x)$ cancel with five out of the siz zeroes of $R^{42}(x)$,
leaving a single zero at $x=1$, hence we infer that $g(x)=(1-x^h)$.
This accounts for the factor $(1-x^h)^{-1}$ on the right-hand side of
(\ref{eq: fusev}). The factor $(1-x^hq^4)^{-1}$ is accounted for,
since it is the 42 correction from Table 1, and $(1-x^hq^2)$ is the 32
correction taken to the other side of the bootstrap equation.

In a similar fashion, it is possible to show for the outstanding $A_n$
cases that the normalisations
arising from fusion agree with the factors thrown up in the
bootstrap on $\tilde{v}^{ij}$, and the factors in Table 1.
\resection{The lowest mass breather bootstrap, and comparison with the Toda
particle S-matrices}

We follow through the bootstrap for the lowest mass breathers and
calculate the lowest mass breather -- lowest mass breather S-matrices 
$S^{b_i b_j}(\theta)$. Recall the notation for the building blocks,
from \cite{BCDS},
$$(x)_\theta={\sinh(\frac\theta 2+{i\pi x\over 2h})\over
\sinh(\frac\theta 2-{i\pi x\over 2h})},$$
and
$$\{x\}_\theta={(x-1)_\theta(x+1)_\theta\over
(x-1+B)_\theta(x+1-B)_\theta},$$
where 
\beq B(\beta)={1\over 2\pi}{\beta^2\over 1+{\beta^2\over 4\pi}}
\label{eq: funbeta},\eeq
the $\beta$ in (\ref{eq: funbeta}) is real for the affine Toda
particles, and purely imaginary for the solitons.

Recall also the general formula for the Toda particle S-matrices \cite{Dorey2},
\beq S^{ij}(\theta)=\prod_{q=1}^h\biggl\{2q-{c(i)+c(j)\over 2}\biggr\}_\theta^
{-{\lambda_i\cdot\sigma^q \gamma_j\over 2}}.\label{eq: Doreys_formula}
\eeq

The lowest mass breather pole in
$S^{i\ib}_{\lambda\bar{\lambda}}(\theta)$ is at 
$$\theta=i\pi-{i\gamma\over 2h}.$$

We always take the transmissive soliton S-matrix elements, rather than
the reflective, since only then does the bootstrap make sense.

\bea S^{s_i b_j}_{\lambda}(\theta)&=&S^{ij}_{\lambda\mu}\biggl(\theta-
i\biggl(\frac\pi 2 -\frac\gamma {4h}\biggr)\biggr)
S^{i\jb}_{\lambda\bar{\mu}}\biggl(\theta+i\biggl(\frac\pi
2 -\frac\gamma {4h}\biggr)\biggr) \cr \cr \cr 
&=&S^{ij}_{\lambda\mu}\biggl(\theta-
i\biggl(\frac\pi 2 -\frac\gamma {4h}\biggr)\biggr)
S^{i{j}}_{\lambda{\mu}}\biggl(-\theta+i\biggl(\frac\pi
2 +\frac\gamma {4h}\biggr)\biggr)  
\label{eq: Ssibj} \eea

\bea S^{\bar{s_i} b_j}_{\lambda}(\theta)&=&S^{\ib j}_{\bar{\lambda}\mu}\biggl(\theta-
i\biggl(\frac\pi 2 -\frac\gamma {4h}\biggr)\biggr)
S^{\ib\jb}_{\bar{\lambda}\bar{\mu}}\biggl(\theta+i\biggl(\frac\pi
2 -\frac\gamma {4h}\biggr)\biggr) \cr \cr \cr
&=&S^{ij}_{\lambda\mu}\biggl(-\theta+i\pi +
i\biggl(\frac\pi 2 -\frac\gamma {4h}\biggr)\biggr)
S^{ij}_{\lambda{\mu}}\biggl(\theta+i\biggl(\frac\pi
2 -\frac\gamma {4h}\biggr)\biggr) . \label{eq: Ssibbj}
\eea
We will ignore all the polynomials in $x^h$ of the form
$\prod_p (1-x^h q^p)$.

With this proviso in mind, using the formula (\ref{eq: vij}), so we
also ignore the corrections of the form $(p)_\theta$, we have
\beq
S^{s_i b_j}_\lambda(\theta)={X_q^{ij}\bigl(\theta-i\bigl(\frac\pi 2
-\frac\gamma{4h}\bigr)\bigr)\over X_q^{ij}\bigl(-\theta+i\bigl(\frac\pi 2
-\frac\gamma{4h}\bigr)\bigr)}
{X_q^{ij}\bigl(-\theta+i\bigl(\frac\pi 2
+\frac\gamma{4h}\bigr)\bigr)\over X_q^{ij}\bigl(\theta-i\bigl(\frac\pi 2
+\frac\gamma{4h}\bigr)\bigr)} \eeq

\begin{eqnarray*}
S^{\bar{s_i} b_j}_{\bar{\lambda}}(\theta)&=&{X_q^{ij}\bigl(-\theta+i\pi+i\bigl(\frac\pi 2
-\frac\gamma{4h}\bigr)\bigr)\over X_q^{ij}\bigl(\theta-i\pi -i\bigl(\frac\pi 2
-\frac\gamma{4h}\bigr)\bigr)}
{X_q^{ij}\bigl(\theta+i\bigl(\frac\pi 2
-\frac\gamma{4h}\bigr)\bigr)\over X_q^{ij}\bigl(-\theta-i\bigl(\frac\pi 2
-\frac\gamma{4h}\bigr)\bigr)} \cr \cr \cr
&=&{1\over\prod_p(1-x^{-h}q^p)\prod_r(1-x^hq^r)}
{X_q^{ij}\bigl(-\theta-i\bigl(\frac\pi 2
+\frac\gamma{4h}\bigr)\bigr)\over X_q^{ij}\bigl(\theta + i\bigl(\frac\pi 2
+\frac\gamma{4h}\bigr)\bigr)}
{X_q^{ij}\bigl(\theta+i\bigl(\frac\pi 2
-\frac\gamma{4h}\bigr)\bigr)\over X_q^{ij}\bigl(-\theta-i\bigl(\frac\pi 2
-\frac\gamma{4h}\bigr)\bigr)} \end{eqnarray*}

We see, with the above proviso again in mind, that 
\beq
S^{\bar{s_i} b_j}_{\bar{\lambda}}(\theta)=S^{s_i
b_j}_\lambda(-\theta)^{-1}. \eeq
Then
$$ S^{s_i b_j}_\lambda(\theta)={X_q^{ij}\bigl(\theta-i\bigl(\frac\pi 2
-\frac\gamma{4h}\bigr)\bigr)\over X_q^{ij}\bigl(\theta-i\bigl(\frac\pi 2
+\frac\gamma{4h}\bigr)\bigr)}
{X_q^{ij}\bigl(-\theta+i\bigl(\frac\pi 2
+\frac\gamma{4h}\bigr)\bigr)\over X_q^{ij}\bigl(-\theta+i\bigl(\frac\pi 2
-\frac\gamma{4h}\bigr)\bigr)} $$
can be written as a product of sinh's:

\noindent with $$X_q^{ij}(\theta)=\prod_{p=0}^{h-1}S_{\qt^{-h}}\biggl(e^{is_p\gamma
\over 4h}e^\theta e^{\frac{i\pi}{2h}(c(i)-c(j))}e^{{2\pi ip\over
h}-i\pi}\biggr)^{\gamma_i\cdot\sigma^p\gamma_j\over 2},$$

$$S^{s_i b_j}_\lambda\biggl(\theta+i\biggl(\frac\pi 2-\frac\gamma{4h}\biggr)\biggr)=$$
$$\prod_{p=0}^{h-1}\sinh\biggl(\frac\theta 2 +\frac{\pi i p}h +
\frac{i\pi}{4h}(c(i)-c(j))+\frac{i\gamma}{8h}(s_p-1)\biggr)^{-{\gamma_i\cdot\sigma^p
\gamma_j\over 2}}\qquad\qquad\qquad\qquad$$
\beq \qquad\qquad\qquad\qquad
\times\prod_{p=0}^{h-1}\sinh\biggl(-\frac\theta 2 +\frac{\pi i p}h +
\frac{i\pi}{4h}(c(i)-c(j))+\frac{i\gamma}{8h}(s_p+1)\biggr)^{-{\gamma_i\cdot\sigma^p
\gamma_j\over 2}}.\label{eq: sibj} \eeq

\vskip 3mm
We now use \cite{Fring}
\beq\gamma_i\cdot\sigma^p\gamma_j=\lambda_i\cdot\sigma^p(1-\sigma)\sigma^{c(i)-
1\over 2}\gamma_j, \label{eq: recurs} \eeq
and then, 
$$S^{s_i b_j}_\lambda\biggl(\theta+i\biggl(\frac\pi 2-\frac\gamma{4h}\biggr)\biggr)=$$
$${\prod_{q=1}^{h}\sinh\biggl(\frac\theta 2 +\frac{\pi i q}h -
\frac{i\pi}{4h}(c(i)+c(j))+\frac{\pi i}{2h} +
\frac{i\gamma}{8h}(s_{q-{c(i)-1\over 2}}-1)\biggr)^{-{\lambda_i\cdot\sigma^q
\gamma_j\over 2}}\over
\prod_{q=1}^{h}\sinh\biggl(\frac\theta 2 +\frac{\pi i q}h -
\frac{i\pi}{4h}(c(i)+c(j))-\frac{\pi i}{2h} +
\frac{i\gamma}{8h}(s_{q-{c(i)+1\over 2}}-1)\biggr)^{-{\lambda_i\cdot\sigma^q
\gamma_j\over 2}}}\qquad\qquad\qquad\qquad
$$
$$ \qquad\qquad\qquad\qquad\times
{\prod_{q=1}^{h}\sinh\biggl(-\frac\theta 2 +\frac{\pi i q}h -
\frac{i\pi}{4h}(c(i)+c(j))+\frac{\pi i}{2h} +
\frac{i\gamma}{8h}(s_{q-{c(i)-1\over 2}}+1)\biggr)^{-{\lambda_i\cdot\sigma^q
\gamma_j\over 2}}\over
\prod_{q=1}^{h}\sinh\biggl(-\frac\theta 2 +\frac{\pi i q}h -
\frac{i\pi}{4h}(c(i)+c(j))-\frac{\pi i}{2h} +
\frac{i\gamma}{8h}(s_{q-{c(i)+1\over 2}}+1)\biggr)^{-{\lambda_i\cdot\sigma^q
\gamma_j\over 2}}}.$$

Now 
\bea S^{b_i b_j}(\theta)&=&
S^{s_i b_j}_\lambda\biggl(\theta+i\biggl(\frac\pi
2-\frac\gamma{4h}\biggr)\biggr)S^{\bar{s_i}
b_j}_{\bar{\lambda}}\biggl(\theta-i\biggl(\frac\pi
2-\frac\gamma{4h}\biggr)\biggr) \cr \cr \cr
&=&{S^{s_i b_j}_\lambda\bigl(\theta+i\bigl(\frac\pi
2-\frac\gamma{4h}\bigr)\bigr)\over
S^{s_i b_j}_\lambda\bigl(-\theta+i\bigl(\frac\pi
2-\frac\gamma{4h}\bigr)\bigr)}. \eea
We observe that 
$$S^{b_i b_j}(\theta)={S(\theta)\over S(-\theta)},$$
where
$$
S(\theta)=\prod_{q=1}^h\Biggl({\sinh(\frac\theta 2 +\frac{\pi i q}h -
\frac{i\pi}{4h}(c(i)+c(j))+\frac{\pi i}{2h} +
\frac{i\gamma}{8h}(s_{q-{c(i)-1\over 2}}-1)) \over
\sinh(\frac\theta 2 +\frac{\pi i q}h -
\frac{i\pi}{4h}(c(i)+c(j))-\frac{\pi i}{2h} +
\frac{i\gamma}{8h}(s_{q-{c(i)+1\over 2}}-1))}\qquad\qquad\qquad\qquad
$$ \beq\qquad\qquad\qquad\qquad\times
{\sinh(\frac\theta 2 +\frac{\pi i q}h -
\frac{i\pi}{4h}(c(i)+c(j))-\frac{\pi i}{2h} +
\frac{i\gamma}{8h}(s_{q-{c(i)+1\over 2}}+1)) \over
\sinh(\frac\theta 2 +\frac{\pi i q}h -
\frac{i\pi}{4h}(c(i)+c(j))+\frac{\pi i}{2h} +
\frac{i\gamma}{8h}(s_{q-{c(i)-1\over 2}}+1))}
\Biggr)^{-{\lambda_i\cdot\sigma^q
\gamma_j\over 2}}. \label{eq: star2} \eeq
We have obtained $S(\theta)$ from $S^{s_i
b_j}_\lambda\bigl(\theta+i\bigl(\frac\pi
2-\frac\gamma{4h}\bigr)\bigr)$, by exchanging the factors of the form
$\sinh(-\theta/2 + A)$ in the second half of $S^{s_i
b_j}_\lambda\bigl(\theta +i\bigl(\frac\pi 2 -\frac\gamma{4h}\bigr)\bigr)$
with $\sinh(\theta/2 + A)^{-1}$.
We write this as
\beq 
S^{b_i b_j}(\theta)=\prod_{q=1}^{h}\biggl [2q-{c(i)+c(j)\over
2}\biggr ]^{-{\lambda_i\cdot\sigma^q\gamma_j\over 2}} \label{eq:
square_brackets}, \eeq
with $\biggl [2q-{c(i)+c(j)\over 2}\biggr ]$ defined in the obvious
way as
\beq
\Biggl({(2q - {c(i)+c(j)\over 2} +1 + \frac{\gamma}{4\pi}
(s_{q-{c(i)-1\over 2}} - 1 ))_\theta\over
(2q - {c(i)+c(j)\over 2} - 1 + \frac{\gamma}{4\pi}
(s_{q-{c(i)+1\over 2}}-1))_\theta}
{(2q - {c(i)+c(j)\over 2} -1 + \frac{\gamma}{4\pi}
(s_{q-{c(i)+1\over 2}}+1))_\theta\over
(2q - {c(i)+c(j)\over 2} + 1 + \frac{\gamma}{4\pi}
(s_{q-{c(i)-1\over 2}}+1))_\theta}\Biggr). \label{eq: square_def} \eeq
Suppose that
\beq
s_{q-{c(i)+1\over 2}}=-1,\qquad s_{q-{c(i)-1\over 2}}=1, 
\label{eq: choice1}  \eeq
for $\lambda_i\cdot\sigma^q\gamma_j\neq 0,$ 
then that value of $q$ contributes the block
$$ \biggl [2q-{c(i)+c(j)\over 2}\biggr ]^{-{\lambda_i\cdot\sigma^q
\gamma_j\over 2}} = \Biggl({(2q-{c(i)+c(j)\over 2}+1)_\theta
(2q-{c(i)+c(j)\over 2}-1)_\theta\over
(2q-{c(i)+c(j)\over 2}+1+{\gamma\over 2\pi})_\theta
(2q-{c(i)+c(j)\over 2}-1-{\gamma\over
2\pi})_\theta}\Biggr)^{-{\lambda_i\cdot\sigma^q \gamma_j\over 2}}$$
to $S^{b_i b_j}(\theta)$.
We see how the correct block structure of 
$\{2q-{c(i)+c(j)\over 2}\}_\theta$ has
emerged, and this factor becomes
$$\biggl [2q-{c(i)+c(j)\over 2}\biggr ]^{-{\lambda_i\cdot\sigma^q
\gamma_j\over 2}}=\biggl\{2q-{c(i)+c(j)\over 2}\biggr\}_\theta^
{-{\lambda_i\cdot\sigma^q \gamma_j\over 2}}, $$
after noting that
$$B(\beta)={1\over 2\pi}{\beta^2\over 1+{\beta^2\over 4\pi}}=
-\frac\gamma{2\pi}.$$
The case (\ref{eq: choice1}) can not occur for all values of $q$,
since there are cases when $\lambda_i\cdot\sigma^q\gamma_j\neq 0$, and
$\lambda_i\cdot\sigma^{q+1}\gamma_j\neq 0$, i.e. consecutive blocks. 

However, for the case of the fundamental element $S^{11}(\theta)$,
there are  consecutive blocks with
$\lambda_1\cdot\sigma^q\gamma_1=\pm 1$, and 
$\lambda_1\cdot\sigma^{q+1}\gamma_1=\mp 1$, which occur when
$\gamma_1\cdot\sigma^p\gamma_1=\pm 2$, that is $p=0$, or $p=\frac h2$
(if $\bar{1}=1$), respectively. It is important to note that 
all the remaining blocks are non-consecutive, this can be shown by
inspecting the data $\lambda_1\cdot\sigma^q\gamma_1$, $q=1,\ldots,h$,
on a case by case basis. This data is available, for example, in
\cite{BCDS}, \cite{Dorey2}, or \cite{thesis}.

Suppose that the block $\lambda_1\cdot\sigma^q\gamma_1\neq 0$, and
that it is isolated. Suppose also that $q$ is in the second half of its
range, noting that $\lambda_i\cdot\sigma^q\gamma_j=-\lambda_i\cdot
\sigma^{q'}\gamma_j$, with $q'=h-q+\frac{c(i)+c(j)}2$. It suffices to
show the result in this range. Also in this range $\lambda_1\cdot
\sigma^q\gamma_1>0$. Then, the only way for this to occur, from the
recursion formula (\ref{eq: recurs}), is if 
$$\gamma_1\cdot\sigma^{q-\frac{c(i)+1}2}\gamma_1=-1,\qquad {\rm and}\qquad 
\gamma_1\cdot\sigma^{q-\frac{c(i)-1}2}\gamma_1=1,$$
with $\lambda_1\cdot\sigma^q\gamma_1=1$. In this range 
$$s_{q-\frac{c(i)+1}2}=-1,\qquad {\rm and}\qquad s_{q-\frac{c(i)-1}2}=1,$$
as required to form a correct block at $q$.

Now consider the situation when $\gamma_1\cdot\sigma^p\gamma_1=2$, at
$p=0$, and suppose that $c(1)=1$. There are two consecutive blocks here,
but these two are isolated. From (\ref{eq: recurs})
$$\lambda_1\cdot\sigma^{h-1}\gamma_1=0,\quad\ 
\lambda_1\cdot\sigma^{h}\gamma_1=1
\quad\ \lambda_1\cdot\sigma^{1}\gamma_1=-1
\quad\ \lambda_1\cdot\sigma^{2}\gamma_1=0,$$
it also follows that (since the two blocks are isolated)
$$s_{h-1}=-1,\quad s_0=1,\quad s_1=1.$$ 
The block at $q=h$ is evidently correctly formed, since it satisfies
the condition (\ref{eq: choice1}). The block at $q=1$ is, from
equation (\ref{eq: square_def}),
$$\biggl [2q-1\biggr ]_{q=1}^{-{\lambda_i\cdot\sigma^q
\gamma_j\over
2}}=\Biggl({(1+1)_\theta(1-1+\frac\gamma{2\pi})_\theta\over
(1-1)_\theta(1+1+\frac\gamma{2\pi})_\theta}\Biggr)^{\frac12},$$
but $(x)_\theta=(-x)_{\theta}^{-1}$, and $(0)_\theta=1$, so 
$$=\Biggl({(1+1)_\theta(1-1)_\theta\over
(1-1-\frac\gamma{2\pi})_\theta(1+1+\frac\gamma{2\pi})_\theta}\Biggr)^{\frac12}
=\{1\}_\theta^{\frac12}.$$
Now consider the situation when $\gamma_1\cdot\sigma^p\gamma_1=-2$, at
$p=\frac h2=H$, if $\bar{1}=1$, and suppose that $c(1)=1$.
Then 
$$\lambda_1\cdot\sigma^{H-1}\gamma_1=0,\quad\ 
\lambda_1\cdot\sigma^{H}\gamma_1=-1
\quad\ \lambda_1\cdot\sigma^{H+1}\gamma_1=1
\quad\ \lambda_1\cdot\sigma^{H+2}\gamma_1=0,$$
and $$s_{H-1}=-1,\quad s_H=1,\quad s_{H+1}=1.$$
The block at $q=H$ is correctly formed, and the block at $q=H+1$ is
$$\biggl [2q-1\biggr ]_{q=H+1}^{-{\lambda_i\cdot\sigma^q
\gamma_j\over
2}}=\Biggl({(2H+1+1)_\theta(2H+1-1+\frac\gamma{2\pi})_\theta\over
(2H+1-1)_\theta(2H+1+1+\frac\gamma{2\pi})_\theta}\Biggr)^{-\frac12}.$$
Now $(x)_\theta=(2h-x)^{-1}_\theta$, so 
$$(2H+1-1+\frac\gamma{2\pi})_\theta=(h-\frac\gamma{2\pi})^{-1}_\theta=
(2H+1-1-\frac\gamma{2\pi})^{-1}_\theta, $$
and $(2H+1-1)_\theta=(2H+1-1)^{-1}_\theta$.
Hence 
$$\biggl [2q-1\biggr ]_{q=H+1}^{-{\lambda_i\cdot\sigma^q
\gamma_j\over 2}}=\Biggl({(2H+1+1)_\theta(2H+1-1)_\theta\over
(2H+1-1-\frac\gamma{2\pi})_\theta(2H+1+1+\frac\gamma{2\pi})_\theta}\Biggr)^
{-\frac12},$$
as required. The case $c(1)=-1$ is left to the reader.

This shows that we get the same as the particle S-matrix, in the case
of the fundamental element, $i=j=1$. For the remaining elements, we
merely remark that we can run through the bootstrap on the soliton
result $v^{ij}(\theta)$, and that this is directly equivalent to
running through the bootstrap on the lowest mass breather S-matrix 
starting from $S^{b_1 b_1}$, since the fusing angles for the 
lowest mass breathers
and solitons are precisely the same. The agreement of the result with
the particle S-matrix for
the fundamental element, guarantees that it will agree for all
$i$, and $j$.

We have previously run through the
soliton bootstrap on the ground state solitons,
and obtained the corrections of the form
$(p)_\theta$. These corrections are actually precisely the corrections needed
to turn the incorrectly formed blocks, which occured when they were
consecutive in (\ref{eq: square_def}), into correctly formed ones.

\resection{Discussion and conclusions}
In this paper, we have given exact expressions for the soliton
S-matrices in the simply-laced affine Toda field theories, found by
`$q$-deforming' the classical time delays, by inserting the
regularised quantum dilogarithms. We have then checked crossing and
unitarity, which essentially follow from analogous properties
satisfied by the time delays. The time delays also satisfy the bootstrap
equation, in these Toda theories, which allows the bootstrap property
of the exact S-matrix to be partially checked. However, in checking
the bootstrap, we throw up two types of factor, one of type
$(p)_\theta$, and the other of type $(1-x^hq^p)^{-1}$, and the
expressions found for the S-matrix have to be corrected by multiplying
by factors of both these types.  These additional corrections
trivially satisfy crossing and unitarity, by themselves, except for
the unitarity of the corrections  $(1-x^hq^p)^{-1}$. These are
precisely required in order to satisfy the unitarity condition.
The corrections introduce poles onto the physical strip which either
enhance poles already present, due to fusing solitons, or they are
new. The new ones should have an interpretation in terms of 
Landau singularities.

For the $A_n$ theories, it is possible to determine all the
corrections, since we have exact R-matrices intertwining between
tensor products of all the fundamental representations, and in
addition it is possible to determine the precise normalisation (or
zeroes) of fused R-matrices. For the $A_n$ case, the bootstrap property
is fully checked in this paper. However the paper is incomplete, since
the corrections of type $(1-x^hq^p)^{-1}$ are not determined for the
$D$ and $E$ series. This is because the fusion procedure for the
R-matrices is much more subtle in these cases, and the precise
normalisation or zeroes of fused R-matrices  has not yet been
determined. This means that the bootstrap has only been partially
checked for these cases. 
A crude way of determining them would be to fuse the R-matrices
explicitly using a computer algebra package. This could be difficult,
however, because of the sizes of the matrices involved.

It is also shown in this paper, that the S-matrix of the lowest mass
breathers $S^{b_i b_j}(\theta)$ is the same as the particle S-matrix
$S^{ij}(\theta)$ known previously in the affine Toda field theories.
In this calculation, we have ignored the factors of the type
$(1-x^hq^p)^{-1}$, these include the individual elements of the
R-matrix. This is an important independent check on the
solutions. The excited solitons are not discussed much, beyond
pointing out that they exist. It would be interesting to try to show
full closure of the bootstrap on all these excited states.

The reader may also have noticed that quantum groups and R-matrices
have not been discussed in any great detail. In particular, we have
avoided all discussion of any deeper relation between affine quantum
groups and the symmetries of quantum integrable models. The author
believes that such a relation may indeed be true, but since the
S-matrix must satisfy the quantum Yang-Baxter equation, the results
presented here are valid, even if the relation is not true, provided
the missing charge problem can be solved.
It will be the subject of future publications to study this relation,
and to see how the quantum dilogarithms appear naturally in a
quantisation of the models. Recall that, classically, the time delays,
$X^{ij}(\theta)$, appeared quite naturally from the vertex operator
representations of the affine Kac-Moody algebras.

\vskip 0.4in
\noindent{\bf Acknowledgements}

Prof. David Olive brought the regularised quantum dilogarithm to my
attention, after seeing it at Faddeev's 60$^{\rm th}$ birthday
celebration. Thanks go to Tim Hollowood for pointing out how the
zeroes which occur in the bootstrap can be computed in the $A_n$
case. Thanks must also go to Edwin Beggs for encouragement.

The author was mainly 
supported by EPSRC, in the form of a studentship under
which most of this work was carried out. Further work was conducted,
and the paper written, whilst the author was supported by PPARC.

\end{document}